\documentclass[a4paper,11pt]{article}
\usepackage[skins]{tcolorbox}
\usepackage{mathtools,slashed}
\usepackage[T1]{fontenc}
\usepackage{slashed,verbatim,subfigure}
\usepackage[numbers,sort&compress]{natbib}
\usepackage{amsmath}
\usepackage{mathrsfs}
\usepackage{amsbsy}
\usepackage{amssymb}
\usepackage{appendix}
\usepackage{graphicx}
\usepackage[final]{pdfpages}
\usepackage{dcolumn}
\usepackage{bm}
\usepackage[colorlinks=true,linktocpage=true,
linkcolor=blue,citecolor=blue]{hyperref}
\usepackage[a4paper]{geometry}
\catcode`@11
\def\seceqaa{\@addtoreset{equation}{section}
	\def\theequation{A\arabic{equation}}}
\def\seceqbb{\@addtoreset{equation}{section}
	\def\theequation{B\arabic{equation}}}
\def\seceqcc{\@addtoreset{equation}{section}
	\def\theequation{C\arabic{equation}}}
\def\seceqdd{\@addtoreset{equation}{section}
	\def\theequation{D\arabic{equation}}}
\def\seceqee{\@addtoreset{equation}{section}
	\def\theequation{E\arabic{equation}}}
\catcode`@11
\newcommand{\be}{\begin{eqnarray}}
\newcommand{\ee}{\end{eqnarray}}
\begin{document}
\large
\title{${\mathscr {M}}$cTEQ (${\mathscr {M}}$ {\bf c}hiral perturbation theory-compatible deconfinement {\bf T}emperature and {\bf E}ntanglement Entropy up to terms {\bf Q}uartic in curvature) and FM ({\bf F}lavor {\bf M}emory) }
\author{Gopal Yadav\footnote{email- gyadav@ph.iitr.ac.in}, ~Vikas Yadav\footnote{email- vyadav@ph.iitr.ac.in} ~~and~~Aalok Misra\footnote{email- aalok.misra@ph.iitr.ac.in}\vspace{0.1in}\\
Department of Physics,\\
Indian Institute of Technology Roorkee, Roorkee 247667, India}
\date{}
\maketitle
\begin{abstract}
A (semiclassical) holographic computation of the deconfinement temperature at {\it intermediate coupling} from (a top-down) $ {\mathscr {M}}$-theory dual of thermal QCD-like theories, has been missing in the literature. In the process of filling this gap,  we demonstrate a novel UV-IR connection, (conjecture and provide evidence for) a non-renormalization beyond one loop of $\pmb {\mathscr {M}}-{\bf c}$hiral perturbation theory \cite{MChPT}-compatible deconfinement {\bf T}emperature, and show equivalence with an {\bf E}ntanglement (as well as Wald) entropy \cite{Tc-EE} computation, up to terms {\bf Q}uartic in curvature ($R$).  We demonstrate a {\bf F}lavor-{\bf M}emory (FM) effect in the $ {\mathscr {M}}$-theory uplifts of the gravity duals,  wherein the no-braner $ {\mathscr {M}}$-theory uplift retains the "memory" of the flavor $D7$-branes of the parent type IIB dual in the sense that a specific combination of the aforementioned quartic corrections to the metric components precisely along the  compact part (given by $S^3$ as an $S^1$-fibration over the vanishing two-cycle $S^2$) of the non-compact four-cycle "wrapped" by the flavor $D7$-branes, is what determines, e.g., the Einstein-Hilbert action at O$(R^4)$.  The aforementioned linear combination of ${\cal O}(R^4)$ corrections to the ${\mathscr {M}}$-theory uplift \cite{MQGP,NPB} metric, upon matching the holographic result from ${\mathscr {M}}\chi$PT \cite{MChPT} with the phenomenological value of the coupling constant of one of the $SU(3)$ NLO $\chi$PT Lagrangian of \cite{GL}, is required to have a definite sign. Interestingly, in the decompactification (or ``$M_{\rm KK}\rightarrow0$") limit of the spatial circle in \cite{MChPT} to recover a QCD-like theory in four dimensions after integrating out the compact directions, we not only derive this, but in fact obtain the values of the relevant  ${\cal O}(R^4)$ metric corrections. Further, equivalence with Wald entropy for the black hole in the high-temperature ${\mathscr {M}}$-theory dual at ${\cal O}(R^4)$ imposes a linear constraint on a similar linear combination of the abovementioned metric corrections. Remarkably, when evaluating the deconfinement temperature from an entanglement entropy computation in the thermal gravity dual, due to a delicate cancellation  between the contributions arising  from the metric corrections at ${\cal O}(R^4)$ in the ${\mathscr{M}}$ theory uplift along the $S^1$-fiber and an  $S^2$ (which too involves a similar $S^1$-fibration) resulting in a non-zero contribution only along the vanishing $S^2$ surviving, one sees that there are consequently no corrections to $T_c$ at quartic order in the curvature supporting the conjecture made on the basis of a semiclassical computation.

\end{abstract}

\newpage
\tableofcontents

\section{Introduction} 
 AdS/CFT correspondence \cite{AdS/CFT}, in its original form,  was a duality between strongly coupled $\cal N$ = 4 $SU(N)$ supersymmetric Yang-Mills theory  and weakly coupled type IIB string theory on $AdS_5 \times S^5$. All the theories are not conformal therefore it is better to call this duality as gauge/gravity duality to include non-conformal theories. Strong coupling dynamics of non-abelian gauge theories at finite temperature have been studied via gauge/gravity duality. In later years, this duality has been generalised to many branches of Physics (e.g. Particle Physics, Condensed Matter Physics, Cosmology etc.). We are interested in thermal QCD, which is a non-conformal theory because its gauge coupling runs with energy. There are various proposals to study QCD using gauge/gravity duality (e.g. AdS/QCD) but almost all involve a conformal AdS background. There are two approaches to construct holographic duals of thermal QCD-like theories - the bottom-up and the top-down approach. In this paper we will work with the latter.\par
Gauge/gravity duality also allows us to compute the corrections to the infinite-'t Hooft-coupling limit as done  in \cite{SGS}, but again working with an AdS background. These authors included higher-derivative terms on gravity side using terms quartic in the Weyl tensor. On the gauge theory side one has $\cal N$ = 4 $SU(N)$  supersymmetric Yang-Mills plasma at intermediate 't Hooft coupling. They explained the transport peak in the
small frequency region of the stress-energy tensor specatral function at zero spatial momentum. which is a generic feature of perturbative plasma.
\par
QCD at strong coupling has  been studied by various bottom-up models constructed from gauge/gravity duality(from both, bottom-up and top-down approaches) and at weak coupling from perturbation theory. A popular top-down holographic type IIA dual, though catering only to the IR, is the Sakai-Sugimoto model \cite{SS}. The only UV-complete (type IIB) top-down holographic dual of QCD-like theories at strong coupling that we are aware of is \cite{metrics}, and its ${\mathscr {M}}$-theory uplift \cite{MQGP}. Authors in \cite{IC} have studied QCD thermodynamic functions at intermediate coupling based on hard-thermal-loop perturbation theory (HTLpt). One can also study intermediate coupling regime of QCD from gauge/gravity duality. In this direction, two of the authors (VY and AM) worked out $O(l_p^6)$ corrections to the ${\mathscr {M}}$-theory metric in \cite{OR4-Yadav+Misra}. Starting with ${\mathscr {M}}$-theory dual of thermal QCD-like theories in the 'MQGP' limit as constructed in \cite{MQGP} and incorporating  higher derivative terms in eleven dimensional supergravity action which are quartic in Riemann curvature tensor i.e. $\cal O$($R^{4}$), the corrections to the supergravity background were then worked out in \cite{OR4-Yadav+Misra} and in fact, successfully used in \cite{MChPT} in obtaining phenomenologically-compatible values of the NLO LECs in $SU(3)\ \chi$PT Lagrangian of \cite{GL}.\par 
Wald proposed a method to calculate black hole entropy in general theories of gravity \cite{Wald-Entropy}. He considered classical theory of gravity in $n$ dimensions which arises from diffeomorphism invariant Lagrangian. In these theories of gravity Noether charge is an $(n-2)$-form and black hole entropy will be given by 2$\pi$ times integral over  bifurcate killing horizon of the $(n-2)$-form Noether charge. Therefore entropy is Noether Charge for stationary black holes with bifurcate killing horizons. For dynamical black holes a proposal to calculate entropy was given in \cite{Wald-Entropy-2}.
 \par Every physical phenomenon is characterised by their energy/distance scale.  One goes from UV to IR via RG group flow. But, e.g., in noncommutative field theories and string theory, short distance physics becomes related to long distance Physics, which is known as UV/IR mixing. The consequence of UV/IR mixing  \cite{UV_IR} is, e.g., that UV divergences of real $\phi^4$ theory defined on commutative space are transformed into infrared poles in the same theory defined on noncommutative space. There are other more examples explained nicely in \cite{UV_IR}. \par
 Gravitational theories are nonlocal. Therefore UV/IR mixing could appear in such theories.  In gauge/gravity duality radial coordinate in gravitational theory corresponds to energy scale in gauge theory side. In \cite{Susskind+Witten}, authors have studied "I.R.-U.V." connection in the context of AdS/CFT correspondance in which they showed that infrared effects in bulk theory are transformed into ultravoilet effects in the boundary theory. In particular, infrared regulator in the bulk theory plays the role of ultraviolet regulator in $\cal N$ = 4 super Yang Mills theory.  It is interesting that a certain form of UV-IR connection  manifests itself in our work when matching at the deconfinement temperature, ${\mathscr {M}}$-theory actions dual to the thermal and black-hole backgrounds at the UV-cut-off and obtain a relationship between the ${\cal O}(R^4)$ metric corrections in the IR\par
In this paper we have calculated deconfinement temperature of QCD-like theory at intermediate coupling from gauge/gravity duality using a semiclassical computation as first discussed in \cite{Witten-Hawking-Page-Tc}. In our case gravity dual is ${\mathscr {M}}$-theory uplift of type IIB string dual \cite{metrics} which includes ${\cal O}(R^4)$ corrections \cite{OR4-Yadav+Misra} to the MQGP background \cite{MQGP}.
One can also discuss confinement deconfinement phase transition in large $N_c$ gauge theories from entanglement entropy point of view based on \cite{Tc-EE}. In this process one is required to calculate entanglement entropy between two regions by dividing one of spatial coordinates into segment of length ${\it l}$ and its complement. There is a prescription to calculate entanglement entropy from AdS/CFT correspondence given by Ryu and Takayanagi in \cite{RT}. As discussed in \cite{Tc-EE}, there are two surfaces - connected and disconnected. There is a critical value of ${\it l}$ which is denoted by ${\it l_{crit}}$. If one is below the critical value of ${\it l}$ i.e. ${\it l}<{\it l_{crit}}$ then it is the connected surface that dominates the entanglement entropy and if one is above the critical value of ${\it l}$ i.e. ${\it l}>{\it l_{crit}}$ then it is the disconnected surface that dominates entanglement entropy; ${\it l}<{\it l_{crit}}$ corresponds to confining phase of large $N_c$ gauge theories whereas ${\it l}>{\it l_{crit}}$ corresponds to deconfining phase of the same. So we can interpret this as confinement deconfinement phase transition in large $N_c$ gauge theories.\par
The rest of the paper is organized as follows. In Section {\bf 2}, we review the basic setup and results of \cite{metrics, MQGP, OR4-Yadav+Misra} inclusive of a summary of the results of \cite{Green and Gutperle, Green and Vanhove} on the conjectured $SL(2,\mathbb{Z})$ completion of the $D=11$ supergravity action by studying four-graviton scattering in a D-instanton background and a non-renormalization beyond one loop at ${\cal O}(R^4)$ in the zero-instanton sector. Section {\bf 3}, divided into two sub-sections ${\bf 3.1}$ and ${\bf 3.2}$, have to do with the holographic renormalization of the on-shell action corresponding respectively to the gravity duals of the black-hole and thermal backgrounds. Section {\bf 4} is on the computation of the deconfinement temperature via a semiclassical computation, as well as on the comparison of the black hole entropy computed  with the Wald entropy up to ${\cal O}(R^4)$. Section {\bf 5} is on the computation of the deconfinement temperature from entanglement entropy between an interval along one of the non-compact non-radial spatial coordinates and its complement, and comparison with the semiclassical computation of the same in Section {\bf 4}. 
Section {\bf 6} is on the derivation of the sign and the individual values of the contributions that figure  in a linear combination of the same arising from the ${\cal O}(R^4)$ corrections to the  
${\mathscr {M}}$-theory uplift \cite{MQGP} of \cite{metrics}. Sections {\bf 4} and {\bf 6} also include a discussion on the "Flavor Memory" effect. Section {\bf 7} has a summary of the results obtained in the paper. There is a supplementary appendix on the EOMs and solutions to the ${\cal O}(R^4)$ metric corrections to the MQGP background of \cite{MQGP,NPB}, as well as a discussion on taking the decompactification-limit of a periodic spatial direction  in the thermal background ${\mathscr {M}}$-theory uplift used in \cite{MChPT} to obtain the ${\mathscr {M}}$-theory thermal background in this paper, and deriving the sign of a linear combination of constants of integration in the solutions to the EOMs of the ${\cal O}(R^4)
 {\mathscr {M}}$-theory metric corrections as required from matching of results of \cite{MChPT} with phenomenological values of the LECs at NLO in the $SU(3)\ \chi$PT Lagrangian of \cite{GL}.

\section{Review of UV Complete Top-Down Type IIB String Dual of \cite{metrics}, Type IIA Mirror of \cite{metrics} and its ${\mathscr {M}}$-Theory Uplift at Intermediate 't Hooft Coupling}

In this section we will start with a brief review type IIB string dual of large $N$ QCD-like theory at finite temperature constructed by McGill group \cite{metrics}. Then we will briefly discuss type IIA SYZ mirror of \cite{metrics} and its ${\mathscr {M}}$-theory uplift as constructed in \cite{MQGP}. We will also discuss the origin of ${\cal O}(R^4)$ terms in eleven dimensional supergravity action and non-renormalization  of type IIB action up to ${\cal O}(R^4)$ beyond one loop in the zero-instanton sector. In the last part of this section we will explain that how the ${\mathscr {M}}$-theory metric of \cite{MQGP,NPB} will be modified by incorporating higher derivative terms in eleven dimensional supergravity action.
\par 
To the best of our knowledge the only UV-complete type IIB string dual of large $N$ thermal QCD-like theories from a top-down approach was given in \cite{metrics} in the infinite-'t Hooft-coupling limit. For strongly coupled systems like QGP, the relavant coupling is finite/intermediate as explained in \cite{Natsuume}. Therefore, it is required to construct string dual(s) of large $N$ thermal QCD-like theories at finite/intermediate coupling. In this direction authors in \cite{MQGP,NPB} considered a particular limit which they called as 'MQGP limit' which uses finite/intermdiate coupling.  One can also study intermediate coupling regime of gauge theory using gauge/gravity duality. To study thermal QCD at intermediate coupling higher derivative corrections on the gravity side, need to be incorporated - \cite{SGS, OR4-Yadav+Misra}. 

{\bf Brane Setup used in \cite{metrics}}
\begin{itemize}
\item The authors considered $N$ $D3$-branes placed at tip of the six-dimensional conifold, $M\ D5$-branes wrapping the vanishing $S^2(\theta_1,\phi_1)$, referred to as fractional $D3$-branes and $M\ \overline{D5}$-branes  distributed along the resolved $S^2_a(\theta_2,\phi_2)$ placed at antipodal points relative to the $M$ $D5$-branes. Let average seperation between $D5/\overline{D5}$-branes will be represented by ${\cal R}_{D5/\overline{D5}}$. Then one can characterise the various regions using radial coordinate on gravity side.
\begin{enumerate}
\item $r_0<r<{\cal R}_{D5/\overline{D5}} $ : IR-UV interpolating region with $r_0<r\ll {\cal R}_{D5/\overline{D5}}$ corresponding to the deep IR.
\item $r>{\cal R}_{D5/\overline{D5}}$ : UV region.
\end{enumerate}
\item To introduce quarks in the fundamental representation of the flavor group, $N_f$ $D7$-branes were introduced via Ouyang embedding \cite{ouyang} in the resolved conifold geometry, ``smeared"/delocalized along the angular directions $\theta_{1,2}$. The flavor $D7$-branes are embedded in the UV all the way into the IR up to a certain minimum separation from the color $D3$-branes determined by the modulus of the Ouyang embedding parameter $|\mu_{\rm Ouyang}|$.
To ensure UV conformality in the theory $N_f\ \overline{D7}$-branes were introduced in the UV and the UV-IR interpolaring region but not the IR (to effect chiral symmetry breaking). The embedding equation for the flavor $D7$-branes in the resolved conifold geometry is:
\begin{equation}
\label{Ouyang-definition}
\left(r^6 + 9 a^2 r^4\right)^{\frac{1}{4}} e^{\frac{i}{2}\left(\psi-\phi_1-\phi_2\right)}\sin\left(\frac{\theta_1}{2}\right)\sin\left(\frac{\theta_2}{2}\right) = \mu_{\rm Ouyang},
\end{equation}
effected by (\ref{small-theta_12}) for vanishingly small $|\mu_{\rm Ouyang}|$.

\item
In the UV, the color gauge group is $SU(N+M)\times SU(N+M)$ and the flavor gauge group is $SU(N_f)\times SU(N_f)$. When one goes RG flows from the UV to the IR, two things happen,
\begin{itemize}
\item 
 The color gauge group $SU(N+M)\times SU(N+M)$ is partially Higgsed down to $SU(N+M)\times SU(N)$ because in the IR $\overline{D5}$-branes are not present due to which rank of one of the product gauge groups($SU(N + {\rm number\ of}\ D5-{\rm branes})\times SU(N + {\rm number\ of}\ \overline{D5}-{\rm branes})$) decreases to ($SU(N + {\rm number\ of}\ D5-{\rm branes})\times SU(N)$.
\item Flavor gauge group $SU(N_f)\times SU(N_f)$ breaks to the diagonal subgroup $SU(N_f)$ because of absense of $\overline{D7}$-branes in the IR. This is the analagoue of chiral symmetry breaking in this brane setup.
\end{itemize}
\item
The pair of couplings corresponding to $SU(N+M)$ and $SU(N)$ gauge groups flow in opposite directions in the IR.
\begin{eqnarray}
\hskip -0.3in 4\pi^2 \left( \frac{1}{g^2_{SU(N+M)}}+\frac{1}{g^2_{SU(N)}}\right)e^\phi \sim \pi ; 
4\pi^2 \left( \frac{1}{g^2_{SU(N+M)}} - \frac{1}{g^2_{SU(N)}}\right)e^\phi \sim \frac{1}{2\pi\alpha^{'}} \int_{S^2} B_2
\end{eqnarray}
 From the above equation it is clear that $\int_{S^2} B_2$ is the reason for introduction of non-conformality. If flux term on r.h.s. vanishes then couplings for the both gauge groups will be same which is indication of a conformal theory. This is the reason to include $M$ $\overline{D5}$-branes in \cite{metrics} to cancel the net $D5$-brane charge in the UV. {\it Therefore we have a holographic  theory which, like QCD, is UV conformal}.
\item Under the UV-to-IR RG flow, the higher rank gauge group($SU(N+M)$) flows towards strong coupling and lower rank gauge group($SU(N)$) flows towards the weak coupling. Since $SU(N+M)_{strong} \xrightarrow{\text {Sieberg-like dual}} SU(N-(M-N_f))_{weak}$, after performing repeated Seiberg-like dualities, in the IR the number of colors $N_c$ gets identified with $M$, which in the `MQGP limit' can be taken to be 3 (See \cite{Misra+Gale}). 'MQGP limit' have been defined later in this paper.
\item
{\bf Holographic gravity dual of the brane setup of \cite{metrics}}: If one is above the deconfinement temperature i.e. $T>T_c$ then finite temperature on the gauge/brane side corresponds to a black hole in the gravitational dual and if one is below the deconfinement temperature i.e. $T<T_c$ then finite temperature on the gauge/brane side corresponds to thermal background in the gravitational dual \cite{Witten-Hawking-Page-Tc}. Finite temperature on the brane/gauge side and finite seperation between the $M\ D5$-branes and $M\ \overline{D5}$-branes (which is denoted by ${\cal R}_{D5/\overline{D5}}$) requires to having a non-zero resolution parameter of the conifold in gravitational dual side. Similarly, IR confinement on the gauge theory side corresponds to deformation of the conifold singular \`{a} la the Klebanov-Strassler model. {\it Therefore holographic dual of thermal QCD-like theories in this model \cite{metrics} involves a resolved warped deformed conifold;  $D3$-branes and the $D5$-branes are replaced by fluxes in the IR, and the back-reactions are included in the warp factor and fluxes}.
\item
{\bf Color-Flavor Enhancement of Length Scale in the IR}:In the MQGP limit (\ref{MQGP_limit}), there is color-flavor enhancement of length scale as compared to a Planckian length scale in Klebanov-Strassler(KS)-like model even for ${\cal O}(1)$ $M$, in the IR . This is true when one includes terms higher order in $g_s N_f$  in the RR and NS-NS three-form fluxes and the NLO terms in $N$ in the metric. which indicates suppression quantum corrections and validity of supergravity calculations. For a detailed discussion of this issue, see \cite{NPB,Misra+Gale}. Now, the effective number of color branes $N_{\rm eff}(r)$ is given by:
\begin{eqnarray}
\label{NeffMeffNfeff}
& & N_{\rm eff}(r) = N\left[ 1 + \frac{3 g_s M_{\rm eff}^2}{2\pi N}\left(\log r + \frac{3 g_s N_f^{\rm eff}}{2\pi}\left(\log r\right)^2\right)\right],\nonumber\\
& & M_{\rm eff}(r) = M + \frac{3g_s N_f M}{2\pi}\log r + \sum_{m\geq1}\sum_{n\geq1} N_f^m M^n f_{mn}(r),\nonumber\\
& & N^{\rm eff}_f(r) = N_f + \sum_{m\geq1}\sum_{n\geq0} N_f^m M^n g_{mn}(r).
\end{eqnarray}
Type IIB axion is $C_0 =N_f^{\rm eff} \frac{\left(\psi - \phi_1-\phi_2\right)}{4\pi}$,
$N_{\rm eff}(r_0\in\rm IR)=0$ and writing the ten-dimensional warp factor $h \sim \frac{L^4}{r^4}$,  the length scale $L$ in the IR will be given by the following equation:
\begin{eqnarray}
\label{length-IR}
& & \hskip -0.9in L\sim\sqrt[4]{M}N_f^{\frac{3}{4}}\sqrt{\left(\sum_{m\geq0}\sum_{n\geq0}N_f^mM^nf_{mn}(r_0)\right)}\left(\sum_{l\geq0}\sum_{p\geq0}N_f^lM^p g_{lp}(r_0)\right)^{\frac{1}{4}} L_{\rm KS},
\end{eqnarray}
$L_{KS}\sim \sqrt[4]{g_sM}\sqrt{\alpha^\prime}$. Equation (\ref{length-IR}) implies enhancement of color-flavor length scale in the IR as compared to KS.
Hence if we consider number of colors $N_c^{\rm IR}=M=3$ and number of flavors $N_f=2(u/d)+1(s)$, inclusion of $n,m>1$  terms in
$M_{\rm eff}$ and $N_f^{\rm eff}$ in (\ref{NeffMeffNfeff}) implies that $L\gg L_{\rm KS}(\sim L_{\rm Planck})$ in the MQGP limit (\ref{MQGP_limit}), which indicates {\it validity of supergravity calculations}.

\item
{\bf Obtaining} ${\bf N_c=3}$: 
Now based on \cite{Misra+Gale}  we will briefly summarize how to identify number of colors $N_c$ with $M$ which in the `MQGP limit' (\ref{MQGP_limit})  can be tuned to equal 3. One can write $N_c$ as sum of the effective number $N_{\rm eff}$ of $D3$-branes and the effective number $M_{\rm eff}$ of $D5$-branes as:
\begin{equation}
\label{N_c}
N_C = N_{\rm eff}(r) + M_{\rm eff}(r)
\end{equation}
where  $N_{\rm eff}(r)$ is defined via the following relation,
\begin{equation}
\tilde{F}_5\equiv dC_4 + B_2\wedge F_3 = {\cal F}_5 + *{\cal F}_5
\end{equation}
where, ${\cal F}_5\equiv N_{\rm eff}\times{\rm Vol}({\rm Base\ of\ Resolved\ Warped\ Deformed\ Conifold})$. Similarly, $M_{\rm eff}$ is defined via the following relation,  
\begin{equation}
M_{\rm eff} = \int_{S^3}\tilde{F}_3
\end{equation}
 In the above equation, $S^3$ being dual to $\ e_\psi\wedge\left(\sin\theta_1 d\theta_1\wedge d\phi_1 - B_1\sin\theta_2\wedge d\phi_2\right)$, $B_1$ is an `asymmetry factor' defined in \cite{metrics}; $e_\psi\equiv d\psi + {\rm cos}~\theta_1~d\phi_1 + {\rm cos}~\theta_2~d\phi_2$) and \cite{M(r)N_f(r)-Dasgupta_et_al}: $\tilde{F}_3 (\equiv F_3 - \tau H_3)\propto M(r)\equiv M\frac{1}{1 + e^{\alpha(r-{\cal R}_{D5/\overline{D5}})}}, \alpha\gg1.$
 
 Let us denote UV and IR values of the resolution parameter "$a$" as $[a_{UV},a_{IR}]$. In this notation,
\begin{equation}
\label{ranges}
N_{\rm eff} \in[N,0] \hskip 0.2in and \hskip 0.2in M_{\rm eff} \in [0,M].
\end{equation}
From equations (\ref{N_c}) and (\ref{ranges}), it is clear that
\begin{equation}
N_c \in [M, N].
\end{equation}
This implies that in the IR number of colors i.e. $N_c$ is $M$ which in the MQGP limit we have taken to equal 3.\par
{\it Therefore, we see that after application of repeated Seiberg-like duality $N\ D3$-branes are cascaded away in the IR and we have finite $M$ corresponding to a strongly coupled IR-confining $SU(M)$ gauge theory}; authors in \cite{metrics} considered finite temperature version this theory. {\it The holographic dual of large $N$ thermal QCD-like theories  constructed in \cite{metrics} exhibits UV conformality (no Landau poles), IR confinement, quarks transform in the fundamental representation of flavor and color groups, and is valid at all temperatures.}
 
\item {\textbf{Type IIA Strominger-Yau-Zaslow (SYZ) Mirror of \cite{metrics}, ${\mathscr {M}}$-Theory Uplift and the MQGP Limit}} 

Type IIA SYZ mirror of \cite{metrics} and its ${\mathscr {M}}$-theory uplift using Witten's prescription at intermediate gauge coupling but large 't Hooft coupling, have been worked out in \cite{MQGP,NPB}. The $O(R^4)$ corrections to the MQGP metric for the black hole background corresponding to intermediate 't Hooft coupling,  were worked out by two of the authors (VY, AM) in \cite{OR4-Yadav+Misra},  and for the thermal background the $O(R^4)$ corrections to the MQGP metric are given in (\ref{solutions-fMN}). Let us discuss how to obtain ${\mathscr {M}}$-theory uplift at intermediate coupling.
\begin{itemize}
\item To implement Strominger-Yau-Zaslow (SYZ) mirror symmetry one needs a (delocalized) special Lagrangian (sLag) $T^3$ $-$ which could be identified with the $T^2$-invariant sLag of \cite{M.Ionel and M.Min-OO (2008)} with a large base ${\cal B}(r,\theta_1,\theta_2)$ (of a $T^3(\phi_1,\phi_2,\psi)$-fibration\footnote{As there is no $\psi$-isometry in the warped resolved deformed conifold, we in fact require a delocalized $T^3(x,y,z)$, where $(x, y, z)$ are given by (\ref{xyz-defs}).} over ${\cal B}(r,\theta_1,\theta_2)$) \cite{NPB,EPJC-2}. SYZ mirror symmetry is triple T duality along three isometry directions (in our case T-duality along $T^3(\phi_1,\phi_2,\psi)$). Let us see the effect of three T-dualities along $(\phi_1,\phi_2,\psi)$ directions.
\begin{itemize} 
\item \textbf{T-duality along $\psi$:} First T-duality along the $\psi$ direction converts $N\ D3$-branes into $N\ D4$-branes wrapping the $\psi$ circle, $M$ fractional $D3(\overline{D3})$-branes into $M\ D4(\overline{D4})$-branes straddling a pair of orthogonal $NS5$-branes and $N_f$ flavor $D7(\overline{D7})$-branes into $N_f$ flavor $D6(\overline{D6})$-branes. Worldvolume coordinates for $NS5$ branes are $NS5_1\ (x^{0,1,2,3},\theta_1,\phi_1)$ and $NS5_2\ (x^{0,1,2,3},\theta_2,\phi_2)$.
\item \textbf{T-duality along $\phi_1$:} Second T-duality along $\phi_1$ direction leaves $NS5_1\ (x^{0,1,2,3},\theta_1,\phi_1)$ invariant but converts $NS5_2\ (x^{0,1,2,3},\theta_2,\phi_2)$ into a Taub-NUT space$(r,\psi,\theta_2,\phi_1)$. Also $N\ D4$-branes, $M\ D4(\overline{D4})$-branes and $N_f$ flavor $D6(\overline{D6})$-branes will be converted into $N\ D5$-branes, $M\ D5(\overline{D5})$-branes and $N_f$ flavor $D5(\overline{D5})$-branes.
\item \textbf{T-duality along $\phi_2$:} Third T-duality along $\phi_2$ direction leaves $NS5_2\ (x^{0,1,2,3},\theta_2,\phi_2)$ invariant but converts $NS5_1\ (x^{0,1,2,3},\theta_1,\phi_1)$ into a Taub-NUT space$(r,\psi,\theta_1,\phi_2)$. Also $N\ D5$-branes, $M\ D5(\overline{D5})$-branes and $N_f$ flavor $D5(\overline{D5})$-branes will be converted into $N\ D6$-branes, $M\ D6(\overline{D6})$-branes and $N_f$ flavor $D6(\overline{D6})$-branes(which are "wrapping" a non-compact three-cycle  $\Sigma^{(3)}(r, \theta_1, \phi_2$)).
\end{itemize}  
Therefore we see that in SYZ type IIA mirror we have $N\ D6$-branes, $M\ D6(\overline{D6})$-branes and $N_f$ flavor $D6(\overline{D6})$-branes. Keep in mind that in type IIA mirror ranges for the presence of anti-branes also matter as in \cite{metrics}. In actual calculation as done in \cite{MQGP}, T-dualities are performed along $T^3(x,y,z)$, where $(x,y,z)$ are toroidal analagoue of $(\phi_1,\phi_2,\psi)$. For the convention we have just written $(\phi_1,\phi_2,\psi)$. Upon uplifting the SYZ type IIA mirror as obtained above to ${\mathscr {M}}$-theory  using Witten's prescription, $D6$-branes will be converted into KK monopoles(variants of Taub-NUT spaces). Hence, all the branes will be converted into geometry and fluxes. One can show that the {\it${\mathscr {M}}$-theory uplift involves a $G_2$-structure manifold \cite{NPB,OR4-Yadav+Misra}.} 
  
 \item \textbf{MQGP Limit}\\
 MQGP limit is defined as \cite{MQGP,NPB}
 \begin{equation}
\label{MQGP_limit}
g_s\sim\frac{1}{{\cal O}(1)}, M, N_f \equiv {\cal O}(1),\ g_sN_f<1,\ N\gg1,\ \frac{g_s M^2}{N}\ll1,
\end{equation}
wherein one considers intermediate/finite string coupling. This hence necessitates addressing the MQGP limit from ${\mathscr {M}}$-theory.
\end{itemize}

\item In \cite{OR4-Yadav+Misra}, two of the authors(VY and AM) worked out $O(l_p^6)$ corrections to the MQGP metric for blackhole background by incorporating $O(R^4)$ terms in eleven dimensional supergravity action. Since in gravitational dual side we have higher derivative corrections which on gauge theory side corresponds to intermediate gauge/'t Hooft coupling. Therefore using \cite{OR4-Yadav+Misra} we can explore the intermediate coupling regime of thermal QCD-like theories. The $SU(3)/G_2/SU(4)/Spin(7)$-structure torsion classes of the relevant six-, seven- and eight-folds associated with the ${\mathscr {M}}$-theory uplift worked out in \cite{OR4-Yadav+Misra}.
\end{itemize}

There are two ways of understanding the origin of the ${\cal O}(R^4)$-corrections to the ${\cal N}=1, D=11$ supergravity action. One is in the context of the effects of $D$-instantons  in IIB supergravity/string theory via the four-graviton scattering amplitude \cite{Green and Gutperle}. The other is $D=10$ supersymmetry \cite{Green and Vanhove}. Let us briefly discuss both.

\begin{itemize}
\item
Let us first look at interactions that are induced at leading order in a $D$-instanton (closed-string states of type IIB superstring theory in which the whole string is localized at a single point in superspace) background   in both, type IIB supergravity and the string descriptions, including a one-instanton correction to the tree-level  as well as one-loop $R^4$ terms \cite{Green and Gutperle} - both having the same tensorial structure. The bosonic zero modes are parameterised by the coordinates corresponding to the position of the $D$-instanton. The fermionic zero modes are generated by the broken supersymmetries.  The Grassmann parameters are fermionic  supermoduli  corresponding to zero modes of  the dilatino  and
must be integrated over together with the bosonic zero modes. The simplest open-string world-sheet that arises in a D-brane
process is the disk diagram.  An instanton carrying some zero  modes corresponds,  at lowest
order, to a disk world-sheet with open-string states attached to the boundary.  The one-instanton terms in the supergravity effective action  can be deduced by considering on-shell amplitudes in the instanton background.      The
integration over the fermionic moduli absorbs  the  independent
fermionic zero modes.  Consider now amplitudes with four external gravitons. The leading
term  in supergravity is one in which each graviton is associated with
four fermionic zero modes. Integration over the bosonic zero modes  generates a nonlocal four-graviton interaction. In the corresponding string calculation  the world-sheet  consists of four disconnected disks to each of which is  attached a single closed-string graviton vertex and four fermionic open-string vertices. Writing the graviton polarization tensor as $\zeta^{\mu_r\nu_r} =
\zeta^{(\mu_r} \tilde \zeta^{\nu_r)}$, and evaluating the fermionic integral, it was shown in  \cite{Green and Gutperle} that the four-graviton scattering amplitude can be expressed in terms of:
\begin{eqnarray}
\label{A}
 &  & C  e^{2i\pi  \tau_0}   \int d^{10}y
e^{i \sum_r
k_r\cdot y}  \nonumber\\
&&\times\left(\hat{t}^{i_1j_1\cdots i_4j_4}\hat{t}_{m_1n_1\cdots
  m_4n_4}-{1\over 4}\epsilon^{i_1j_1\cdots j_4j_4}\epsilon_{m_1n_1\cdots
  m_4n_4}\right)  R_{i_1j_1}^{m_1n_1}
R_{i_2j_2}^{m_2n_2}R_{i_3j_3}^{m_3n_3}
R_{i_4j_4}^{m_4n_4};\nonumber\\
& &
\end{eqnarray}
$t_8$ symbol defined as:
{\footnotesize
\begin{eqnarray}
\label{t_8}
& &\hat{t}_8^{N_1\dots N_8}   = \frac{1}{16} \biggl( -  2 \left(   G^{ N_1 N_3  }G^{  N_2  N_4  }G^{ N_5   N_7  }G^{ N_6 N_8  }
 + G^{ N_1 N_5  }G^{ N_2 N_6  }G^{ N_3   N_7  }G^{  N_4   N_8   }
 +  G^{ N_1 N_7  }G^{ N_2 N_8  }G^{ N_3   N_5  }G^{  N_4 N_6   }  \right) \nonumber \\
 & &  +
 8 \left(  G^{  N_2     N_3   }G^{ N_4    N_5  }G^{ N_6    N_7  }G^{ N_8   N_1   }
  +G^{  N_2     N_5   }G^{ N_6    N_3  }G^{ N_4    N_7  }G^{ N_8   N_1   }
  +   G^{  N_2     N_5   }G^{ N_6    N_7  }G^{ N_8    N_3  }G^{ N_4  N_1   }
\right) \nonumber \\
& &  - (N_1 \leftrightarrow  N_2) -( N_3 \leftrightarrow  N_4) - (N_5 \leftrightarrow  N_6) - (N_7 \leftrightarrow  N_8) \biggr)
\end{eqnarray}
} ($N_i$ being valued in the ${\bf 8}_v$ of $SO(8)$ or covariantized to $10D$ (or $11D$ to be used later in ${\mathscr {M}}$-theory)), from \cite{Tseytlin} the light-cone
8D ``zero mode" tensor $t_8$ is generalized to 10D: $\hat{t}_8 = t_8 - 1/4 B \epsilon_{10}$ wherein assuming $B_{\rm LC directions}=1$, yields $\hat{t}^{i_1j_1...i_4j_4} = t^{i_1j_1...i_4j_4} - \frac{1}{2}\epsilon^{i_1j_1...i_4j_4}$, and the overall factor of $e^{2\pi i  \tau_0}$,
which  is characteristic of the stringy D-instanton, is evaluated at $\chi =\Re \tau_0 =0$ in the stringy calculation.

\item
In \cite{Green and Vanhove}, it is shown that the eleven-dimensional ${\cal O}(R^4)$ corrections have an independent motivation based on supersymmetry in ten dimensions.
This was shown to follow from its relation to the term  $C^{(3)}\wedge X_8$ in the ${\mathscr {M}}$-theory
effective action which is known to arise from a variety of
arguments, e.g. anomaly cancellation \cite{Duff, Horava and Witten}\footnote{We thank M.J.Duff for bringing \cite{Duff} to our attention.}. The expression $X_8$ is the eight-form in the curvatures that is inherited from the term in type IIA superstring theory \cite{Vafa and Witten} which is given by
\begin{equation}-   \int d^{10}x  B\wedge  X_8 =  -
{1\over 2}
\int  d^{10}x \sqrt{-g^{A(10)}}\epsilon_{10} B X_8,
\end{equation}
where
\begin{equation}
\label{X_8-def}
X_8 = {1 \over 192} \left( {\rm tr}\ R^4 -
{1\over 4} ({\rm tr}\ R^2)^2\right).
\end{equation}

There are two independent ten-dimensional $N=1$ super-invariants which contain an odd-parity term
(\cite{Tseytlin} and previous authors):
$
I_3= t_8 {\rm tr}\ R^4 - {1\over 4} \epsilon_{10}B {\rm tr}\ R^4
$
 and:
$
I_4= t_8 ({\rm tr}\ R^2)^2 - {1\over 4} \epsilon_{10}B ({\rm tr}\ R^2)^2.
$
 Using  that
$
t_8t_8R^4=24t_8{\rm tr}(R^4)-6t_8({\rm tr}\ R^2)^2,
$
 it follows that  the particular linear combination,
\begin{equation}
I_3 - {1\over 4}I_4 = {1\over 24} t_8 t_8 R^4 - 48 \epsilon_{10}B\ X_8
\end{equation}
contains both the ten-form $B\wedge X_8$ and $t_8 t_8 R^4$.

\item
We will now summarize the argument of \cite{Green and Gutperle} that an $SL(2,\mathbb{Z})$ completion of the effective $R^4$ interactionleads to an interesting non-renormalization theorem that forbids
perturbative corrections to this term beyond one loop in the zero-instanton sector.
The term  in
(\ref{A}) bilinear in the tensor $\hat t$  has precisely the
same form as
 terms
 that arise in the zero instanton sector that come both from the one-loop
four-graviton amplitude  and from an
$(\alpha')^3$ effect
at  tree level \cite{wittengross}.   In the Einstein frame one hence obtains the following expression
 for the complete  effective $R^4$ action in the Einstein's frame, that can be expressed in the form,
\begin{equation}
\label{modular-completion-i}
S_{R^4} = (\alpha')^{-1} \left[ a \zeta(3)\tau_2^{3/2}   + b
\tau_2^{-1/2} + c
e^{2\pi i  \tau} +\cdots \right] R^4 \equiv  (\alpha')^{-1}
f(\tau,\bar{\tau})R^4,
\end{equation}
where $a, b$ are known numerical constants, $R^4$ denotes the contractions $\hat t \hat t R^4$ in
(\ref{A}) and
$\cdots$ indicates possible perturbative and nonperturbative
corrections to the
coefficient of $R^4$; the `constant' $c$ can depend on $\tau = C_{\rm RR} + i e^{-\phi}$ (type IIB) and $\bar \tau$. The first term in (\ref{modular-completion-i}) represents the $(\alpha^\prime)^3$ tree-level contribution and the second the one-loop relative to the former.

However, the complete expression for $S_{R^4}$ must be invariant under
$SL(2,\mathbb{Z})$ transformations:  $\tau\to (a\tau+b)(c\tau+d)^{-1}$ ($a,b,c,d\in\mathbb{Z}: ad-bc=1$), which provides very strong constraints on its structure. As the $R^4$ factor is already invariant, $f(\tau,\bar \tau)$
in (\ref{modular-completion-i}) is a scalar under the $SL(2,\mathbb{Z})$ transformations, which hence implies a sum over  all instantons and anti-instantons.  

There are some strong constraints which this term must satisfy and are spelt out in \cite{Green and Gutperle}, in which there is a simple function proposed by the authors that satisfies all these
criteria,  namely,
\begin{equation}
\label{modular-completion-ii}
  f(\tau,\bar{\tau}) = \sum_{(p,n )\neq
(0,0)}{\tau_2^{3/2}\over |p+n\tau|^3},
\end{equation}
where the sum indicates the sum is over all positive and negative values of $p,n$ except $p=n=0$.   Now:
\begin{equation}\label{modular-completion-iii}
  f=2\zeta(3)\tau_2^{3/2} +{\tau_2^{3/2}\over \Gamma(3/2)}\sum_{n\neq 0,p }\int_0^{\infty}dy
y^{1/2}e^{-y|p+n\tau|^2}.
\end{equation}
The sum over $p$, using  the   Poisson
resummation
formula\\ $\left(\sum_{n\rightarrow-\infty}^\infty f(n) = \sum_{m\rightarrow-\infty}^\infty {\rm FT}[f](m); {\rm FT}[e^{-\pi A(p+x)^2}] = \frac{e^{-\frac{M^2}{4 A\pi} - i M x}}{2\pi A}, M=2\pi m\right)$,
gives,
\begin{eqnarray}
\label{modular-completion-iii}
& &  \hskip -0.5in   f(\tau,\bar \tau) = 2\zeta(3)\tau_2^{3/2} + {2\pi^{2}\over 3}\tau_2^{-1/2} +
2\tau_2^{3/2}\sum_{m,n\neq 0}\int_0^{\infty}dy
  \exp\left(-{\pi^2 m^2\over y}+2\pi i m n\tau_1 - yn^2 \tau_2^2
\right),\nonumber\\
& & \hskip -0.5in = 2\zeta(3)\tau_2^{3/2} +
  {2\pi^2\over
3} \tau_2^{-1/2}  +8 \pi  \tau_2^{ 1/2}  \sum_{m \ne 0 n\ge 1   }
\left|{m\over n}\right|
e^{2\pi i mn\tau_1} K_1 (2\pi |mn|\tau_2)\nonumber\\
&&  \hskip -0.5in = 2\zeta(3)\tau_2^{3/2} + {2\pi^2\over  3} \tau_2^{-1/2}   \nonumber\\
&&  \hskip -0.5in +4\pi^{3/2}  \sum_{m,n \ge 1} \left({m\over n^3}\right)^{1/2}
(e^{2\pi i mn
\tau} + e^{-2\pi i mn \bar  \tau} ) \left(1 + \sum_{k=1}^\infty  (4\pi mn
\tau_2)^{-k} {\Gamma(  k -1/2)\over \Gamma(- k -1/2) } \right) ,
\end{eqnarray}
where performing a perturbative expansion in $\frac{1}{\tau_2}$ of the non-perturbative instanton contribution of charge $mn$, one uses the asymptotic expansion for $K_1(z)$ for
large $z$  in (\ref{modular-completion-iii}).

The perturbative terms in (\ref{modular-completion-iii}) terminate after the one-loop term as
suggested earlier.  The non-perturbative terms have the form of a
sum over
single multiply-charged instantons and anti-instantons with action
proportional
to $|mn|$.  Identifying $p$  with the discrete momentum (euclidean energy)  of a compactified
$D$-particle of charge $n$ then the Poisson resummation exchanges it with the
winding number of the world-line and the result is what is  expected by T-duality from
type IIA in nine dimensions.  The terms in parenthesis in (\ref{modular-completion-iii})
represent the infinite sequence of  perturbative corrections around the
instantons of  charge $mn$.   It is hence evident that in the zero-instanton sector, there are no perturbative corrections figuring in the action up to ${\cal O}(R^4)$, i.e., $S_{R^4}$, beyond one loop. This will be important in demonstrating the "non-renormalization" beyond one-loop of the deconfinement temperature $T_c$ later in the paper.

\item
The ${\cal N}=1, D=11$ supergravity action inclusive of ${\cal O}(l_p^6)$ terms (promoting the 10D $\hat{t}$ to 11D), is hence given by:
\begin{eqnarray}
\label{D=11_O(l_p^6)}
& & \hskip -0.8inS = \frac{1}{2\kappa_{11}^2}\int_M\left[  {\cal R} *_{11}1 - \frac{1}{2}G_4\wedge *_{11}G_4 -
\frac{1}{6}C\wedge G\wedge G\right] + \frac{1}{\kappa_{11}^2}\int_{\partial M} d^{10}x \sqrt{h} K \nonumber\\
& & \hskip -0.8in+ \frac{1}{(2\pi)^43^22^{13}}\left(\frac{2\pi^2}{\kappa_{11}^2}\right)^{\frac{1}{3}}\int d^{11}x\sqrt{-g}\left( J_0 - \frac{1}{2}E_8\right) + \left(\frac{2\pi^2}{\kappa_{11}^2}\right)\int C_3\wedge X_8,
\end{eqnarray}
where:
\begin{eqnarray}
\label{J0+E8-definitions}
& & \hskip -0.8inJ_0  =3\cdot 2^8 (R^{HMNK}R_{PMNQ}{R_H}^{RSP}{R^Q}_{RSK}+
{1\over 2} R^{HKMN}R_{PQMN}{R_H}^{RSP}{R^Q}_{RSK})\nonumber\\
& & \hskip -0.8inE_8  ={ 1\over 3!} \epsilon^{ABCM_1 N_1 \dots M_4 N_4}
\epsilon_{ABCM_1' N_1' \dots M_4' N_4' }{R^{M_1'N_1'}}_{M_1 N_1} \dots
{R^{M_4' N_4'}}_{M_4 N_4},\nonumber\\
& & \hskip -0.8in\kappa_{11}^2 = \frac{(2\pi)^8 l_p^{9}}{2};
\end{eqnarray}
$\kappa_{11}^2$ being related to the eleven-dimensional Newtonian coupling constant, and $G=dC$ with $C$ being the ${\mathscr {M}}$-theory three-form potential with the four-form $G$ being the associated four-form field strength.  The EOMS are:
\begin{eqnarray}
\label{eoms}
& & R_{MN} - \frac{1}{2}g_{MN}{\cal R} - \frac{1}{12}\left(G_{MPQR}G_N^{\ PQR} - \frac{g_{MN}}{8}G_{PQRS}G^{PQRS} \right)\nonumber\\
 & &  = - \beta\left[\frac{g_{MN}}{2}\left( J_0 - \frac{1}{2}E_8\right) + \frac{\delta}{\delta g^{MN}}\left( J_0 - \frac{1}{2}E_8\right)\right],\nonumber\\
& & d*G = \frac{1}{2} G\wedge G +3^22^{13} \left(2\pi\right)^{4}\beta X_8,\nonumber\\
& &
\end{eqnarray}
where \cite{Becker-sisters-O(R^4)}:
\begin{equation}
\label{beta-def}
\beta \equiv \frac{\left(2\pi^2\right)^{\frac{1}{3}}\left(\kappa_{11}^2\right)^{\frac{2}{3}}}{\left(2\pi\right)^43^22^{12}} \sim l_p^6,
\end{equation}
$R_{MNPQ}, R_{MN}, {\cal R}$  in  (\ref{D=11_O(l_p^6)})/(\ref{eoms}) being respectively the elven-dimensional Riemann curvature tensor, Ricci tensor and the Ricci scalar.

Now, one sees that if one makes an ansatz:
\begin{eqnarray}
\label{ansaetze}
& & \hskip -0.8ing_{MN} = g_{MN}^{(0)} +\beta g_{MN}^{(1)},\nonumber\\
& & \hskip -0.8inC_{MNP} = C^{(0)}_{MNP} + \beta C_{MNP}^{(1)},
\end{eqnarray}
then symbolically, one obtains:
\begin{eqnarray}
\label{deltaC=0consistent}
& & \beta \partial\left(\sqrt{-g}\partial C^{(1)}\right) + \beta \partial\left[\left(\sqrt{-g}\right)^{(1)}\partial C^{(0)}\right] + \beta\epsilon_{11}\partial C^{(0)} \partial C^{(1)} = {\cal O}(\beta^2) \sim 0 [{\rm up\ to}\ {\cal O}(\beta)].
\nonumber\\
& & \end{eqnarray}
One can see that one can find a consistent set of solutions to (\ref{deltaC=0consistent}) wherein $C^{(1)}_{MNP}=0$ up to ${\cal O}(\beta)$.  Assuming that one can do so, henceforth we will  define:
\begin{eqnarray}
\label{fMN-definitions}
\delta g_{MN} =\beta g^{(1)}_{MN} = G_{MN}^{\rm MQGP} f_{MN}(r),
\end{eqnarray}
no summation implied.

\item
Let us discuss how one can controls the higher derivative contributions to the MQGP limit result of \cite{MQGP}. This was explained in \cite{OR4-Yadav+Misra} and we review the same here.  It is evident from (\ref{ M-theory-metric-psi=2npi-patch}) in the $\psi=2n\pi,n=0, 1, 2$-coordinate patch  that in the IR: $r = \chi r_h, \chi\equiv {\cal O}(1)$, and up to ${\cal O}(\beta)$:
\begin{equation}
\label{IR-beta-N-suppressed-logrh-rh-neg-exp-enhanced}
f_{MN} \sim \beta\frac{\left(\log {\cal R}_h\right)^{m}}{{\cal R}_h^n N^{\beta_N}},\ m\in\left\{0,1,3\right\},\ n\in\left\{0,2,5,7\right\},\
\beta_N>0,
\end{equation}
where ${\cal R}_h\equiv\frac{r_h}{{\cal R}_{D5/\overline{D5}}}$. Now, $|{\cal R}_h|\ll1$. As estimated in \cite{Bulk-Viscosity-McGill-IIT-Roorkee}, $|\log {\cal R}_h|\sim N^{\frac{1}{3}}$, implying there is a competition between Planckian and large-$N$ suppression and infra-red enhancement arising from $m,n\neq0$ in (\ref{IR-beta-N-suppressed-logrh-rh-neg-exp-enhanced}). Choosing a heirarchy: $\beta\sim e^{-\gamma_\beta N^{\gamma_N}}, \gamma_\beta,\gamma_N>0: \gamma_\beta N^{\gamma_N}>7N^{\frac{1}{3}} + \left(\frac{m}{3} - \beta_N\right)\log N$ (ensuring that the IR-enhancement does not overpower Planckian suppression - we took the ${\cal O}(\beta)$ correction to $G^{\cal M}_{yz}$, which had the largest IR enhancement, to set a lower bound on $\gamma_{\beta,N}$/Planckian suppression). If $\gamma_\beta N^{\gamma_N}\sim7N^{\frac{1}{3}}$, then one will be required to go to a higher order in $\beta$. This hence answers the question, when one can truncate at ${\cal O}(\beta)$.

\end{itemize}

\section{${\mathscr {M}}$-Theory Duals at Low ($T<T_c$ on the gauge theory side) and High ($T>T_c$ on the gauge theory side) Temperatures}

In this section we will explain how to calculate deconfinement temperature in QCD-like theory via gauge/gravity duality and holographic renormalization of the eleven dimensional supergravity action using a semiclassical computation as advocated by Witten \cite{Witten-Hawking-Page-Tc}. We have partitioned the calculation in two subsections. In subsection {\bf 3.1}, we will calculate various terms appearing in on-shell supergravity action for blackhole background and discuss holographic renormalization of the same. We do the same thing in subsection {\bf 3.2} for the thermal background.
\par
The ${\mathscr {M}}$-theory  dual corresponding to high temperatures, i.e., $T>T_c$, will involve a black hole with the metric given by
\begin{eqnarray}
\label{TypeIIA-from-M-theory-Witten-prescription-T>Tc}
\hskip -0.1in ds_{11}^2 & = & e^{-\frac{2\phi^{\rm IIA}}{3}}\Biggl[\frac{1}{\sqrt{h(r,\theta_{1,2})}}\left(-g(r) dt^2 + \left(dx^1\right)^2 +  \left(dx^2\right)^2 +\left(dx^3\right)^2 \right)
\nonumber\\
& & \hskip -0.1in+ \sqrt{h(r,\theta_{1,2})}\left(\frac{dr^2}{g(r)} + ds^2_{\rm IIA}(r,\theta_{1,2},\phi_{1,2},\psi)\right)
\Biggr] + e^{\frac{4\phi^{\rm IIA}}{3}}\left(dx^{11} + A_{\rm IIA}^{F_1^{\rm IIB} + F_3^{\rm IIB} + F_5^{\rm IIB}}\right)^2,
\end{eqnarray}
where $A_{\rm IIA}^{F^{\rm IIB}_{i=1,3,5}}$ are the type IIA RR 1-forms obtained from the triple T/SYZ-dual of the type IIB $F_{1,3,5}^{\rm IIB}$ fluxes in the type IIB holographic dual of \cite{metrics}, and $g(r) = 1 - \frac{r_h^4}{r^4}$. For low temperatures, i.e., $T<T_c$, is given by the thermal gravitational dual:
\begin{eqnarray}
\label{TypeIIA-from-M-theory-Witten-prescription-T<Tc}
\hskip -0.1in ds_{11}^2 & = & e^{-\frac{2\phi^{\rm IIA}}{3}}\Biggl[\frac{1}{\sqrt{h(r,\theta_{1,2})}}\left(-dt^2 + \left(dx^1\right)^2 +  \left(dx^2\right)^2 + \tilde{g}(r)\left(dx^3\right)^2 \right)
\nonumber\\
& & \hskip -0.1in+ \sqrt{h(r,\theta_{1,2})}\left(\frac{dr^2}{\tilde{g}(r)} + ds^2_{\rm IIA}(r,\theta_{1,2},\phi_{1,2},\psi)\right)
\Biggr] + e^{\frac{4\phi^{\rm IIA}}{3}}\left(dx^{11} + A_{\rm IIA}^{F_1^{\rm IIB} + F_3^{\rm IIB} + F_5^{\rm IIB}}\right)^2,
\end{eqnarray}
where $\tilde{g}(r) = 1 - \frac{r_0^4}{r^4}$. One notes that $t\rightarrow x^3,\ x^3\rightarrow t$ in (\ref{TypeIIA-from-M-theory-Witten-prescription-T>Tc}) following by a Double Wick rotation in the new $x^3, t$ coordinates obtains (\ref{TypeIIA-from-M-theory-Witten-prescription-T<Tc}); $h(r,\theta_{1,2})$ is the ten-dimensional warp factor \cite{metrics, MQGP}. This amounts to:  $-g_{tt}^{\rm BH}(r_h\rightarrow r_0) = g_{x^3x^3}\ ^{\rm Thermal}(r_0),$ $ g_{x^3x^3}^{\rm BH}(r_h\rightarrow r_0) = -g_{tt}\ ^{\rm Themal}(r_0)$ in the results of
\cite{VA-Glueball-decay, OR4-Yadav+Misra} (See \cite{Kruczenski et al-2003} in the context of Euclidean/black $D4$-branes in type IIA).  

In (\ref{TypeIIA-from-M-theory-Witten-prescription-T<Tc}), we will assume the spatial part of the solitonic $M3$ brane (which, locally, could be interpreted as solitonic $M5$-brane wrapped around a homologous sum of two-spheres \cite{DM-transport-2014}) world volume to be given by $\mathbb{R}^2(x^{1,2})\times S^1(x^3)$ where the period of $S^1(x^3)$ is given by a very large: $\frac{2\pi}{M_{\rm KK}}$, where the very small $M_{\rm KK} = \frac{2r_0}{ L^2}\left[1 + {\cal O}\left(\frac{g_sM^2}{N}\right)\right]$, $r_0$ being the very small IR cut-off in the thermal background (See also \cite{Armoni et al-2020}) and $L = \left( 4\pi g_s N\right)^{\frac{1}{4}}$. So, $\lim_{M_{\rm KK}\rightarrow0}\mathbb{R}^2(x^{1,2})\times S^1(x^3) = \mathbb{R}^3(x^{1,2,3})$, thereby recovering 4D Physics. The working metric used in this section and section {\bf 4} for the thermal background corresponding to $T<T_c$ will involve setting $\tilde{g}(r)$ to unity in (\ref{TypeIIA-from-M-theory-Witten-prescription-T<Tc}).

Now, if $\beta_{\rm BH,Th}$ are respectively the periodicities of the thermal circle in the black and thermal ${\mathscr {M}}$-theory backgrounds then at $r={\cal R}_{\rm UV}$, $\beta_{\rm BH}\sqrt{ G^{\rm BH}_{tt}} = \beta_{\rm Th} \sqrt{G^{\rm Th}_{tt}}$. Now at $T=T_c$ \cite{Witten-Hawking-Page-Tc}, 
\begin{equation}
\beta_{\rm BH}\slashed{\int}_{M_{11}}\left({\cal L}^{\rm BH}_{\rm EH} + {\cal L}^{\rm BH}_{\rm GHY}\delta(r-{\cal R}_{\rm UV}) 
+ {\cal L}^{\rm BH}_{{\cal O}(R^4)}\right) = \beta_{\rm Th}\slashed{\int}_{M_{11}}\left({\cal L}^{\rm Th}_{\rm EH} + {\cal L}^{\rm Th}_{\rm GHY}\delta(r-{\cal R}_{\rm UV}) 
+{\cal L}^{\rm Th}_{{\cal O}(R^4)}\right),
\end{equation}
where $\slashed{\int}$ excludes the coordinate integral with respect to  $x^{0}$,
 implying: 
\begin{eqnarray}
\label{actions-equal-Tc-ii}
& & \left(\sqrt{1-\frac{r_h^4}{{\cal R}_{\rm UV}^4}}\right)^{-1}\int_{M_{10}}\left({\cal L}^{\rm BH}_{\rm EH}+{\cal L}^{\rm BH}_{\rm GHY}\delta(r-{\cal R}_{\rm UV}) + {\cal L}^{\rm BH}_{{\cal O}(R^4)}\right) 
\nonumber\\
& & = \int_{\tilde{M}_{10}}\left({\cal L}^{\rm Th}_{\rm EH} + {\cal L}^{\rm Th}_{\rm GHY}\delta(r-{\cal R}_{\rm UV}) 
+ {\cal L}^{\rm Th}_{{\cal O}(R^4)}\right).
\end{eqnarray}     

\subsection{Black Hole Background Uplift (Relevant to $T>T_c$) and Holographic Renormalization of On-Shell $D=11$ Action}

The ${\cal O}(\beta)$-corrected ${\mathscr {M}}$-theory metric of \cite{MQGP} in the MQGP limit near the $\psi=2n\pi, n=0, 1, 2$-branches (whereat there is a decoupling of $M_5(t,x^{1,2,3},r)$ and $M_6(\theta_{1,2},x,y,z,x^{11})$) up to ${\cal O}((r-r_h)^2)$ [and up to ${\cal O}((r-r_h)^3)$ for some of the off-diagonal components along the delocalized $T^3(x,y,z)$] - the components which do not receive an ${\cal O}(\beta)$ corrections, are not listed in (\ref{ M-theory-metric-psi=2npi-patch}) - was worked out in \cite{OR4-Yadav+Misra} and is given in (\ref{ M-theory-metric-psi=2npi-patch}).

Restricting to the Ouyang embedding of the flavor $D7$-branes in type IIB can, for an extremely small (modulus of the) Ouyang embedding parameter, may be effected by working near \cite{NPB}:
\begin{equation}
\label{small-theta_12}
\theta_1 = \frac{\alpha_{\theta_1}}{N^{\frac{1}{5}}},\ \theta_2 = \frac{\alpha_{\theta_2}}{N^{\frac{3}{10}}}.
\end{equation}
Einstein-Hilbert  term in the IR will be given by the following expression:
\begin{eqnarray}
\label{EH-BH-IR}
& & \left.\sqrt{-G^{\mathscr {M}}}R\right|_{\rm Ouyang}^{\rm IR}\nonumber\\
& &  \sim  -\frac{b^2 {g_s}^{3/4} {\log N}^3 M {N_f}^3 {r_h}^4 \log
   \left(\frac{{r_h}}{{{\cal R}_{D5/\overline{D5}}^{\rm bh}}}\right) \log \left(1-\frac{{r_h}}{{{\cal R}_{D5/\overline{D5}}^{\rm bh}}}\right) (1-\beta 
   ({\cal C}_{zz}^{\rm bh}-2 {\cal C}_{\theta_1z}^{\rm bh}+3 {\cal C}_{\theta_1x}^{\rm bh}))}{{N^{\frac{5}{4}}} {{\cal R}_{D5/\overline{D5}}^{\rm bh}}^3(r-r_h)
   \sin^3{\theta _1} \sin^2{\theta _2}}.\nonumber\\
\end{eqnarray}
Now performing angular integration of the above equation,
{\footnotesize
\begin{equation}
\label{EH-global-i}
\int\int d\theta_1 d\theta_2 \frac{1}{\sin^3\theta_1\sin^2\theta_2} = \frac{1}{8} \cot ({\theta_2}) \left(\csc ^2\left(\frac{{\theta_1}}{2}\right)-\sec
   ^2\left(\frac{{\theta_1}}{2}\right)-4 \log \left(\sin \left(\frac{{\theta_1}}{2}\right)\right)+4 \log
   \left(\cos \left(\frac{{\theta_1}}{2}\right)\right)\right)
\end{equation}
}
implying:
\begin{equation}
\label{EH-global-ii}
\int_{\theta_1={\alpha_{\theta_1}}{N^{-\frac{1}{5}}}}^{\pi-{\alpha_{\theta_1}}{N^{-\frac{1}{5}}}}\int_{\theta_2= {\alpha_{\theta_2}}{N^{-\frac{3}{10}}}}^{\pi- {\alpha_{\theta_2}}{N^{-\frac{3}{10}}}}
 \frac{d\theta_1 d\theta_2}{\sin^3\theta_1\sin^2\theta_2} \sim \frac{N^{\frac{7}{10}}}{\alpha_{\theta_1}^2\alpha_{\theta_2}} + {\cal O}\left(N^{\frac{3}{10}}\right),
\end{equation}
Using equation (\ref{EH-global-ii}) and performing radial integration of equation (\ref{EH-BH-IR}), we obtained the following expression for the action density (for Einstein-Hilbert  term) in the IR:
\begin{eqnarray}
\label{int-EH-BH-IR}
& & \frac{\left.\int_{r=r_h}^{{\cal R}_{D5-\overline{D5}}^{\rm bh}}\int_{\theta_1={\alpha_{\theta_1}}{N^{-\frac{1}{5}}}}^{\pi-{\alpha_{\theta_1}}{N^{-\frac{1}{5}}}}\int_{\theta_2= {\alpha_{\theta_2}}{N^{-\frac{3}{10}}}}^{\pi- {\alpha_{\theta_2}}{N^{-\frac{3}{10}}}}\sqrt{-G^{\mathscr {M}}}R\right|_{\rm Ouyang}}{{\cal V}_4} \sim \nonumber\\
& &  -\frac{b^2 {g_s}^{3/4} {\log N}^3 M {N_f}^3 {r_h}^4 \log
   \left(\frac{{r_h}}{{{\cal R}_{D5/\overline{D5}}^{\rm bh}}}\right) \log \left(1-\frac{{r_h}}{{{\cal R}_{D5/\overline{D5}}^{\rm bh}}}\right) (1-\beta 
   ({\cal C}_{zz}^{\rm bh}-2 {\cal C}_{\theta_1z}^{\rm bh}+3 {\cal C}_{\theta_1x}^{\rm bh}))}{{N}^{\frac{11}{20}} {{\cal R}_{D5/\overline{D5}}^{\rm bh}}^4
   \alpha _{\theta _1}^2 \alpha _{\theta _2}},
\end{eqnarray}
where ${\cal R}_{D5-\overline{D5}}^{\rm bh}\equiv \sqrt{3}a^{\rm bh}$, $a^{\rm bh} = \left(\frac{1}{\sqrt{3}} + \epsilon^{\rm bh} + {\cal O}\left(\frac{g_sM^2}{N}\right)\right)r_h$ being the resolution parameter of the blown up $S^2$ \cite{OR4-Yadav+Misra}.
Further, there is one more term appearing in on-shell action as given in equation (\ref{on-shell-D=11-action-up-to-beta}), simplified form of that term is:
\begin{eqnarray}
\label{sqrtGbetaRbeta0-IR-bh}
& & \frac{\int_{r_h}^{{\cal R}_{D5-\overline{D5}}^{\rm bh}}\int_{\theta_1={\alpha_{\theta_1}}{N^{-\frac{1}{5}}}}^{\pi-{\alpha_{\theta_1}}{N^{-\frac{1}{5}}}}\int_{\theta_2= {\alpha_{\theta_2}}{N^{-\frac{3}{10}}}}^{\pi- {\alpha_{\theta_2}}{N^{-\frac{3}{10}}}}
\left(\sqrt{-G^{\mathscr {M}}}\right)^{(1)}R^{(0)}}{{\cal V}_4}\nonumber\\
& & \sim \beta {\cal C}_{\theta_1x}^{\rm bh}\frac{b^2g_s^{\frac{3}{4}}\log^3N M N_f^3r_h^4
\log\left(\frac{r_h}{{\cal R}_{D5/\overline{D5}}^{\rm bh}}\right)\log\left(1 - \frac{r_h}{{\cal R}_{D5/\overline{D5}}^{\rm bh}}\right)}{{{\cal R}_{D5/\overline{D5}}^{\rm bh}}^4N^{\frac{11}{20}}\alpha_{\theta_1}^2\alpha_{\theta_2}}.
\end{eqnarray}
Similarly, Einstein-Hilbert  term in the UV is:
{\footnotesize
\begin{eqnarray}
\label{EH-BH-UV}
& & \hskip -0.6in \left.\sqrt{-G^{\mathscr {M}}}R\right|_{\rm Ouyang}^{\rm UV} = -\frac{4 \left(34 {\log r}^2+14 {\log r}-1\right) {M_{\rm UV}} r^3}{9 \sqrt{3} \sqrt[4]{\pi } {g_s}^{9/4}
  N^{\frac{5}{4}}
   \sin^3{\theta _1} \sin^2{\theta _2} } \nonumber\\
   & & \times \frac{1}{\left(-{g_s} {N_f^{\rm UV}} (2 {\log N}
   {\log r}+{\log N}+3 (1-6 {\log r}) {\log r})+2 {g_s} (2 {\log r}+1) {N_f^{\rm UV}} \log
   \left(\frac{1}{4} \alpha _{\theta _1} \alpha _{\theta _2}\right)+8 \pi  {\log r}\right)}
   \nonumber\\
& & \hskip 0.5in \approx \frac{4 {M_{\rm UV}} r^3 \left(34 \log ^2(r)+14 \log (r)-1\right)}{9 \sqrt{3} \sqrt[4]{\pi } {g_s}^{13/4}
    {N_f^{\rm UV}}  N^{\frac{5}{4}}
   \sin^3{\theta _1} \sin^2{\theta _2} \log N  (2 \log (r)+1)},
\end{eqnarray}
}
where, for simplification of the computation, we have dropped the ${\cal O}\left(\frac{1}{\log N}\right)$-correction terms. As $\int \frac{r^3 \left(34 \log ^2(r)+14 \log (r)-1\right)}{2 \log (r)+1} = \frac{1}{16} \left(\frac{4 {Ei}(4 \log (r)+2)}{e^2}+r^4 (68 \log (r)-23)\right)$ + constant. 

Hence, one obtains:
\begin{eqnarray}
\label{int-EH-BH-UV}
& &\left. \left(1+\frac{r_h^4}{2{\cal R}_{\rm UV}^4}\right)\frac{\int_{{\cal R}_{D5-\overline{D5}}}^{{\cal R}_{\rm UV}}\int_{\theta_1={\alpha_{\theta_1}}{N^{-\frac{1}{5}}}}^{\pi-{\alpha_{\theta_1}}{N^{-\frac{1}{5}}}}\int_{\theta_2= {\alpha_{\theta_2}}{N^{-\frac{3}{10}}}}^{\pi- {\alpha_{\theta_2}}{N^{-\frac{3}{10}}}}
\sqrt{-G^{\mathscr {M}}}R}{{\cal V}_4}\right|_{\rm Ouyang}^{\rm UV-Finite}\nonumber\\
& & \sim 
\frac{ {M_{\rm UV}} {r_h}^4 \log \left(\frac{{{\cal R}_{\rm UV}}}{{{\cal R}_{D5/\overline{D5}}^{\rm bh}}}\right) }{ {g_s}^{13/4} {N}^{\frac{11}{20}}  {{\cal R}_{D5/\overline{D5}}^{\rm bh}}^4
   {N_f^{\rm UV}} \alpha _{\theta _1}^2 \alpha _{\theta _2} \log N }.
\end{eqnarray}
The UV-divergent contribution of the EH action is:
\begin{eqnarray}
\label{EH-div}
S_{\rm UV-div}^{\rm EH} \sim \frac{ {M_{\rm UV}} {\cal R}_{\rm UV}^4 \log\left(\frac{{{\cal R}_{\rm UV}}}{{{\cal R}_{D5/\overline{D5}}^{\rm bh}}}\right) }{ {g_s^{\rm UV}}^{13/4}{{\cal R}_{D5/\overline{D5}}^{\rm bh}}^4 {\log N}
    {N_f^{\rm UV}}N^{\frac{11}{20}} \alpha _{\theta _1}^2 \alpha _{\theta _2}}.
\end{eqnarray}
Now for the boundary metric:
\begin{eqnarray}
\label{sqrtminush}
& & \left.\int_{\theta_1={\alpha_{\theta_1}}{N^{-\frac{1}{5}}}}^{\pi-{\alpha_{\theta_1}}{N^{-\frac{1}{5}}}}\int_{\theta_2= {\alpha_{\theta_2}}{N^{-\frac{3}{10}}}}^{\pi- {\alpha_{\theta_2}}{N^{-\frac{3}{10}}}} \sqrt{-h}\right|_{r={\cal R}_{\rm UV}} \sim \frac{
   \left(\frac{1}{{g_s^{\rm UV}}}\right)^{7/3} {M_{\rm UV}} {\cal R}_{\rm UV}^4\log\left(\frac{{{\cal R}_{\rm UV}}}{{{\cal R}_{D5/\overline{D5}}^{\rm bh}}}\right)}{N^{\frac{3}{10}}  \alpha _{\theta _1}^2
   \alpha _{\theta _2}{{\cal R}_{D5/\overline{D5}}^{\rm bh}}^4},
\end{eqnarray}
and counter term to cancel the UV divergence appearing in Einstein-Hilbert  term is,
\begin{eqnarray}
\label{EH-BH-boundary}
\frac{\left.\int_{\theta_1={\alpha_{\theta_1}}{N^{-\frac{1}{5}}}}^{\pi-{\alpha_{\theta_1}}{N^{-\frac{1}{5}}}}\int_{\theta_2= {\alpha_{\theta_2}}{N^{-\frac{3}{10}}}}^{\pi- {\alpha_{\theta_2}}{N^{-\frac{3}{10}}}}\sqrt{-h}R\right|_{r=r_\Lambda}}{{\cal V}_4} \sim \frac{\left(\frac{1}{{g_s^{\rm UV}}}\right)^{2/3} {M_{\rm UV}} {\cal R}_{\rm UV}^4 \log\left(\frac{{{\cal R}_{\rm UV}}}{{{\cal R}_{D5/\overline{D5}}^{\rm bh}}}\right) }{ {g_s^{\rm UV}}^{3/2} N^{1/5}{{\cal R}_{D5/\overline{D5}}^{\rm bh}}^4 \alpha
   _{\theta _1}^2 \alpha _{\theta _2}}.
\end{eqnarray}

The UV-finite boundary Gibbons-Hawking-York contribution  up to ${\cal O}(\beta)$ turns to be:
\begin{eqnarray}
\label{GHY-BH}
& & \left(1 + \frac{r_h^4}{2r^4}\right)\left.\frac{\left.\int_{\theta_1={\alpha_{\theta_1}}{N^{-\frac{1}{5}}}}^{\pi-{\alpha_{\theta_1}}{N^{-\frac{1}{5}}}}\int_{\theta_2= {\alpha_{\theta_2}}{N^{-\frac{3}{10}}}}^{\pi- {\alpha_{\theta_2}}{N^{-\frac{3}{10}}}} \sqrt{-h^{\mathscr {M}}}K\right|_{\rm Ouyang}}{{\cal V}_4}\right|^{r={\cal R}_{\rm UV}}\nonumber\\
& & \sim \frac{{M_{\rm UV}} {r_h}^4 \log \left(\frac{{{\cal R}_{\rm UV}}}{{{\cal R}_{D5/\overline{D5}}^{\rm bh}}}\right)}{{g_s}^{9/4} {{\cal R}_{D5/\overline{D5}}^{\rm bh}}^4 {N}^{\frac{11}{20}}
   \alpha _{\theta _1}^2 \alpha _{\theta _2}}.\nonumber\\
& & 
\end{eqnarray}
The UV-divergent contribution to the GHY term is:
\begin{equation}
\label{GHY-UV-div}
S_{\rm UV-div}^{\rm GHY} \sim\frac{ \log\left(\frac{{{\cal R}_{\rm UV}}}{{{\cal R}_{D5/\overline{D5}}^{\rm bh}}}\right) {M_{\rm UV}} {{\cal R}_{\rm UV}}^4}{ \left(g_s^{\rm UV}\right)^{9/4}  {{\cal R}_{D5/\overline{D5}}^{\rm bh}}^4 {N}^{\frac{11}{20}} \alpha _{\theta _1}^2 \alpha _{\theta _2}}.
\end{equation}

One can further show that up to ${\cal O}(\beta^0)$ contribution from higher derivative term will be:
{\footnotesize
\begin{eqnarray}
\label{int-iGdeltaJ0-BH-IR}
& & \hskip -0.4in \frac{\int_{r_0}^{{\cal R}_{D5/\overline{D5}}}\int_{\theta_1={\alpha_{\theta_1}}{N^{-\frac{1}{5}}}}^{\pi-{\alpha_{\theta_1}}{N^{-\frac{1}{5}}}}\int_{\theta_2= {\alpha_{\theta_2}}{N^{-\frac{3}{10}}}}^{\pi- {\alpha_{\theta_2}}{N^{-\frac{3}{10}}}}\sqrt{-G^{\mathscr {M}}}\left.G^{MN}\frac{\delta J_0}{\delta G^{MN}}\right|_{\rm Ouyang}}{{\cal V}_4}
\nonumber\\
& & \hskip -0.4in \sim \frac{ M {N_f} {r_h} \left(1-\frac{{r_h}}{{{\cal R}_{D5/\overline{D5}}^{\rm bh}}}\right)^3
    \left(\frac{{r_h}}{{{\cal R}_{D5/\overline{D5}}^{\rm bh}}}\right)  \left({\log N}-9  \left(\frac{{r_h}}{{{\cal R}_{D5/\overline{D5}}^{\rm bh}}}\right) \right) \left[\log N -3  \left(\frac{{r_h}}{{{\cal R}_{D5/\overline{D5}}^{\rm bh}}}\right) \right]^2}{\epsilon ^5
   {g_s} N^{39/20} \log ^2(N)}\nonumber\\
& & \hskip -0.4in \times\int_{\theta_1={\alpha_{\theta_1}}{N^{-\frac{1}{5}}}}^{\pi-{\alpha_{\theta_1}}{N^{-\frac{1}{5}}}}\int_{\theta_2= {\alpha_{\theta_2}}{N^{-\frac{3}{10}}}}^{\pi- {\alpha_{\theta_2}}{N^{-\frac{3}{10}}}}\frac{19683 \sqrt{6} \sin^6{\theta _1}+6642 \sin^2{\theta _2} \sin^3{\theta _1}-40 \sqrt{6}
   \sin^4{\theta _2}}{\sin^7{\theta _1}\sin^4{\theta _2}};\nonumber\\
& & \nonumber\\
\end{eqnarray}
}
using:
{\footnotesize
\begin{eqnarray}
\label{integrals}
& & -\int dr\frac{r^3 \log (r)}{(2 \log (r)+1)^7}\nonumber\\
& &  = \frac{1}{360} \left(\frac{r^4 \left(256 \log ^5(r)+704 \log ^4(r)+800 \log ^3(r)+488 \log ^2(r)+184 \log
   (r)+19\right)}{(2 \log (r)+1)^6}-\frac{16 {Ei}(4 \log (r)+2)}{e^2}\right);
\nonumber\\
& & -\int dr\frac{r \log (r) \left(4 \log ^2(r)+15 \log (r)+9\right)}{(2 \log (r)+1)^7} \nonumber\\
& & = \frac{2 r^2 \left(7504 \log ^5(r)+22512 \log ^4(r)+30016 \log ^3(r)+23384 \log ^2(r)+8799 \log (r)+833\right)}{2880(2
   \log (r)+1)^6}-\frac{469 {Ei}(2 \log (r)+1)}{5760e};\nonumber\\
& & -\int dr\frac{\log (r) \left(45 a^4 \left(64 \log ^3(r)+208 \log ^2(r)+212 \log (r)+57\right)-8 {r_h}^4 (2 \log
   (r)+5)\right)}{r (2 \log (r)+1)^7}\nonumber\\
& &  = \frac{15 \left(3915 a^4-4 {r_h}^4\right) \log ^2(r)+12 \left(1620 a^4-11 {r_h}^4\right) \log (r)+21600 a^4 \log
   ^4(r)+61200 a^4 \log ^3(r)+1620 a^4-11 {r_h}^4}{30 (2 \log (r)+1)^6},\nonumber\\
& & 
\end{eqnarray}
}
one can show that UV finite part coming from higher derivative term will be:
{\footnotesize
\begin{eqnarray}
\label{int-iGdeltaJ0--UV}
& & \hskip -0.6in  \left.\left(1 + \frac{r_h^4}{2{\cal R}_{\rm UV}^4}\right)\frac{\int_{\theta_1={\alpha_{\theta_1}}{N^{-\frac{1}{5}}}}^{\pi-{\alpha_{\theta_1}}{N^{-\frac{1}{5}}}}\int_{\theta_2= {\alpha_{\theta_2}}{N^{-\frac{3}{10}}}}^{\pi- {\alpha_{\theta_2}}{N^{-\frac{3}{10}}}}\int_{{\cal R}_{D5/\overline{D5}}}^{{\cal R}_{\rm UV}}\sqrt{-G^{\mathscr {M}}}\left.G^{MN}\frac{\delta J_0}{\delta G^{MN}}\right|_{\rm Ouyang}}{{\cal V}_4}\right|^{\rm UV-finite}
\nonumber\\
& & \hskip -0.6in = \left.\frac{\int_{\theta_1={\alpha_{\theta_1}}{N^{-\frac{1}{5}}}}^{\pi-{\alpha_{\theta_1}}{N^{-\frac{1}{5}}}}\int_{\theta_2= {\alpha_{\theta_2}}{N^{-\frac{3}{10}}}}^{\pi- {\alpha_{\theta_2}}{N^{-\frac{3}{10}}}}\int_{{\cal R}_{D5/\overline{D5}}}^{{\cal R}_{\rm UV}}\sqrt{-G^{\mathscr {M}}}\left.G^{MN}\frac{\delta J_0}{\delta G^{MN}}\right|_{\rm Ouyang}}{{\cal V}_4}\right|^{\rm UV-finite}\nonumber\\
 & & \hskip -0.6in \sim -\frac{ {M_{\rm UV}}}{{g_s^{\rm UV}}^{14/3}
   {\log N}^{11/3} N^{39/20} {N_f^{\rm UV}}^{8/3}}\left(\frac{
   {r_h}^2}{{{\cal R}_{D5/\overline{D5}}^{\rm bh}}^2}+1\right)\nonumber\\
   & & \times \int_{\theta_1={\alpha_{\theta_1}}{N^{-\frac{1}{5}}}}^{\pi-{\alpha_{\theta_1}}{N^{-\frac{1}{5}}}}\int_{\theta_2= {\alpha_{\theta_2}}{N^{-\frac{3}{10}}}}^{\pi- {\alpha_{\theta_2}}{N^{-\frac{3}{10}}}} \frac{19683 \sqrt{6} \sin^6{\theta _1}+6642 \sin^2{\theta _2} \sin^3{\theta _1}-40 \sqrt{6}
   \sin^4{\theta _2}}{\sin^7{\theta _1}\sin^4{\theta _2}}.\nonumber\\
& & 
\end{eqnarray}
}
Use has been made of the following.  The most dominant term in $\sqrt{-G^{\mathscr {M}}}G^{MN}\frac{\delta J_0}{\delta G^{MN}}$ is \\ $-\frac{8 {M_{\rm UV}} r^3 \log (r)}{177147 \pi  {g_s^{\rm UV}}^4 (2 {\log r}+1)^7 N^{39/20} {N_f^{\rm UV}}^2 \log ^3(N)}\frac{19683 \sqrt{6} \sin^6{\theta _1}+6642 \sin^2{\theta _2} \sin^3{\theta _1}-40 \sqrt{6}
   \sin^4{\theta _2}}{\sin^7{\theta _1}\sin^4{\theta _2}}$. As: \\
$\int dr \frac{r^3 \log (r)}{(2 \log (r)+1)^7} = \frac{2 {Ei}(4 \log (r)+2)}{45 e^2}-\frac{r^4 \left(256 \log ^5(r)+704 \log ^4(r)+800 \log ^3(r)+488 \log ^2(r)+184
   \log (r)+19\right)}{360 (2 \log (r)+1)^6}$,\\ one therefore notes that the UV-finite contribution to: $\left(1 + \frac{r_h^4}{2{\cal R}_{\rm UV}^4}\right)\sqrt{-G^{\mathscr {M}}}G^{MN}\frac{\delta J_0}{\delta G^{MN}}$ is:\\
$-\frac{ {M_{\rm UV}}}{{g_s^{\rm UV}}^{14/3}
   {\log N}^{11/3} N^{39/20} {N_f^{\rm UV}}^{8/3}} \frac{r_h^4}{\log\left(\frac{{{\cal R}_{\rm UV}}}{{{\cal R}_{D5/\overline{D5}}^{\rm bh}}}\right)}\left(\frac{
   {r_h}^2}{{{\cal R}_{D5/\overline{D5}}^{\rm bh}}^2}+1\right)\frac{19683 \sqrt{6} \sin^6{\theta _1}+6642 \sin^2{\theta _2} \sin^3{\theta _1}-40 \sqrt{6}
   \sin^4{\theta _2}}{\sin^7{\theta _1}\sin^4{\theta _2}}$,\\ which turns out to be ${\cal O}\left(\frac{r_h^4}{\log\left(\frac{{{\cal R}_{\rm UV}}}{{{\cal R}_{D5/\overline{D5}}^{\rm bh}}}\right)}\right)$-suppressed as compared to the UV-finite contribution
from $\left.\int_{{\cal R}_{D5/\overline{D5}}}^{{\cal R}_{\rm UV}} \sqrt{-G^{\mathscr {M}}}G^{MN}\frac{\delta J_0}{\delta G^{MN}}\right|^{\rm UV-finite}$. \\
 Also UV divergent contribution from higher derivative term will be:
\begin{eqnarray}
\label{int-sqrtGiGdeltaJ0-beta0-UV-div}
& & \hskip -0.3in \frac{\left.\int \sqrt{-G^{\mathscr {M}}}G^{MN}\frac{\delta J_0}{\delta G^{MN}}\right|_{\rm UV-div}}{{\cal V}_4}
\nonumber\\
& & \hskip -0.3in  \sim \int_{\theta_1={\alpha_{\theta_1}}{N^{-\frac{1}{5}}}}^{\pi-{\alpha_{\theta_1}}{N^{-\frac{1}{5}}}}\int_{\theta_2= {\alpha_{\theta_2}}{N^{-\frac{3}{10}}}}^{\pi- {\alpha_{\theta_2}}{N^{-\frac{3}{10}}}}\frac{19683 \sqrt{6} \sin^6{\theta _1}+6642 \sin^2{\theta _2} \sin^3{\theta _1}-40 \sqrt{6}
   \sin^4{\theta _2}}{\sin^7{\theta _1}\sin^4{\theta _2}}\nonumber\\
& & \times \frac{ {M_{\rm UV}} {{\cal R}_{\rm UV}}^4 }{ {{\cal R}_{D5/\overline{D5}}^{\rm bh}}^4 {g_s^{\rm UV}}^{14/3} N^{39/20} {N_f^{\rm UV}}^{8/3} \alpha _{\theta _1}^7
   \alpha _{\theta _2}^4 \log ^{\frac{11}{3}}(N) \log\left(\frac{{{\cal R}_{\rm UV}}}{{{\cal R}_{D5/\overline{D5}}^{\rm bh}}}\right)}.\nonumber\\
& & 
\end{eqnarray}
We further note that at constant $r$ surface counter term to cancel UV divergence appearing from higher derivative term will be given by the following expression:
{\footnotesize
\begin{eqnarray}
\label{isqrthinvGdeltaJ0-constr}
& & \hskip -0.3in  \beta\left.\frac{\left.\int \sqrt{-h}h^{mn}\frac{\delta J_0}{\delta h^{mn}}\right|_{\rm UV-div}}{{\cal V}_4}\right|_{r=\rm constant} \sim -\beta\frac{\left(\frac{1}{{g_s}^{\rm UV}}\right)^{2/3} {M_{\rm UV}} {{\cal R}_{\rm UV}}^4 }{{g_s^{\rm UV}}^{13/4}{{\cal R}_{D5/\overline{D5}}^{\rm bh}}^4 {\log N}^3 \left\{\log\left(\frac{{{\cal R}_{\rm UV}}}{{{\cal R}_{D5/\overline{D5}}^{\rm bh}}}\right)\right\}^6 N^{11/5} {N_f^{\rm UV}}^2}\nonumber\\
& & \hskip -0.3in \times\int_{\theta_1={\alpha_{\theta_1}}{N^{-\frac{1}{5}}}}^{\pi-{\alpha_{\theta_1}}{N^{-\frac{1}{5}}}}\int_{\theta_2= {\alpha_{\theta_2}}{N^{-\frac{3}{10}}}}^{\pi- {\alpha_{\theta_2}}{N^{-\frac{3}{10}}}} \frac{19683 \sqrt{6} \sin^6{\theta _1}+6642 \sin^2{\theta _2} \sin^3{\theta _1}-40 \sqrt{6}
   \sin^4{\theta _2}}{\sin^7{\theta _1}\sin^4{\theta _2}} + {\cal O}(\beta^2).
\end{eqnarray}
}
From:
{\footnotesize
\begin{eqnarray}
\label{theta1-theta2-global-intsqrtGiGdeltaJ0-i}
& & \hskip -0.3in \frac{1}{N^{\frac{7}{5}}}\int\int d\theta_1 d\theta_2 \frac{19683 \sqrt{6} \sin^6{\theta _1}+6642 \sin^2{\theta _2} \sin^3{\theta _1}-40 \sqrt{6}
   \sin^4{\theta _2}}{\sin^7{\theta _1}\sin^4{\theta _2}}\nonumber\\
& & \hskip -0.3in  =  \frac{1}{N^{\frac{7}{5}}}\Biggl[-2214 \cot ^3({\theta_1}) \cot ({\theta_2})+2214 \cot ({\theta_1}) \left(2 \csc
   ^2({\theta_1})+1\right) \cot ({\theta_2})+\frac{1}{32 \sqrt{6}}\Biggl\{20 {\theta_2} \csc
   ^6\left(\frac{{\theta_1}}{2}\right)+120 {\theta_2} \csc
   ^4\left(\frac{{\theta_1}}{2}\right)\nonumber\\
& & \hskip -0.3in  +600 {\theta_2} \csc
   ^2\left(\frac{{\theta_1}}{2}\right)+5 {\theta_2} \sec
   ^6\left(\frac{{\theta_1}}{2}\right) \Biggl(-150 \log \left(\sin
   \left(\frac{{\theta_1}}{2}\right)\right)+15 \cos (3 {\theta_1}) \log \left(\cos
   \left(\frac{{\theta_1}}{2}\right)\right)+150 \log \left(\cos
   \left(\frac{{\theta_1}}{2}\right)\right)\nonumber\\
& & \hskip -0.3in +9 \cos ({\theta_1}) \left(-25 \log
   \left(\sin \left(\frac{{\theta_1}}{2}\right)\right)+25 \log \left(\cos
   \left(\frac{{\theta_1}}{2}\right)\right)-8\right)+15 \cos (2 {\theta_1}) \left(-6
   \log \left(\sin \left(\frac{{\theta_1}}{2}\right)\right)+6 \log \left(\cos
   \left(\frac{{\theta_1}}{2}\right)\right)-1\right)\nonumber\\
& & \hskip -0.3in -15 \cos (3 {\theta_1}) \log
   \left(\sin \left(\frac{{\theta_1}}{2}\right)\right)-61\Biggr)-1259712 (\cos (2
   {\theta_2})-2) \cot ({\theta_2}) \csc ^2({\theta_2}) \left(\log \left(\cos
   \left(\frac{{\theta_1}}{2}\right)\right)-\log \left(\sin
   \left(\frac{{\theta_1}}{2}\right)\right)\right)\Biggr\}\Biggr],\nonumber\\
   & &
\end{eqnarray}
}
one notes that the dominant contribution arises from small values of $\theta_{1,2}$, i.e., along Ouyang embedding for small Ouyang embedding paramter('s modulus). Hence, 
\begin{eqnarray}
\label{theta1-theta2-global-intsqrtGiGdeltaJ0-ii}
& &  \int_{\theta_1={\alpha_{\theta_1}}{N^{-\frac{1}{5}}}}^{\pi-{\alpha_{\theta_1}}{N^{-\frac{1}{5}}}}\int_{\theta_2= {\alpha_{\theta_2}}{N^{-\frac{3}{10}}}}^{\pi- {\alpha_{\theta_2}}{N^{-\frac{3}{10}}}} \frac{19683 \sqrt{6} \sin^6{\theta _1}+6642 \sin^2{\theta _2} \sin^3{\theta _1}-40 \sqrt{6}
   \sin^4{\theta _2}}{\sin^7{\theta _1}\sin^4{\theta _2}}\nonumber\\
& &  \sim N^{\frac{9}{10}}\frac{ \left(-6642 {\alpha_{\theta_1}}^3 {\alpha_{\theta_2}}^2+19683 \sqrt{6}
   {\alpha_{\theta_1}}^6 \log \left(\frac{{\alpha_{\theta_1}}}{2 \sqrt[5]{N}}\right)-20
   \sqrt{6} {\alpha_{\theta_2}}^4\right)}{3 {\alpha_{\theta_1}}^6 {\alpha_{\theta_2}}^3}
\sim -\frac{\log N}{\alpha_{\theta_2}^3}N^{\frac{9}{10}}.
\end{eqnarray}
Therefore simplified form of the counter term is: 
{\footnotesize
\begin{eqnarray}
\label{ihdeltaJ0-r=rUV}
& & \hskip -0.3in  \beta\left.\frac{\left.\int \sqrt{-h}h^{mn}\frac{\delta J_0}{\delta h^{mn}}\right|_{\rm UV-div}}{{\cal V}_4}\right|_{r=\rm constant} \sim \beta\frac{\left(\frac{1}{{g_s}^{\rm UV}}\right)^{2/3} {M_{\rm UV}} {{\cal R}_{\rm UV}}^4 \log N}{{g_s^{\rm UV}}^{13/4}{{\cal R}_{D5/\overline{D5}}^{\rm bh}}^4 {\log N}^3 \left\{\log\left(\frac{{{\cal R}_{\rm UV}}}{{{\cal R}_{D5/\overline{D5}}^{\rm bh}}}\right)\right\}^6 N^{13/10} {N_f^{\rm UV}}^2\alpha_{\theta_2}^3}.
\end{eqnarray}
}
 
Also,
\begin{eqnarray}
\label{sqrthJ0-constr}
& & \beta\left.\frac{\int_{r={\cal R}_{\rm UV}} \sqrt{-h} J_0}{{\cal V}_4}\right|^{\rm UV}\nonumber\\
& & \sim \frac{M_{\rm UV}}{g_s^{\rm UV}\log^{\frac{8}{3}}N^{3}N_f^{\rm UV}\ ^{\frac{8}{3}}}\left(\frac{{{\cal R}_{\rm UV}}}{{{\cal R}_{D5/\overline{D5}}^{\rm bh}}}\right)^4\log\left(\frac{{{\cal R}_{\rm UV}}}{{{\cal R}_{D5/\overline{D5}}^{\rm bh}}}\right)\int_{\theta_1={\alpha_{\theta_1}}{N^{-\frac{1}{5}}}}^{\pi-{\alpha_{\theta_1}}{N^{-\frac{1}{5}}}}\int_{\theta_2= {\alpha_{\theta_2}}{N^{-\frac{3}{10}}}}^{\pi- {\alpha_{\theta_2}}{N^{-\frac{3}{10}}}}
 \frac{d\theta_1 d\theta_2}{\sin^3\theta_1\sin^2\theta_2}\nonumber\\
& & \sim \frac{M_{\rm UV}}{g_s^{\rm UV}\log^{\frac{8}{3}}N^{\frac{23}{10}}N_f^{\rm UV}\ ^{\frac{8}{3}}}\left(\frac{{{\cal R}_{\rm UV}}}{{{\cal R}_{D5/\overline{D5}}^{\rm bh}}}\right)^4\log\left(\frac{{{\cal R}_{\rm UV}}}{{{\cal R}_{D5/\overline{D5}}^{\rm bh}}}\right). 
\end{eqnarray}
By comparing (\ref{int-sqrtGiGdeltaJ0-beta0-UV-div}) with (\ref{isqrthinvGdeltaJ0-constr}), one imposes:
\begin{equation}
\label{Nf_UV}
\log ^3N\left\{\log\left(\frac{{{\cal R}_{\rm UV}}}{{{\cal R}_{D5/\overline{D5}}^{\rm bh}}}\right)\right\}^5  {N_f^{\rm UV}}^2 \sim 
 {N_f^{\rm UV}}^{8/3}\log ^{\frac{11}{3}}(N),
\end{equation}
implying:
\begin{equation}
\label{Nf-LogRUV}
N_f^{\rm UV}\sim\frac{\left(\log\left(\frac{{{\cal R}_{\rm UV}}}{{{\cal R}_{D5/\overline{D5}}^{\rm bh}}}\right)\right)^{\frac{15}{2}}}{\log N}.
\end{equation}

In $\kappa_{11}^2=1$-units, the $D=11$ supergravity action inclusive of ${\cal O}(R^4)$ terms is given by:
\begin{eqnarray}
& & S_{D=11} = \frac{1}{2}\Biggl[\int_{M_{11}} \sqrt{-G^{\mathscr {M}}}\left(R - \frac{G_4^2}{2. 4!}
+ \beta\left(J_0 - \frac{E_8}{2}\right) \right) \nonumber\\
& & + 2\int_{\partial M_{11}}\sqrt{-h}K - \frac{1}{6}\int_{M_{11}}C_3\wedge G_4\wedge G_4
+ 4\pi^2\int_{M_{11}}C_3\wedge X_8\Biggr],
\end{eqnarray}
which yields as an EOM:
\begin{eqnarray}
\label{GMN_EOM}
& & R_{MN} - \frac{G_{MN}}{2}R - \frac{1}{12}\left(G_{MPQR}G_N^{\ \ PQR} - \frac{G_{MN}}{8}G_4^2\right) \nonumber\\
& & = -\beta\left[\frac{G_{MN}}{2}\left(J_0 - \frac{E_8}{2}\right) + \frac{\delta}{\delta G^{MN}}\left(J_0 - \frac{E_8}{2}\right)\right].
\end{eqnarray}
Taking the trace of (\ref{GMN_EOM}) one obtains:
\begin{eqnarray}
\label{trace}
& & -\frac{9}{2}R + \frac{G_4^2}{32} = -\beta\left[\frac{11}{2}\left(J_0 - \frac{E_8}{2}\right) + G^{MN}
\frac{\delta}{\delta G^{MN}}\left(J_0 - \frac{E_8}{2}\right)\right]\nonumber\\
& & \approx -\beta\left[\frac{11}{2}J_0+ G^{MN}
\frac{\delta J_0}{\delta G^{MN}}\right],
\end{eqnarray}
where the approximation is justified in \cite{OR4-Yadav+Misra}. Writing $R = R^{(0)} + \beta R,
K = K^{(0)} + \beta K, C_3 = C_3^{(0)} + \beta C_3$, etc., in \cite{OR4-Yadav+Misra} it is also shown that one can self consistently set $C_3=0$. One thus notes:
\begin{eqnarray}
\label{on-shell-relations}
& & G_4^2 = 144 R^{(0)},\nonumber\\
& &  J_0 \approx \frac{9}{11} R^{(1)} - \frac{2}{11}G^{MN} \frac{\delta J_0}{\delta G^{MN}}.
\end{eqnarray}
Therefore, the on-shell action up to ${\cal O}(\beta)$ using $X_8(M_{11})=0$ \cite{MQGP}, becomes:
{\footnotesize
\begin{equation}
\label{on-shell-D=11-action-up-to-beta}
 S_{D=11}^{\rm on-shell} =- \frac{1}{2}\Biggl[-2 S_{\rm EH}^{(0)} + 2 S_{\rm GHY}^{(0)} 
+ \beta \left(\frac{20}{11}S_{\rm EH} - 2 \int_{M_{11}}\left(\sqrt{-G^{\mathscr {M}}}\right)^{(1)}R^{(0)}
+ 2 S_{\rm GHY} - \frac{2}{11}\int_{M_{11}}\sqrt{-G^{\mathscr {M}}}G^{MN}\frac{\delta J_0}{\delta G^{MN}}\right)\Biggr],
\end{equation}
}
where $\left(\sqrt{-G^{\mathscr {M}}}\right)^{(1)}=\frac{G_{\mathscr {M}}^{(1)}}{2\sqrt{-G_{\mathscr {M}}^{(0)}} }$. To see the holographic renormalization of (\ref{on-shell-D=11-action-up-to-beta}), its UV-divergent portion is:
{\footnotesize
\begin{eqnarray}
\label{on-shell-UV-divergent-BH}
 S_{\rm D=11}^{\rm on-shell UV-divergent} =- \frac{1}{2}\Biggl[\left(-2\kappa_{\rm EH}^{(0)}(g_s^{\rm UV},N,M_{\rm UV},N_f^{\rm UV};\alpha_{\theta_{1,2}}) + 2\kappa^{(0)}_{\rm GHY}(g_s^{\rm UV},N,M_{\rm UV},N_f^{\rm UV};\alpha_{\theta_{1,2}})\right)\nonumber\\
& &  \hskip -5in \times
\left(\frac{{{\cal R}_{\rm UV}}}{{{\cal R}_{D5/\overline{D5}}^{\rm bh}}}\right)^4\log\left(\frac{{{\cal R}_{\rm UV}}}{{{\cal R}_{D5/\overline{D5}}^{\rm bh}}}\right) + \beta\Biggl\{ -\frac{2}{11}\kappa^{(0)}_{\sqrt{-g}g^{MN}\frac{\delta J_0}{\delta g^{MN}}}(g_s^{\rm UV},N,M_{\rm UV},N_f^{\rm UV};\alpha_{\theta_{1,2}})\frac{\left(\frac{{{\cal R}_{\rm UV}}}{{{\cal R}_{D5/\overline{D5}}^{\rm bh}}}\right)^4}{\log\left(\frac{{{\cal R}_{\rm UV}}}{{{\cal R}_{D5/\overline{D5}}^{\rm bh}}}\right)}\Biggr\}\Biggr].\nonumber\\
& &
\end{eqnarray}
}
Therefore, the counter terms required to cancel the UV-divergent terms are:
\begin{eqnarray}
\label{CT-BH-UV-divergent}
& & \hskip -0.3in S_{\rm CT}^{\rm BH}  = - \frac{1}{2}\Biggl[-\frac{\kappa^{(0)}_{\rm EH+GHY}}{\kappa^{(0)}_{{\rm EH}@{\cal R}_{\rm UV}}}\int_{r={\cal R}_{\rm UV}}\sqrt{-h}R\nonumber\\
& & \hskip -0.3in  + \beta N^{\frac{1}{4}} \frac{\left(\frac{2}{11}\kappa^{(0)}_{\sqrt{-g}g^{MN}\frac{\delta J_0}{\delta g^{MN}}}(g_s^{\rm UV},M_{\rm UV},N_f^{\rm UV};\alpha_{\theta_{1,2}})\right)}{\kappa^{(0)}_{\sqrt{-h}g^{mn}\frac{\delta J_0}{\delta g^{mn}}@{\cal R}_{\rm UV}}} \int_{r={\cal R}_{\rm UV}}\sqrt{-h}g^{mn}\frac{\delta J_0}{\delta g^{mn}}\Biggr]_{\log\left(\frac{{{\cal R}_{\rm UV}}}{{{\cal R}_{D5/\overline{D5}}^{\rm bh}}}\right) = \left(N_f^{\rm UV}\log N\right)^{\frac{2}{15}}}.\nonumber\\
& & 
\end{eqnarray}

\subsection{Thermal Background Uplift (Relevant to $T<T_c$) and Holographic Renormalization of On-Shell $D=11$ Action}

In the large-$N$ limit and in the IR, the $f_{MN}$ EOMs for the thermal background are algebraic. Writing below only those components which pick up a non-trivial ${\cal O}(\beta)$ corrections, here is the ${\cal O}(R^4)$-corrected MQGP metric for thermal background in the IR in the $\psi=2n\pi,n=0,1,2$-coordinate patches:
\begin{eqnarray}
\label{MQGP-th-OR4}
& & G^{\mathscr {M}}_{rr} = G^{\rm MQGP}_{rr}\left(1-\frac{99 \sqrt{\frac{3}{2}} \beta  {g_s}^{3/2} M \sqrt[5]{\frac{1}{N}} N_f  {r_0} \alpha _{\theta _1}^6
   f_{x^{10}x^{10}}({r_0}) \log ^2({r_0})}{2 \pi ^{3/2} \alpha _{\theta _2}^5}\right),\nonumber\\
& & G^{\mathscr {M}}_{x\theta_1} = G^{\rm MQGP}_{x\theta_1}\left(1 - \beta f_{x^{10}x^{10}}(r_0)\right),\nonumber\\
& & G^{\mathscr {M}}_{y\theta_1} = G^{\rm MQGP}_{y\theta_1}\left(1 + \beta f_{\theta_1y}(r_0)\right),\nonumber\\
& & G^{\mathscr {M}}_{z\theta_1} = G^{\rm MQGP}_{z\theta_1}\left(1+\frac{539 \pi ^3 \beta  N^{2/5} \alpha _{\theta _2}^2 f_{x^{10}x^{10}}({r_0})}{1728 {g_s}^3 M^2 N_f ^2 \log^2({r_0})}\right),\nonumber\\
& & G^{\mathscr {M}}_{y\theta_2} = G^{\rm MQGP}_{y\theta_2}\left(1 - 2\beta f_{x^{10}x^{10}}(r_0)\right),\nonumber\\
& & G^{\mathscr {M}}_{yy} = G^{\rm MQGP}_{yy}\left(1-\frac{\pi ^{3/2} \beta  N^{2/5} \alpha _{\theta _2}^2 f_{x^{10}x^{10}}({r_0}) \left(1617 \sqrt{3} \pi ^{3/2} {r_0}
   \alpha _{\theta _1}^4-32 \sqrt{2} {g_s}^{3/2} M N_f  \alpha _{\theta _2}\right)}{2592 \sqrt{3} {g_s}^3
   M^2 N_f ^2 {r_0} \alpha _{\theta _1}^4 \log ^2({r_0})}\right),\nonumber\\
& & G^{\mathscr {M}}_{yz} = G^{\rm MQGP}_{yz}\left(1 + \frac{\pi ^{3/2} \beta  N^{2/5} \alpha _{\theta _2}^3 f_{x^{10}x^{10}}({r_0})}{81 \sqrt{6} {g_s}^{3/2} M N_f 
   {r_0} \alpha _{\theta _1}^4 \log ^2({r_0})}\right),\nonumber\\
& & G^{\mathscr {M}}_{zz} = G^{\rm MQGP}_{zz}\left(1+\frac{539 \pi ^3 \beta  N^{2/5} \alpha _{\theta _2}^2 f_{x^{10}x^{10}}({r_0})}{864 {g_s}^3 M^2 N_f ^2 \log
   ^2({r_0})}\right),\nonumber\\
& & G^{\mathscr {M}}_{x^{10}x^{10}} = G^{\rm MQGP}_{x^{10}x^{10}}\left(1 + \beta f_{x^{10}x^{10}}(r_0)\right).
\end{eqnarray}

Working in the IR and near (\ref{small-theta_12}), Einstein-Hilbert  term for thermal background is:
\begin{eqnarray}
\label{EH-thermal-IR}
& & \left.\sqrt{-G^{\mathscr {M}}}R\right|_{\rm Ouyang}^{\rm IR}\sim  -\frac{{g_s}^{3/4} {\log N}^2 \log \left(\frac{r_0}{{\cal R}_{D5/\overline{D5}}^{\rm th}}\right) M {N_f}^3 }{{{\cal R}_{D5/\overline{D5}}^{\rm th}}^3 N^{\frac{5}{4}} \sin^3\theta_1\sin^2\theta_2}\sum_{m=0}^2\kappa_{\rm EH,\ IR}^{\beta^0,m}\ {r_0}^{3-m}\left(r-r_0\right)^{m}\nonumber\\
& &  -\beta  \frac{
   {\log N}  \left(243 \sqrt{6} \sin^3{\theta _1}-8 \sin^2{\theta _2}\right)
   f_{x^{10}x^{10}}({r_0})}{ N^{\frac{1}{4}}\sin^5{\theta _1}{{\cal R}_{D5/\overline{D5}}^{\rm th}}^3 {g_s}^{9/4}  \log \left(\frac{r_0}{{\cal R}_{D5/\overline{D5}}^{\rm th}}\right) M  ({\log N}
   {N_f})^{5/3}}\sum_{m=0}^2\kappa_{\rm EH,\ IR}^{\beta^1,m}\ {r_0}^{3-m}\left(r-r_0\right)^{m},
\end{eqnarray} 
where ${\cal R}_{D5/\overline{D5}}^{\rm th} = \sqrt{3}a^{\rm th}$, $a^{\rm th} = \left(\frac{1}{\sqrt{3}} + \epsilon + {\cal O}\left(\frac{g_sM^2}{N}\right)\right)r_0$ \cite{OR4-Yadav+Misra} being the resolution parameter for the blown up $S^2$ in the thermal background, and $\kappa_{\rm EH,\ IR}^{\beta^m,n}$ are the numerical prefactors - which we have worked out but have not given here as they are not particularly illuminating in the $\beta^m(r-r_0)^n$ terms in $\sqrt{-G^{\mathscr {M}}}R$ restricted to the Ouyang embedding of the flavor $D7$-branes in the type IIB dual and in the IR. From (\ref{EH-thermal-IR}), one can evaluate $\frac{\left.\int_{r=r_0}^{{\cal R}_{D5-\overline{D5}}}\sqrt{-G^{\mathscr {M}}}R\right|_{\rm Ouyang}}{{\cal V}_4}$
where ${\cal V}_4$ is the coordinate volume of $S^1(x^0)\times\mathbb{R}^3(x^{1,2,3})$.

Also additional term that appear in on-shell action (\ref{on-shell-relations}) at $O(\beta)$ is:
\begin{eqnarray}
\label{intsqrtGbetaRbeta0}
& & \frac{\int_{r_h}^{{\cal R}_{D5-\overline{D5}}^{\rm bh}}\int_{\theta_1={\alpha_{\theta_1}}{N^{-\frac{1}{5}}}}^{\pi-{\alpha_{\theta_1}}{N^{-\frac{1}{5}}}}\int_{\theta_2= {\alpha_{\theta_2}}{N^{-\frac{3}{10}}}}^{\pi- {\alpha_{\theta_2}}{N^{-\frac{3}{10}}}}
\left(\sqrt{-G^{\mathscr {M}}}\right)^{(1)}R^{(0)}}{{\cal V}_4}\nonumber\\
& & \sim \frac{\beta  {g_s}^{3/4} {\log N}^2 M \left(\frac{1}{N}\right)^{11/20} {N_f}^3 {r_0}
  {f_{x^{10}x^{10}}}({r_0}) \left(1-\frac{{r_0}}{{\cal R}_{D5/\overline{D5}}^{\rm th}}\right) \log
   \left(\frac{{r_0}}{{\cal R}_{D5/\overline{D5}}^{\rm th}}\right)}{{{\cal R}_{D5/\overline{D5}}^{\rm th}} \alpha _{\theta _1}^2 \alpha
   _{\theta _2}}.
\end{eqnarray}

Similarly Einstein-Hilbert  term in the UV is:
{\footnotesize
\begin{eqnarray}
\label{EH-thermal-UV}
& & \left.\sqrt{-G^{\mathscr {M}}}R\right|_{\rm Ouyang}^{\rm UV} =  \kappa_{\rm EH,\ UV}^{\beta^0,0}\ \frac{\Biggl[11-8 {\log \left(\frac{{r}}{{{\cal R}_{D5/\overline{D5}}^{\rm th}}}\right)}\Biggr] {M_{\rm UV}} {N_f^{\rm UV}} {r_0}^4}{{{\cal R}_{D5/\overline{D5}}^{\rm th}}^3 {g_s}^{5/4}  r
  N^{\frac{5}{4}} \sin^3\theta_1\sin^2\theta_2}\nonumber\\
& & +\kappa_{\rm EH,\ UV}^{\beta^0,1}\ \frac{\Biggl[4 {\log \left(\frac{{r}}{{{\cal R}_{D5/\overline{D5}}^{\rm th}}}\right)}-1\Biggr] {M_{\rm UV}} {N_f^{\rm UV}} r {r_0}^2}{{{\cal R}_{D5/\overline{D5}}^{\rm th}}^3 {g_s}^{5/4} N^{\frac{5}{4}} \sin^3\theta_1\sin^2\theta_2}+\kappa_{\rm EH,\ UV}^{\beta^0,2}\ \frac{{M_{\rm UV}}
   {N_f^{\rm UV}} r^3}{18 \sqrt{3} \pi ^{9/4}{{\cal R}_{D5/\overline{D5}}^{\rm th}}^3 {g_s}^{5/4}N^{\frac{5}{4}} \sin^3\theta_1\sin^2\theta_2},\nonumber\\
& & 
\end{eqnarray}
}
implying: 
\begin{eqnarray}
\label{int-EH-thermal-UV}
& &  \frac{\left.\int_{\theta_1={\alpha_{\theta_1}}{N^{-\frac{1}{5}}}}^{\pi-{\alpha_{\theta_1}}{N^{-\frac{1}{5}}}}\int_{\theta_2= {\alpha_{\theta_2}}{N^{-\frac{3}{10}}}}^{\pi- {\alpha_{\theta_2}}{N^{-\frac{3}{10}}}} \int_{{\cal R}_{D5-\overline{D5}}}^{{\cal R}_{\rm UV}}dr \sqrt{-G^{\mathscr {M}}}R\right|_{\rm Ouyang}^{\rm UV-Finite}}{{\cal V}_4}\nonumber\\
& & \sim 
 -\kappa_{\rm EH,\ UV}^{\beta^0,0}\ \int_{\theta_1={\alpha_{\theta_1}}{N^{-\frac{1}{5}}}}^{\pi-{\alpha_{\theta_1}}{N^{-\frac{1}{5}}}}\int_{\theta_2= {\alpha_{\theta_2}}{N^{-\frac{3}{10}}}}^{\pi- {\alpha_{\theta_2}}{N^{-\frac{3}{10}}}} \frac{{M_{\rm UV}} {N_f^{\rm UV}}
   \left(-\frac{121 {r_0}^4}{16 {{\cal R}_{D5/\overline{D5}}^{\rm th}}^4}-\frac{6 {r_0}^2}{{{\cal R}_{D5/\overline{D5}}^{\rm th}}^2}+2\right)}{ {g_s^{\rm UV}}^{5/4} N^{\frac{5}{4}} \sin^3\theta_1\sin^2\theta_2}\nonumber\\
\end{eqnarray}
Therefore, after performing radial as well as angular integral of the above equation UV finite Einstein-Hilbert  term is:
\begin{eqnarray}
\label{EH-thermal-global-iii}
& &  \frac{\left.\int_{\theta_1={\alpha_{\theta_1}}{N^{-\frac{1}{5}}}}^{\pi-{\alpha_{\theta_1}}{N^{-\frac{1}{5}}}}\int_{\theta_2= {\alpha_{\theta_2}}{N^{-\frac{3}{10}}}}^{\pi- {\alpha_{\theta_2}}{N^{-\frac{3}{10}}}} \int_{{\cal R}_{D5-\overline{D5}}}^{{\cal R}_{\rm UV}}dr \sqrt{-G^{\mathscr {M}}}R\right|_{\rm Ouyang}^{\rm UV-Finite}}{{\cal V}_4}\nonumber\\
& & \sim 
 -\kappa_{\rm EH,\ UV}^{\beta^0,0}\ \frac{{M_{\rm UV}} {N_f^{\rm UV}}
   \left(-\frac{121 {r_0}^4}{16 {{\cal R}_{D5/\overline{D5}}^{\rm th}}^4}\frac{1}{\alpha_{\theta_1}^2\alpha_{\theta_2}}-\frac{6 {r_0}^2}{{{\cal R}_{D5/\overline{D5}}^{\rm th}}^2}+2\right)}{ {g_s^{\rm UV}}^{5/4} N^{\frac{11}{20}} }\frac{1}{\alpha_{\theta_1}^2\alpha_{\theta_2}} \nonumber\\
\end{eqnarray}
UV divergent Einstein-Hilbert  term for thermal background is:
\begin{eqnarray}
\label{EH-th-UV-div}
& & \frac{S_{\rm EH - thermal}^{\rm UV-divergent}}{{\cal V}_4} \sim \frac{\kappa_{\rm EH,\ UV}^{\beta^0,2}}{4}\frac{{M_{\rm UV}} {N_f^{\rm UV}} {\cal R}_{\rm UV}^4}{{{\cal R}_{D5/\overline{D5}}^{\rm th}}^4{g_s^{\rm UV}}^{5/4}N^{\frac{11}{20}} }\frac{1}{\alpha_{\theta_1}^2\alpha_{\theta_2}}. \nonumber\\
\end{eqnarray}

The UV-finite part of the boundary Gibbons-Hawking-York term for the thermal background up to ${\cal O}(\beta)$ turns to be:
\begin{eqnarray}
\label{GHY-thermal}
& & \hskip -0.5in \frac{\left.\int \sqrt{-h^{\mathscr {M}}}K\right|_{\rm Ouyang}^{r={\cal R}_{\rm UV}\ {\rm UV-finite}}}{{\cal V}_4}\nonumber\\
& & \hskip -0.5in \sim\frac{ {r_0}^4 \left( \kappa_{\rm GHY}^{\beta^0, {\rm UV-finite}}\ \frac{{g_s}^3 {M_{\rm UV}}^2 \log
   \left(\frac{{{\cal R}_{\rm UV}}}{{{\cal R}_{D5/\overline{D5}}^{\rm th}}}\right)}{N^{\frac{11}{20}}}\frac{1}{\alpha_{\theta_1}^2\alpha_{\theta_2}}\right)}{{{\cal R}_{D5/\overline{D5}}^{\rm th}}^4 {g_s}^{21/4}
   {M_{\rm UV}} }.\nonumber\\
& & 
\end{eqnarray}
Further UV divergent part of the Gibbons-Hawking-York term for the thermal background is:
\begin{eqnarray}
\label{GHY-UV-divergent}
& & \frac{S_{\rm GHY}^{\rm UV-divergent}}{{\cal V}_4} = -\kappa_{\rm GHY}^{\beta^0,\ {\rm UV-div}}\ \frac{ \log
   \left(\frac{{{\cal R}_{\rm UV}}}{{{\cal R}_{D5/\overline{D5}}^{\rm th}}}\right) {M_{\rm UV}} {{\cal R}_{\rm UV}}^4}{{{\cal R}_{D5/\overline{D5}}^{\rm th}}^4
   {g_s^{\rm UV}}^{9/4} N^{\frac{11}{20}} \alpha_{\theta_1}^2\alpha_{\theta_2}}.\nonumber\\
\end{eqnarray}

One can further show that  up to ${\cal O}(\beta^0)$ contribution from higher derivative term in the IR is:
\begin{eqnarray}
\label{iGdeltaJ0-thermal-IR}
& & \left.G^{MN}\frac{\delta J_0}{\delta G^{MN}}\right|_{\rm Ouyang}^{\rm IR}\sim\frac{ \left(19683 \sqrt{6} \sin^6{\theta _1}+6642 \sin^2{\theta _2} \sin^3{\theta
   _1}-40 \sqrt{6} \sin^4{\theta _2}\right)}{ \epsilon ^8 {g_s}^{9/4} {\log N}^2 N^{\frac{6}{5}}
   \sin^4\theta_1 \sin^2\theta_2{N_f}^2  \Biggl[{N_f} \left\{{\log N}-3 \log \left(\frac{{r_0}}{{{\cal R}_{D5/\overline{D5}}^{\rm th}}}\right)\right\}\Biggr]^{2/3}},\nonumber\\
& & 
\end{eqnarray}
implying:
{\footnotesize
\begin{eqnarray}
\label{int-iGdeltaJ0-thermal-IR}
& & \hskip -0.6in \frac{\int_{r_0}^{{\cal R}_{D5/\overline{D5}}}\sqrt{-G^{\mathscr {M}}}\left.G^{MN}\frac{\delta J_0}{\delta G^{MN}}\right|_{\rm Ouyang}}{{\cal V}_4}
\sim\nonumber\\
& & \hskip -0.6in  \kappa_{\footnotesize G^{MN}\frac{\delta J_0}{\delta G^{MN}}}^{\beta^0,\ \rm IR}\ \frac{M {N_f} {r_0}^2 \left(1-\frac{{r_0}}{{{\cal R}_{D5/\overline{D5}}^{\rm th}}}\right)^2 \log \left(\frac{{r_0}}{{{\cal R}_{D5/\overline{D5}}^{\rm th}}}\right)
  }{ \epsilon ^8 {g_s}
   {\log N}^2 N^{21/20} {{\cal R}_{D5/\overline{D5}}^{\rm th}}^2 }\frac{ \left(-6642 {\alpha_{\theta_1}}^3 {\alpha_{\theta_2}}^2+19683 \sqrt{6}
   {\alpha_{\theta_1}}^6 \log \left(\frac{{\alpha_{\theta_1}}}{2 \sqrt[5]{N}}\right)-20
   \sqrt{6} {\alpha_{\theta_2}}^4\right)}{3 {\alpha_{\theta_1}}^6 {\alpha_{\theta_2}}^3}\nonumber\\
& & \times  \left\{{\log N}-9 \log \left(\frac{{r_0}}{{{\cal R}_{D5/\overline{D5}}^{\rm th}}}\right)\right\} \Biggl[{\log N}-3 \log\left(\frac{{r_0}}{{{\cal R}_{D5/\overline{D5}}^{\rm th}}}\right)\Biggr]^2.\nonumber\\
& & 
\end{eqnarray}
}
Further, near (\ref{small-theta_12}) contribution from higher derivative term in the UV is:
\begin{eqnarray}
\label{iGdeltaJ0-thermal-UV}
& & \left.G^{MN}\frac{\delta J_0}{\delta G^{MN}}\right|_{\rm Ouyang}^{\rm UV}=  {g_s^{\rm UV}}^{1/4} {\log \tilde{r}} {M_{\rm UV}} \frac{\left(19683 \sin^6{\theta _1}+1107 \sqrt{6}
   \sin^2{\theta _2}\sin^3{\theta _1}-40 \sin^4{\theta _2}\right)}{ {g_s^{\rm UV}}^{17/4} {\log N}^3 (2
   {\log  \tilde{r}}+1)^7 N^{39/20} {N_f^{\rm UV}}^2  \tilde{r} \sin^7{\theta _1}\sin^4{\theta _2}}\nonumber\\
& & \times \Biggl(-\kappa_{\footnotesize G^{MN}\frac{\delta J_0}{\delta G^{MN}}}^{\beta^0,0}\ { a^4 \left(64
   {\log  \tilde{r}}^3+208 {\log  \tilde{r}}^2+212 {\log  \tilde{r}}+57\right)}
\nonumber\\
& & -\kappa_{\footnotesize G^{MN}\frac{\delta J_0}{\delta G^{MN}}}^{\beta^0,1}\ {8 a^2 \left(4
   {\log  \tilde{r}}^2+15 {\log \tilde{r}}+9\right)  \tilde{r}} -\kappa_{\footnotesize G^{MN}\frac{\delta J_0}{\delta G^{MN}}}^{\beta^0,2}\ { \tilde{r}^3}\Biggr),\nonumber\\
& & 
\end{eqnarray}
where $\tilde{r} \equiv \frac{{r}}{{{\cal R}_{D5/\overline{D5}}^{\rm th}}}$,
thus after performing angular and radial integration of the above equation the UV-finite part will be:
\begin{eqnarray}
\label{int-iGdeltaJ0-thermal-UV}
& &  \frac{\int_{\theta_1={\alpha_{\theta_1}}{N^{-\frac{1}{5}}}}^{\pi-{\alpha_{\theta_1}}{N^{-\frac{1}{5}}}}\int_{\theta_2= {\alpha_{\theta_2}}{N^{-\frac{3}{10}}}}^{\pi- {\alpha_{\theta_2}}{N^{-\frac{3}{10}}}}\int_{{\cal R}_{D5/\overline{D5}}}^{{\cal R}_{\rm UV}}\sqrt{-G^{\mathscr {M}}}\left.G^{MN}\frac{\delta J_0}{\delta G^{MN}}\right|_{\rm Ouyang}^{\rm UV-finite}}{{\cal V}_4}\nonumber\\
& & \sim  - \kappa_{\footnotesize G^{MN}\frac{\delta J_0}{\delta G^{MN}}}^{\beta^0,\ \rm UV-finite}\frac{  {M_{\rm UV}}}{{g_s^{\rm UV}}^{4}
   {\log N}^3 N^{21/20} {N_f^{\rm UV}}^2 }\nonumber\\
& & \times\left(\frac{a^2}{{{\cal R}_{D5/\overline{D5}}^{\rm th}}^2}+1\right) \left(\frac{ \left(-6642 {\alpha_{\theta_1}}^3 {\alpha_{\theta_2}}^2+19683 \sqrt{6}
   {\alpha_{\theta_1}}^6 \log \left(\frac{{\alpha_{\theta_1}}}{2 \sqrt[5]{N}}\right)-20
   \sqrt{6} {\alpha_{\theta_2}}^4\right)}{3 {\alpha_{\theta_1}}^6 
{\alpha_{\theta_2}}^3}\right).
\end{eqnarray}
Further, UV-divergent part from equation (\ref{iGdeltaJ0-thermal-UV}) will be:
\begin{eqnarray}
\label{int-iGdeltaJ0-thermal-UV-div}
& &  \frac{\int_{\theta_1={\alpha_{\theta_1}}{N^{-\frac{1}{5}}}}^{\pi-{\alpha_{\theta_1}}{N^{-\frac{1}{5}}}}\int_{\theta_2= {\alpha_{\theta_2}}{N^{-\frac{3}{10}}}}^{\pi- {\alpha_{\theta_2}}{N^{-\frac{3}{10}}}}\int_{{\cal R}_{D5/\overline{D5}}}^{{\cal R}_{\rm UV}}\sqrt{-G^{\mathscr {M}}}\left.G^{MN}\frac{\delta J_0}{\delta G^{MN}}\right|_{\rm Ouyang}^{\rm UV-divergent}}{{\cal V}_4}\nonumber\\
& & =\kappa_{\footnotesize G^{MN}\frac{\delta J_0}{\delta G^{MN}}}^{\beta^0,\ \rm UV-div}\ \frac{{\frac{1}{{g_s^{\rm UV}}}} {M_{\rm UV}} {{\cal R}_{\rm UV}}^4}{{{\cal R}_{D5/\overline{D5}}^{\rm th}}^4{g_s^{\rm UV}}^{9/2} N^{21/20} {N_f^{\rm UV}}^{8/3} \log ^{\frac{11}{3}}(N) \log
   \left(\frac{{{\cal R}_{\rm UV}}}{{{\cal R}_{D5/\overline{D5}}^{\rm th}}}\right)}\nonumber\\
& & \times \frac{ \left(-6642 {\alpha_{\theta_1}}^3 {\alpha_{\theta_2}}^2+19683 \sqrt{6}
   {\alpha_{\theta_1}}^6 \log \left(\frac{{\alpha_{\theta_1}}}{2 \sqrt[5]{N}}\right)-20
   \sqrt{6} {\alpha_{\theta_2}}^4\right)}{3 {\alpha_{\theta_1}}^6 {\alpha_{\theta_2}}^3}
\end{eqnarray}

We note that:
\begin{eqnarray}
\label{Boundary-CT-th}
& &\left.\frac{\int \sqrt{-h}}{{\cal V}_4}\right|_{r={\cal R}_{\rm UV}} = \kappa_{ \sqrt{-h}}^{\beta^0}\ \frac{ \left(\frac{1}{{g_s^{\rm UV}}}\right)^{7/3} {M_{\rm UV}}  {{\cal R}_{\rm UV}}^4 \log
  \left(\frac{{\cal R}_{\rm UV}}{{\cal R}_{D5/\overline{D5}}}\right)}{N {{\cal R}_{D5/\overline{D5}}^{\rm th}}^4  }\nonumber\\
  & & \hskip 1.2in \times \int_{\theta_1={\alpha_{\theta_1}}{N^{-\frac{1}{5}}}}^{\pi-{\alpha_{\theta_1}}{N^{-\frac{1}{5}}}}\int_{\theta_2= {\alpha_{\theta_2}}{N^{-\frac{3}{10}}}}^{\pi- {\alpha_{\theta_2}}{N^{-\frac{3}{10}}}}
 \frac{d\theta_1 d\theta_2}{\sin^3\theta_1\sin^2\theta_2}\nonumber\\
& & = \kappa_{ \sqrt{-h}}^{\beta^0}\ \frac{ \left(\frac{1}{{g_s^{\rm UV}}}\right)^{7/3} {M_{\rm UV}}  {{\cal R}_{\rm UV}}^4 \log
  \left(\frac{{\cal R}_{\rm UV}}{{\cal R}_{D5/\overline{D5}}}\right)}{N^{\frac{3}{10}}\alpha^2_{\theta_1}\alpha_{\theta_2} {{\cal R}_{D5/\overline{D5}}^{\rm th}}^4  }   \equiv \kappa^{(0)}_{\sqrt{-h}@\partial M_{11}}\ \left(\frac{{\cal R}_{\rm UV}}{{\cal R}_{D5/\overline{D5}}^{\rm th}}\right)^4\log\left(\frac{{\cal R}_{\rm UV}}{{\cal R}_{D5/\overline{D5}}^{\rm th}}\right)\nonumber\\
& & \left.\frac{\int \sqrt{-h}R}{{\cal V}_4}\right|_{r={\cal R}_{\rm UV}}^{\rm UV-div} 
\sim \frac{\left(\frac{1}{{g_s}}\right)^{2/3} {M_{\rm UV}}  {N_f^{\rm UV}} {{\cal R}_{\rm UV}}^4}{N^{\frac{11}{10}}{{\cal R}_{D5/\overline{D5}}^{\rm th}}^4  \sqrt{{g_s}^{\rm UV}} \alpha _{\theta _1}^2 \alpha _{\theta _2}}\equiv\kappa_{\sqrt{-h}R@\partial M_{11}}^{\beta^0}\ {\cal R}_{\rm UV}^4\nonumber\\
& & \left.\frac{\int \sqrt{-h}K}{{\cal V}_4}\right|_{r={\cal R}_{\rm UV}}^{\rm UV-div} =-\kappa_{\rm GHY@\partial M_{11}}^{\beta^0,\ {\rm UV-div}}\ \frac{ \log
   \left(\frac{{{\cal R}_{\rm UV}}}{{{\cal R}_{D5/\overline{D5}}^{\rm th}}}\right) {M_{\rm UV}} {{\cal R}_{\rm UV}}^4}{{{\cal R}_{D5/\overline{D5}}^{\rm th}}^4
   {g_s^{\rm UV}}^{9/4} N^{\frac{11}{20}} \alpha_{\theta_1}^2\alpha_{\theta_2}}\nonumber\\
& & 
\nonumber\\
& & \beta\frac{\int\sqrt{-h}\left.G^{mn}\frac{\delta J_0}{\delta G^{mn}}\right|^{\rm Ouyang}_{r={\cal R}_{\rm UV}}}{{\cal V}_4}  \sim {\footnotesize -\frac{ {M_{\rm UV}} {{\cal R}_{\rm UV}}^4 \beta\left(-19683 \sqrt{6} \right)}{{{\cal R}_{D5/\overline{D5}}^{\rm th}}^4 {g_s^{\rm UV}}^{47/12} {\log N}^2 \left\{\log
   \left(\frac{{{\cal R}_{\rm UV}}}{{{\cal R}_{D5/\overline{D5}}^{\rm th}}}\right)\right\}^6 N^{13/10} {N_f^{\rm UV}}^2 
   \alpha _{\theta _2}^3}}\nonumber\\
& & \equiv \beta \kappa^{(0)}_{\sqrt{-h}G^{mn}@\partial M_{11}\frac{\delta J_0}{\delta G^{mn}}}\frac{\left(\frac{{\cal R}_{\rm UV}}{{\cal R}_{D5/\overline{D5}}}\right)^4}{\log^6\left(\frac{{{\cal R}_{\rm UV}}}{{{\cal R}_{D5/\overline{D5}}^{\rm th}}}\right)}.\nonumber\\
\end{eqnarray}

One should note that in (\ref{int-EH-BH-IR}), (\ref{sqrtGbetaRbeta0-IR-bh}), (\ref{int-EH-BH-UV}), (\ref{sqrtminush}), (\ref{EH-BH-boundary}), (\ref{GHY-BH}), (\ref{int-iGdeltaJ0-BH-IR}), (\ref{int-iGdeltaJ0--UV}), (\ref{int-sqrtGiGdeltaJ0-beta0-UV-div}), (\ref{isqrthinvGdeltaJ0-constr}), (\ref{ihdeltaJ0-r=rUV}), (\ref{sqrthJ0-constr}) (as well as (\ref{Wald-i}) and (\ref{S-J0-Wald})) for the supergravity dual with a BH corresponding to $T>T_c$, and similarly  (\ref{intsqrtGbetaRbeta0}), (\ref{int-EH-thermal-UV}), (\ref{EH-thermal-global-iii}), (\ref{GHY-thermal}), (\ref{int-iGdeltaJ0-thermal-IR}), (\ref{int-iGdeltaJ0-thermal-UV}), (\ref{int-iGdeltaJ0-thermal-UV-div}) and (\ref{Boundary-CT-th}) for the supergravity dual with a thermal background corresponding to $T<T_c$, given that the delocalized toroidal coordinates $(x, y, z)$ are defined as \cite{MQGP}: 
\begin{equation}
\label{xyz-defs}
x\sim h^{\frac{1}{4}}(r_0,\theta_{10,20})r_0\sin\theta_{10}\phi_1,\  y\sim h^{\frac{1}{4}}(r_0,\theta_{10,20})r_0\sin\theta_{20}\phi_2,\  z\sim h^{\frac{1}{4}}(r_0,\theta_{10,20})r_0\psi,
\end{equation}
where the 10-D warp factor is given by:\\ $h(r,\theta_{1,2})
= \frac{L^4}{r^4}\left[1 + \frac{3g_sM_{\rm eff}^2}{N}\log r\left\{1 + \frac{3g_sN_f^{\rm eff}}{2\pi}\left(\log r + \frac{1}{2}\right) + \frac{g_sN_f^{\rm eff}}{4\pi}\log\left(\sin \frac{\theta_1}{2}\sin \frac{\theta_2}{2}\right)\right\}\right]$,
upon promoting to global coordinates, and noting that all the aforementioned equations already have an $N^{-\zeta}, \zeta>0$ suppression implying that up to 
${\cal O}\left(\frac{1}{N}\right)$, one can approximate $h\approx \frac{L^4}{r^4}$, one sees that one will generate a $\sin\theta_1\sin\theta_2$ contribution upon integration w.r.t. $x, y, z$. As this is not accompanied with an $r$-dependence, it will only provide a common multiplicative $N$-suppressed  contribution to all the abovementioned integrals. We have hence not considered the same.

Also, note there is a multiplicative factor of $\beta_{\rm Th}\sim\frac{\sqrt{g_sN}}{r_0}$ in all terms in the action corresponding to the thermal background, e.g., (\ref{intsqrtGbetaRbeta0}), (\ref{int-EH-thermal-UV}),  (\ref{GHY-thermal}), 
(\ref{int-iGdeltaJ0-thermal-IR}), (\ref{int-iGdeltaJ0-thermal-UV-div}) and (\ref{Boundary-CT-th}).

With reference to (\ref{int-iGdeltaJ0-thermal-UV-div}) and the fourth equation in (\ref{Boundary-CT-th}), let us assume:
\begin{equation}
\label{UV-div-CT-iGdeltaJ0-1}
\left[\log
   \left(\frac{{{\cal R}_{\rm UV}}}{{{\cal R}_{D5/\overline{D5}}^{\rm th}}}\right)\right]^6 N^{-\upsilon + \frac{4}{5}} = \log
   \left(\frac{{{\cal R}_{\rm UV}}}{{{\cal R}_{D5/\overline{D5}}^{\rm th}}}\right) N^{\frac{11}{20}},
\end{equation}
or
\begin{equation}
\label{UV-div-CT-iGdeltaJ0-2}
\log
   \left(\frac{{{\cal R}_{\rm UV}}}{{{\cal R}_{D5/\overline{D5}}^{\rm th}}}\right) = N^{\frac{4\upsilon-1}{20}}.
\end{equation}
Further assuming $\upsilon = \frac{1}{4} + \epsilon$,
\begin{equation}
\label{UV-div-CT-iGdeltaJ0-3}
\log
   \left(\frac{{{\cal R}_{\rm UV}}}{{{\cal R}_{D5/\overline{D5}}^{\rm th}}}\right) \sim 1 +\frac{\epsilon }{5}\log N.
\end{equation}
From (\ref{Nf_UV}), assuming,
\begin{equation}
\label{NfLogN-more-than-1}
N_f^{\rm UV}=\frac{1}{\log N} + \epsilon_1: 0<\epsilon_1\ll1,
\end{equation}
\begin{equation}
\label{UV-div-CT-iGdeltaJ0-4}
\log
   \left(\frac{{{\cal R}_{\rm UV}}}{{{\cal R}_{D5/\overline{D5}}^{\rm th}}}\right) \sim 1 + \frac{2\epsilon_1}{15}\log N.
\end{equation}
From (\ref{UV-div-CT-iGdeltaJ0-3}) and (\ref{UV-div-CT-iGdeltaJ0-4}), $\epsilon_1  = \frac{3}{2}\epsilon$. 

One hence sees that the UV-divergent contribution of  $\frac{1}{2}\Biggl(-\frac{7}{2}S_{\rm EH}^{(0)} + 2 S_{\rm GHY}^{(0)}\Biggr)$ is given by:
\begin{eqnarray}
\label{UV-div-on-shell-thermal}
& & \tilde{\kappa}_{\rm UV-div}^{\beta^0,1}\ \left(\frac{{{\cal R}_{\rm UV}}}{{{\cal R}_{D5/\overline{D5}}^{\rm th}}}\right)^4 + \tilde{\kappa}_{\rm UV-div}^{\beta^0,2}\ \left(\frac{{{\cal R}_{\rm UV}}}{{{\cal R}_{D5/\overline{D5}}^{\rm th}}}\right)^4 \log \left(\frac{{{\cal R}_{\rm UV}}}{{{\cal R}_{D5/\overline{D5}}^{\rm th}}}\right).\nonumber\\
\end{eqnarray}
From (\ref{Boundary-CT-th}), one notices the following:
\begin{itemize}
\item
$\tilde{\kappa}_{\rm UV-div}^{\beta^0,1}\ \left(\frac{{{\cal R}_{\rm UV}}}{{{\cal R}_{D5/\overline{D5}}^{\rm th}}}\right)^4$ is canceled off by the counter term: $\int_{r={\cal R}_{\rm UV}} \sqrt{-h^{(0)}}R^{(0)}$;
\item
$ \tilde{\kappa}_{\rm UV-div}^{\beta^0,2}\ \left(\frac{{{\cal R}_{\rm UV}}}{{{\cal R}_{D5/\overline{D5}}^{\rm th}}}\right)^4 \log \left(\frac{{{\cal R}_{\rm UV}}}{{{\cal R}_{D5/\overline{D5}}^{\rm th}}}\right) $ is canceled off by the counter term $\int_{r={\cal R}_{\rm UV}}\sqrt{-h}$.
\end{itemize}

\section{$T_c$ Inclusive of ${\cal O}(R^4)$ Corrections}

In this section using the results from {\bf 3.1, 3.2}, we will calculate deconfinement temperature by equating on-shell actions for blackhole and thermal backgrounds. In the same process we will explain a version of UV-IR connection  which is arising from the ${\cal O}(R^4)$ terms. We will also calculate entropy of the black hole and discuss its consistentency with the Wald entropy up to ${\cal O}(R^4)$. 
\par
The ${\cal O}(\beta^{0,1})$ term in (\ref{on-shell-D=11-action-up-to-beta}) $\left(1+\frac{r_h^4}{2{\cal R}_{\rm UV}^4}\right)S_{D=11,\ {\rm on-shell UV-finite}}^{BH,\  \beta^{0,1}}$
and similarly
$ S_{D=11,\ {\rm on-shell UV-finite}}^{{\rm thermal},\ \beta^{0,1}}$ can now be easily computed. 

One can show that the LO terms in $N, \log\left(\frac{{\cal R}_{\rm UV}}{{\cal R}_{D5/\overline{D5}}}\right)$ and $\frac{r_h}{{\cal R}_{D5/\overline{D5}}}$ of the aforementioned BH and thermal actions are given as under:
\begin{eqnarray}
\label{bh-action-LO}
& & \left(1+\frac{r_h^4}{2{\cal R}_{\rm UV}^4}\right)S_{D=11,\ {\rm on-shell UV-finite}}^{BH} \sim
\frac{2 \kappa_{\rm GHY}^{\rm bh} {M_{\rm UV}} {r_h}^4 \log \left(\frac{{{\cal R}_{\rm UV}}}{{\cal R}_{D5/\overline{D5}}^{\rm bh}}\right)}{{g_s}^{9/4} N^{11/20}
   \alpha _{\theta _1}^2 \alpha _{\theta _2}}\nonumber\\
& & +\Biggl[{2   {\cal C}_{\theta_1x}^{\rm bh} \kappa_{\left(\sqrt{-G^{\mathscr {M}}}\right)^{(1)}R^{(0)}}^{\rm IR} }+\frac{20  \left(-{\cal C}_{zz}^{\rm bh} + 2 {\cal C}_{\theta_1z}^{\rm bh} - 3 {\cal C}_{\theta_1x}^{\rm bh}\right) \kappa_{\rm EH}^{\beta,\ \rm IR} }{11 }\Biggr]\nonumber\\
& & \times\frac{ b^2   {g_s}^{3/4}  M {N_f}^3 {r_h}^4 \log ^3(N) \log
   \left(\frac{{r_h}}{{{\cal R}_{D5/\overline{D5}}}}\right) \log
   \left(1 - \frac{{r_h}}{{{\cal R}_{D5/\overline{D5}}}}\right)}{ N^{11/20} {{\cal R}_{D5/\overline{D5}}}^4 \alpha _{\theta _1}^2 \alpha _{\theta _2}}\beta.
\nonumber\\
\end{eqnarray}
Writing ${\cal R}_{D5/\overline{D5}}^{\rm bh} = \left(1+{\sqrt{3}}\epsilon\right)r_h$, and using (\ref{alphaepsilon}), $ \log\left(1 - \frac{{r_h}}{{{\cal R}_{D5/\overline{D5}}}}\right) = \log\left(\frac{\sqrt{3}\epsilon}{1 + \epsilon}\right)\\ \sim-\left|\log \left(M_{\rm UV}\left(N_f^{\rm UV}\right)^{\frac{2}{15}}\right)\right|\sim-\log\log N$. 

Let us show the consistency of the BH entropy calculated via (\ref{bh-action-LO}) and the Wald's method. One notes from (\ref{bh-action-LO}) that up to ${\cal O}(\beta^0)$, the BH entropy ${\cal S}_{\rm BH}$ is given by:
\begin{eqnarray}
\label{entropy-semi-classical}
& & {\cal S}_{\rm BH} = \beta_{\rm BH} \frac{\partial S^E_{\rm BH}}{\partial\beta} - S^E_{\rm BH} \sim\frac{\sqrt{N}{M_{\rm UV}} {r_h}^3 \log \left(\frac{{{\cal R}_{\rm UV}}}{{\cal R}_{D5/\overline{D5}}^{\rm bh}}\right)}{{g_s}^{7/4} N^{11/20}
   \alpha _{\theta _1}^2 \alpha _{\theta _2}}\nonumber\\
& & \sim \frac{M_{\rm UV}\left(N_f^{\rm UV}\log N\right)^{\frac{2}{15}}r_h^3}{N^{\frac{1}{20}}}.
\end{eqnarray}
Note, there is a multiplicative factor of $\beta_{\rm BH}\sim\frac{\sqrt{g_sN}}{r_h}$ in all terms in the action corrsponding to the BH gravity dual, e.g., (\ref{int-EH-BH-IR}), (\ref{int-EH-BH-IR}), (\ref{sqrtGbetaRbeta0-IR-bh}), (\ref{int-EH-BH-UV}), 
(\ref{GHY-BH}), (\ref{GHY-UV-div}), (\ref{int-iGdeltaJ0-BH-IR}), (\ref{int-sqrtGiGdeltaJ0-beta0-UV-div}), 
(\ref{isqrthinvGdeltaJ0-constr}), (\ref{ihdeltaJ0-r=rUV}) and (\ref{sqrthJ0-constr}). 

Interestingly, the Wald entropy density arising from the ${\cal O}(\beta^0)$ action is given by:
\begin{eqnarray}
\label{Wald-i}
& & \frac{{\cal S}_{\rm BH}^{\rm Wald}}{{\cal V}_3} = \oint_{M_6(\theta_{1,2},x,y,z,x^{10})} \frac{\partial R}{\partial R_{mnpq}}\epsilon^{mn}\epsilon^{pq} \sqrt{g_{x^{1,2,3},\theta_{1,2},x,y,z,x^{10}}}d\theta_1d\theta_2 dx dy dz dx^{10}
\nonumber\\
& & \sim -\frac{\left(3 b^2-1\right) \left(9 b^2+1\right) {g_s}^{5/4} M N^{-1/20} {N_f} {r_h}^3 \log \left(\frac{r_h}{{\cal R}_{D5/\overline{D5}}^{\rm bh}}\right)
   \left(\log N -9 \log \left(\frac{r_h}{{\cal R}_{D5/\overline{D5}}^{\rm bh}}\right)\right) }{ \left(6 b^2+1\right) \alpha _{\theta _1}^2 \alpha _{\theta _2}}\nonumber\\
& & \times \left[{N_f} \left(\log N -3 \log \left(\frac{r_h}{{\cal R}_{D5/\overline{D5}}^{\rm bh}}\right)\right)\right]^{2/3},
\nonumber\\
& & 
\end{eqnarray}
where ${\cal V}_3$ is the $\mathbb{R}^3 (x^{1,2,3}$ coordinate volume). Upon substituting $b = \frac{1}{\sqrt{3}} + \epsilon, \epsilon\sim \left(|\log r_h\right|)^{\frac{9}{2}} N^{-\frac{9}{10} - \alpha_\epsilon}$ yields:
\begin{eqnarray}
\label{Wald-ii}
& & \frac{{g_s}^{5/4} M {N_f} {r_h}^3 N^{-\frac{19}{20}-{\alpha_\epsilon}}\left(|\log r_h\right|)^{\frac{11}{2}}
   \left(\log N -9 \log \left(\frac{r_h}{{\cal R}_{D5/\overline{D5}}^{\rm bh}}\right)\right)}{\alpha _{\theta _1}^2 \alpha
   _{\theta _2}}\nonumber\\
& &  \times \left[{N_f} \left(\log N -3 \log \left(\frac{r_h}{{\cal R}_{D5/\overline{D5}}^{\rm bh}}\right)\right)\right]^{2/3}.
\end{eqnarray}
Both methods yield that the BH entropy goes like $r_h^3$. The results match exactly for (for simplicity, we have disregarded in (\ref{alphaepsilon}) numerical factors and $M, N_f$ which in the MQGP limit are ${\cal O}(1)$):
\begin{eqnarray}
\label{alphaepsilon}
& & \alpha_\epsilon = - \frac{9}{10} 
+\frac{\left|\log \left(M_{\rm UV}\left(N_f^{\rm UV}\right)^{\frac{2}{15}}\right)\right|}{\log N}
+ \frac{211\log\log N}{\log N}\sim \frac{\left|\log \left(M_{\rm UV}\left(N_f^{\rm UV}\right)^{\frac{2}{15}}\right)\right|}{\log N}
- \frac{9}{10};
\end{eqnarray}
$\alpha_\epsilon>0$ if $M_{\rm UV}\left(N_f^{\rm UV}\right)^{\frac{2}{15}}<N^{-\frac{9}{10}}$ .

At ${\cal O}(R^4)$, one can show that to calculate one needs to evaluate the following four class of terms while calculating
$\frac{\partial J_0}{\partial R_{trtr}}$:
\begin{eqnarray}
\label{Wald-J0-i}
& & \hskip -0.3in (i) \left(G^{rr}\right)^2 \left(G^{tt}\right)^2\left(R_{PrtQ} + \frac{1}{2}R_{PQtr}\right)R_t^{\ \ RSP}R^Q_{\ \ RSr}\nonumber\\
& & \hskip -0.3in \sim \left(G^{rr}\right)^2 \left(G^{rr}\right)^2R_{trrt}\left(R_t^{\ \ z\theta_1t}R^r_{\ \ z\theta_1r} + R_t^{\ \ \theta_1zt}R^r_{\ \ \theta_1zr}\right)\nonumber\\
& & \hskip -0.3in \sim-\frac{\left(\log N -9 \log \left(\frac{r_h}{{\cal R}_{D5/\overline{D5}}^{\rm bh}}\right)\right)^2 \sqrt[3]{{N_f} \left(\log N -3 \log \left(\frac{r_h}{{\cal R}_{D5/\overline{D5}}^{\rm bh}}\right)\right)}}{{g_s}^{3/2} N^{3/2}
   {N_f}^5 \left(\log N -3 \log \left(\frac{r_h}{{\cal R}_{D5/\overline{D5}}^{\rm bh}}\right)\right)^5};\nonumber\\
& & \hskip -0.3in (ii) R^{HrtK}R_H^{\ \ RSt}R^r_{\ \ RSK} + \frac{1}{2}R^{HKtr}R_H^{\ \ RSt}R^Q_{\ \ RSK}\sim R^{trrt} \left(R_t^{\ \ z\theta_1t} R^r_{z\theta_1r} + R_t^{\theta_1zt} R^r_{\ \ \theta_1zr}\right)\nonumber\\
& & \hskip -0.3in \sim -\frac{(\log N -9 \log \left(\frac{r_h}{{\cal R}_{D5/\overline{D5}}^{\rm bh}}\right))^2}{{g_s}^{3/2} N^{3/2} {N_f}^{14/3} \left(\log N -3 \log \left(\frac{r_h}{{\cal R}_{D5/\overline{D5}}^{\rm bh}}\right)\right)^{14/3}};
\nonumber\\
& & \hskip -0.3in (iii)  \left(G^{rr}\right)^2 G^{tt}\left(R_{PrtQ} + \frac{1}{2}R_{PQtr}\right)R_t^{\ \ RSP}R^Q_{\ \ RSr}
\sim \frac{{g_s}^{9/2} M^6 \log ^3({r_h}) (\log N -12 \log \left(\frac{r_h}{{\cal R}_{D5/\overline{D5}}^{\rm bh}}\right))^3}{{N_f}^{5/3} \left(\log N -3 \log \left(\frac{r_h}{{\cal R}_{D5/\overline{D5}}^{\rm bh}}\right)\right)^{14/3}};\nonumber\\
& & \hskip -0.3in (iv)  G^{tt} \left(R^{HMNr}R_{PMNt} + \frac{1}{2}R^{HrMN}R_{PtMN}\right)R_H^{\ \ rtP}
\sim \frac{\sqrt{{g_s}} M^2 \log \left(\frac{r_h}{{\cal R}_{D5/\overline{D5}}^{\rm bh}}\right)}{{N_f}^{11/3} \left(\log N -3  \log \left(\frac{r_h}{{\cal R}_{D5/\overline{D5}}^{\rm bh}}\right)\right)^{14/3}}\nonumber\\
& & \hskip 3.5in \times  \left(\log N -12 \log \left(\frac{r_h}{{\cal R}_{D5/\overline{D5}}^{\rm bh}}\right)\right).\nonumber\\
& & 
\end{eqnarray}
One hence sees that it is the contribution (iii) which dominates in the MQGP limit, and obtains:
\begin{eqnarray}
\label{S-J0-Wald}
& & \frac{{\cal S}_E^{{\cal O}(R^4)}}{{\cal V}_3}\sim \oint_{\theta_{1,2},x,y,z,x^{10})} \frac{\partial J_0}{\partial R_{rtrt}}\sqrt{-G_9} d\theta_1 d\theta_2 dx dy dz dx^{10}\nonumber\\
& &  \sim \frac{{g_s}^{23/4} M^8 {N_f}^{4/3} \left(\frac{r_h}{{\cal R}_{D5/\overline{D5}}^{\rm bh}}\right)^3 \log ^4\left(\frac{r_h}{{\cal R}_{D5/\overline{D5}}^{\rm bh}}\right) \left(\log N -12 \log \left(\frac{r_h}{{\cal R}_{D5/\overline{D5}}^{\rm bh}}\right)\right)^3}{\sqrt[20]{N} \alpha _{\theta _1}^2 \alpha _{\theta _2} \left(\log N -3 \log \left(\frac{r_h}{{\cal R}_{D5/\overline{D5}}^{\rm bh}}\right)\right)^{8/3}}.
\end{eqnarray}
One notes the identical $r_h^3$ and $N^{-\frac{1}{20}}$ dependence in (\ref{S-J0-Wald}) and the corresponding semiclassical result at ${\cal O}(\beta)$. For the latter and the Wald entropy result to match exactly at ${\cal O}(\beta)$ imposes the following constraint:
\begin{eqnarray}
\label{Wald-matching}
& & 
\Biggl[{2    {\cal C}_{\theta_1x}^{\rm bh} \kappa_{\left(\sqrt{-G^{\mathscr {M}}}\right)^{(1)}R^{(0)}}^{\rm IR} } +\frac{20  \left(-{\cal C}_{zz}^{\rm bh} + 2 {\cal C}_{\theta_1z}^{\rm bh} - 3 {\cal C}_{\theta_1x}^{\rm bh}\right) \kappa_{\rm EH}^{\beta,\ \rm IR} }{11 }\Biggr]
\log ^3(N) \left|\log \left(M_{\rm UV}\left(N_f^{\rm UV}\right)^{\frac{2}{15}}\right)\right|\nonumber\\
& &  \sim \frac{ \log ^3 \left(\frac{r_h}{{\cal R}_{D5/\overline{D5}}^{\rm bh}}\right) \left(\log N -12 \log \left(\frac{r_h}{{\cal R}_{D5/\overline{D5}}^{\rm bh}}\right)\right)^3}{ \left(\log N -3 \log \left(\frac{r_h}{{\cal R}_{D5/\overline{D5}}^{\rm bh}}\right)\right)^{8/3}}.
\end{eqnarray} 
Hence, one sees that consistency with the Wald entropy computation up to ${\cal O}(R^4)$, imposes a linear constraint on the constants of integration appearing in the solutions to the EOMs of the metric corrections at ${\cal O}(R^4)$ along the directions that encode the "memory" of the compact part of the non-compact four-cycle "wrapped" by the flavor $D7$-branes in the type IIB dual   
\cite{MQGP} of the large-$N$ thermal QCD-like theories. This "Flavor Memory"-effect is also displayed by the ${\mathscr {M}}$-theory uplift of the corresponding thermal background, as seen in {\bf 3.2}.

After that brief detour, returning back to the computation of $T_c$, the on-shell action corresponding to the thermal background uplift is given below:
\begin{eqnarray}
\label{th-action-LO}
& & S_{D=11,\ {\rm on-shell UV-finite}}^{\rm thermal} \sim \frac{2 \kappa_{\rm GHY}^{{\rm th},\ \beta^0} {M_{\rm UV}} {r_0}^4 \log \left(\frac{{{\cal R}_{\rm UV}}}{{\cal R}_{D5/\overline{D5}}^{\rm th}}\right)}{{g_s^{\rm UV}}^{9/4}
   N^{11/20} {{\cal R}_{D5/\overline{D5}}^{\rm th}}^4 \alpha _{\theta _1}^2 \alpha _{\theta _2}}\nonumber\\
& & +\frac{2{g_s}^{3/4} \kappa_{\rm EH, IR}^{{\rm th},\ \beta^0} M {N_f}^3
   {r_0}^2 \log ^2(N) \log \left(\frac{{r_0}}{{\cal R}_{D5/\overline{D5}}^{\rm th}}\right)}{ N^{11/20} {{\cal R}_{D5/\overline{D5}}^{\rm th}}^2 \alpha _{\theta _1}^2 \alpha
   _{\theta _2}}\nonumber\\
& & -\frac{20 \beta  \kappa_{\rm EH, th}^{{\rm IR},\ \beta} {r_0}^3 \left(2 \alpha _{\theta _2}^3-729 \sqrt{6} \alpha _{\theta _1}^3 \alpha _{\theta _2}\right)
   {f_{x^{10}x^{10}}}({r_0})}{11{g_s}^{9/4} M N^{7/20} {N_f}^{5/3} {{\cal R}_{D5/\overline{D5}}^{\rm th}}^3 \alpha _{\theta _1}^4 \log ^{\frac{2}{3}}(N) \log
   \left(\frac{{r_0}}{{\cal R}_{D5/\overline{D5}}^{\rm th}}\right)}\beta.  
\end{eqnarray}

Now, equality at ${\cal O}(\beta^0)$ in (\ref{actions-equal-Tc-ii})
 requires the following equation to be solved:
\begin{eqnarray}
\label{rh-r0-beta0}
& & \frac{2 \kappa_{\rm GHY}^{\rm bh} {M_{\rm UV}} {r_h}^4 \log \left(\frac{{{\cal R}_{\rm UV}}}{{\cal R}_{D5/\overline{D5}}^{\rm bh}}\right)}{{g_s}^{9/4} } = \frac{2 \kappa_{\rm GHY}^{{\rm th},\ \beta^0} {M_{\rm UV}} {r_0}^4 \log \left(\frac{{{\cal R}_{\rm UV}}}{{\cal R}_{D5/\overline{D5}}^{\rm th}}\right)}{{g_s^{\rm UV}}^{9/4}
    {{\cal R}_{D5/\overline{D5}}^{\rm th}}^4}\nonumber\\
& & +\frac{2{g_s}^{3/4} \kappa_{\rm EH, IR}^{{\rm th},\ \beta^0} M {N_f}^3
   {r_0}^2 \log ^2(N) \log \left(\frac{{r_0}}{{\cal R}_{D5/\overline{D5}}^{\rm th}}\right)}{  {{\cal R}_{D5/\overline{D5}}^{\rm th}}^2}.
\end{eqnarray}
Now, one can show that near, e.g., $(\theta_1,\theta_2)\sim \left(\frac{\alpha_{\theta_1}}{N^{\frac{1}{5}}},\frac{\alpha_{\theta_1}}{N^{\frac{3}{10}}}\right)$, $\frac{\kappa_{\rm GHY}^{{\rm th},\ \beta^0}}{\kappa_{\rm EH, IR}^{{\rm th},\ \beta^0}}\sim10^5$, and hence, one can drop the $\kappa_{\rm EH, IR}^{{\rm th},\ \beta^0}$ term.  Therefore,
\begin{equation}
\label{r_h-r_0-relation}
r_h = \frac{\sqrt[4]{\frac{\kappa_{\rm GHY}^{{\rm th},\ \beta^0}}{\kappa_{\rm GHY}^{{\rm bh},\ \beta^0}}} {r_0} {{\cal R}_{D5/\overline{D5}}^{\rm bh}} \sqrt[4]{\frac{\log
   \left(\frac{{{\cal R}_{\rm UV}}}{{{\cal R}_{D5/\overline{D5}}^{\rm th}}}\right)}{\log
   \left(\frac{{{\cal R}_{\rm UV}}}{{{\cal R}_{D5/\overline{D5}}^{\rm bh}}}\right)}}}{{{\cal R}_{D5/\overline{D5}}^{\rm th}}}
.
\end{equation}

Identifying $\frac{r_0}{L^2}$ with $\frac{m^{0^{++}}}{4}$ \cite{Glueball-Roorkee} and using $T_c=r_h/\pi L^2$ \cite{NPB}, deconfinement temperature will be given by the following expression:
\begin{equation}
\label{rh-r0-beta0}
T_c = \frac{\sqrt[4]{\frac{\kappa_{\rm GHY}^{{\rm th},\ \beta^0}}{\kappa_{\rm GHY}^{{\rm bh},\ \beta^0}}} {m^{0^{++}}} {{\cal R}_{D5/\overline{D5}}^{\rm bh}} \sqrt[4]{\frac{\log
   \left(\frac{{{\cal R}_{\rm UV}}}{{{\cal R}_{D5/\overline{D5}}^{\rm th}}}\right)}{\log
   \left(\frac{{{\cal R}_{\rm UV}}}{{{\cal R}_{D5/\overline{D5}}^{\rm bh}}}\right)}}}{4 \pi{{\cal R}_{D5/\overline{D5}}^{\rm th}}}.
\end{equation}
Using (\ref{bh-action-LO}) and (\ref{th-action-LO}), the matching at ${\cal O}(\beta)$  yields:
\begin{eqnarray}
\label{rh-r0-beta-1}
& & \Biggl[{2    {\cal C}_{\theta_1x}^{\rm bh} \kappa_{\left(\sqrt{-G^{\mathscr {M}}}\right)^{(1)}R^{(0)}}^{\rm IR} } +\frac{20  \left(-{\cal C}_{zz}^{\rm bh} + 2 {\cal C}_{\theta_1z}^{\rm bh} - 3 {\cal C}_{\theta_1x}^{\rm bh}\right) \kappa_{\rm EH}^{\beta,\ \rm IR} }{11 }\Biggr]\nonumber\\
& & \times\frac{ b^2   {g_s}^{3/4}  M {N_f}^3 {r_h}^4 \log ^3(N) \log
   \left(\frac{{r_h}}{{{\cal R}_{D5/\overline{D5}}}}\right) \log
   \left(1 - \frac{{r_h}}{{{\cal R}_{D5/\overline{D5}}}}\right)}{ N^{11/20} {{\cal R}_{D5/\overline{D5}}}^4 \alpha _{\theta _1}^2 \alpha _{\theta _2}}\nonumber\\
& & = \frac{20 \kappa_{\rm EH, th}^{{\rm IR},\ \beta} {r_0}^3 \left(2 \alpha _{\theta _2}^3-729 \sqrt{6} \alpha _{\theta _1}^3 \alpha _{\theta _2}\right)
   {f_{x^{10}x^{10}}}({r_0})}{11{g_s}^{9/4} M N^{7/20} {N_f}^{5/3} {{\cal R}_{D5/\overline{D5}}^{\rm th}}^3 \alpha _{\theta _1}^4 \log ^{\frac{2}{3}}(N) \log
   \left(\frac{{r_0}}{{\cal R}_{D5/\overline{D5}}^{\rm th}}\right)},
\end{eqnarray}
which yields:
\begin{eqnarray}
\label{rh-r0-beta-3}
& & \hskip -0.5in f_{x^{10}x^{10}}({r_0})
\nonumber\\
& & \hskip -0.5in \sim \left.-\frac{b^2 {g_s}^3 M^2 {N_f}^{14/3}  \left(\frac{r_h}{{\cal R}_{D5/\overline{D5}}^{\rm bh}}\right)^4 \alpha _{\theta _1}^2 \log ^{\frac{2}{3}}(N) \log
   \left(\frac{{r_0}}{{\cal R}_{D5/\overline{D5}}^{\rm th}}\right) \log \left(\frac{{r_h}}{{\cal R}_{D5/\overline{D5}}^{\rm bh}}\right) \log
   \left(1-\frac{{r_h}}{{\cal R}_{D5/\overline{D5}}^{\rm bh}}\right) }{ \beta  {\kappa_{\rm EH}^{\beta,\ \rm IR}} \sqrt[5]{N} \left(\frac{r_0}{{\cal R}_{D5/\overline{D5}}^{\rm th}}\right)^3     \alpha _{\theta _2}^2 \left(729 \sqrt{6} \alpha _{\theta _1}^3-2 \alpha _{\theta _2}^2\right)}\right|_{(\ref{r_h-r_0-relation})}\nonumber\\
& & \hskip -0.5in \times \left(11 \pi ^{17/4} {\cal C}_{\theta_1x}^{\rm bh} {\kappa_{\left(\sqrt{-G^{\mathscr {M}}}\right)^{(1)}R^{(0)}}^{\rm IR}}
    \log ^3(N)-10 {\kappa_{\rm EH}^{\beta,\ \rm IR}} \log^3(N)  \left(-{\cal C}_{zz}^{\rm bh} + 2 {\cal C}_{\theta_1z}^{\rm bh} - 3 {\cal C}_{\theta_1x}^{\rm bh}\right)\right).\nonumber\\
& & 
\end{eqnarray}

We thus see that by matching the thermal and black-hole ${\mathscr {M}}$-theory background actions at the UV cut-off, respectively dual to large-$N$ QCD-like theories at low and high temperatures, (\ref{rh-r0-beta-3}) gives a relationship between the constants of integration in the IR in the respective metric perturbations as ${\cal O}(l_p^6)$ corrections to the ${\mathscr {M}}$-theory uplift worked out in \cite{MQGP} along respectively the ${\mathscr {M}}$-theory circle (for thermal background dual to $T<T_c$) and the $S^3$-portion of the four cycle (locally) $\Sigma_{(4)} = \mathbb{R}_+\times S^3$ "wrapped" by the flavor $D7$-branes of the type IIB conifold geometry of \cite{metrics} (for black hole background dual to $T>T_c$)!! The equations (\ref{rh-r0-beta-3}) along with (\ref{rh-r0-beta0}), are what we refer to as  UV-IR connection  in this context. We also note that the contributions arising from the ${\cal O}(R^4)$ corrections to the BH background metric retains the "Flavor Memory", referred to earlier, of the compact three-cycle which is part of the non-compact four-cycle "wrapped" by the parent type IIB flavor $D7$-branes.

\section{Deconfinement from Entanglement Entropy}

In this section we will discuss confinement-deconfinement phase transition in QCD$_{2+1}$-like theory from entanglement entropy point of view based on \cite{Tc-EE}. We will calculate entanglement entropy for "connected" and "disconnected" regions - suitably defined. We will show that at a critical value of ${\it l}$ of a spatial interval which we are denoting here by ${\it l_{crit}}$, phase transition will occur from confined phase to the deconfined phase.  Contact with results of the previous section is made by looking at the $M_{KK}\rightarrow0$- or equivalently $r_0\rightarrow0$-limit, i.e., the 4D-limit of results of this section.
\par
Authors in \cite{Tc-EE} have investigated entanglement entropy between an interval and its complement in the gravity dual of large $N_c$ gauge theories and they found that there are two RT surfaces - disconnected and connected. Below a critical value of ${\it  l}$ connected surface dominates while above that critical value of ${\it  l}$ disconnected surface dominates. This is analogous to finite temperature deconfinement transition in dual theories. \par
For $AdS_{d+2}/CFT_{d+1}$ correspondence, quantum entanglement entropy between the regions $A \equiv \mathbb{R}^{d-1}\times l$ and $B \equiv \mathbb{R}^{d-1}\times(\mathbb{R}-l)$, $l$ being an interval of length $l$, is given by the following expression\cite{RT},
\begin{equation}
\label{EE-def}
S_A=\frac{1}{4 G_N^{(d+2)}} \int_{\gamma} d^d\sigma\sqrt{G_{ind}^{(d)}} .
\end{equation}
where, $G_N^{(d+2)}$ is Newton constant in $(d+2)$ dimensions and $G_{ind}^{(d)}$ is the determinant of the induced string frame metric on co-dimension 2 minimal surface $\gamma$. Equation $(\ref{EE-def})$ can be generalised to non-conformal theories as below \cite{Tc-EE}:
\begin{equation}
\label{EE-def-NCT}
S_A=\frac{1}{4 G_N^{(d+2)}} \int_{\gamma} d^d\sigma e^{-2\phi}\sqrt{G_{ind}^{(d)}},
\end{equation}
where $\phi$ is the dilaton profile. In an ${\mathscr {M}}$-theory dual since there is no dilaton, therefore equation (\ref{EE-def-NCT}) will be assumed to be replaced by the following expression,
\begin{equation}
\label{EE-def-M Theory}
S_A=\frac{1}{4 G_N^{(11)}} \int_{\gamma} d^9\sigma \sqrt{G_{ind}^{(9)}} .
\end{equation}  
Now, consider gravity dual's  string frame metric (\ref{TypeIIA-from-M-theory-Witten-prescription-T<Tc}), which can written as:
\begin{equation}
\label{metric}
ds^2=\alpha(r)[\sigma(r) dr^2+dx_{\mu}dx^{\mu}]+g_{mn}dx^mdx^n ,
\end{equation}
where $x_\mu(\mu=0,1,2)$ represents $(2+1)$ Minkowskian coordinates, $r$ is the radial coordinate and $x^m(m=3,5,6,7,8,9,10)$ corresponds to $x^3$ and six angular coordinates$(\theta_{1,2},\phi_{1,2},\psi,x^{10})$. Noting $g_{x^3x^3}(r=r_0)=0$, $(x^3,r)$ form a cigar-like geometry. The volume of the seven-fold is given by:
\begin{equation}
\label{V_int}
{\cal V}_{\rm int}=\int \prod_m dx^m \sqrt{g} .
\end{equation}
Here we have $(2+1)$ dimensional QFT defined on ${\rm I\!R}^{2+1}$. Let us define two regions $A$ and $B$ in two dimensions as below,
\begin{eqnarray}
&&
A={\rm I\!R}\times l , \\
&& B={\rm I\!R} \times ({\rm I\!R}-l).
\end{eqnarray}
Now, we are going to calculate enganglement entropy between  $A$ and $B$ for the metric $(\ref{metric})$ using equation $(\ref{EE-def-M Theory})$. Induced metric on $\gamma$ is given below (here we have represented $x_1$ by $x$):.
\begin{equation}
\label{induced metric}
ds^2|_\gamma=\alpha(r)\left[\left(\sigma(r) + \left(\frac{dx}{dr}\right)^2 \right) dr^2+ dx_2^2\right]+g_{mn}dx^mdx^n .
\end{equation}
Therefore,
\begin{equation}
\label{EEC}
\frac{S_A}{{\cal V}_1} =\frac{1}{4 G_N^{(11)}} \int dr\sqrt{H(r)}\sqrt{\sigma(r)+ (\partial_rx(r))^2} ,
\end{equation}
where,
\begin{equation}
H(r)= {\cal V}_{int}^2\alpha(r)^2 .
\end{equation}
As in \cite{Tc-EE}, $H(r=r_0)=0$. 
Equation of motion for $x(r)$ can be obtained from equation $(\ref{EEC})$ as, 
\begin{equation}
\label{dx/dr}
\frac{dx}{dr}=\pm \frac{\sqrt{\sigma(r)}\sqrt{H(r_{*})}}{\sqrt{H(r)-H(r_*)}} ,
\end{equation}
where, $r_*$ is the value of radial coordinate at which $dr/dx$ vanishes. Integrating above equation we obtain,
\begin{equation}
\label{lrstar}
l(r_*)=2\sqrt{H(r_*)}\int_{r_*}^\infty dr\frac{\sqrt{\sigma(r)}}{\sqrt{H(r)-H(r_*)}} .
\end{equation}
From Equations $(\ref{EEC})$ and $(\ref{dx/dr})$, entanglement entropy can be simplified as,
\begin{equation}
\label{connected-EE}
\frac{S_A}{{\cal V}_1} =\frac{1}{2 G_N^{(11)}} \int_{r_*}^{r_\infty} dr\frac{\sqrt{\sigma(r)} H(r)}{\sqrt{H(r)-H(r_*)}} .
\end{equation}
Here, $r_\infty$ is the UV cutoff. 

For the disconnected surface we have $r_*=r_0$ as explained in \cite{Tc-EE}, when $l>l_{max}$ entanglement entropy for the disconnected surface will be given by the following expression,
\begin{equation}
\label{Disconnected Surface}
\frac{S_A}{{\cal V}_1} =\frac{1}{2 G_N^{(11)}} \int_{r_0}^{r_\infty} dr\sqrt{\sigma(r)H(r)} .
\end{equation}
For $l<l_{max}$, difference of entanglement entropy for connected and disconnected surface have the following form,
\begin{equation}
\frac{2 G_N^{(11)}(S_A^{(conn)}-S_A^{(disconn)})}{{\cal V}_1} = \int_{r_*}^{\infty} dr\left(\frac{\sqrt{\sigma(r)} H(r)}{\sqrt{H(r)-H(r_*)}} - \sqrt{\sigma(r)H(r)}\right)- \int_{r_0}^{r_*} dr\sqrt{\sigma(r)H(r)} .
\end{equation}
\begin{itemize}
\item Connected solution with large $r_*$ has the lower entropy when ${\it  l}$ is small.
\item When ${\it  l}$ starts increasing, two things can happen. First, connected solution can still remain the lower untill ${\it  l}$ reaches it's maximum value i.e. ${\it  l}_{max}$. Second, there could be a critical value of ${\it  l}$ which is denoted by ${\it  l}_{crit}$ and ${\it  l}_{crit}<{\it  l}_{max}$ above which disconnected solution becomes dominant one.
\item There will be a phase transition at ${\it  l}={\it  l}_{max}$ in the first case and at ${\it  l}={\it  l}_{crit}<{\it  l}_{max}$ in the second case.
\end{itemize}
From equations (\ref{TypeIIA-from-M-theory-Witten-prescription-T<Tc}) and (\ref{metric}) we have,
\begin{equation}
\alpha(r)=\frac{e^{\frac{-2\phi^{IIA}}{3}}}{\sqrt{h(r,\theta_{1,2})}}
\end{equation}
and
\begin{equation}
\sigma(r)=\frac{e^{\frac{2\phi^{IIA}}{3}}}{\tilde{g(r)}},
\end{equation}
where $\phi^{IIA}$ is type IIA dilaton profile, which can be read off from ${\mathscr {M}}$-theory metric component as,
\begin{equation}
\label{type-IIA-dilaton}
G^{\mathscr {M}}_{x_{10}x_{10}}=e^{\frac{4\phi^{IIA}}{3}}.
\end{equation}
From the equation (\ref{V_int}), simplified form of the ${\cal V}_{int}$ for the metric given in equation (\ref{TypeIIA-from-M-theory-Witten-prescription-T<Tc}) is:
\begin{eqnarray}
\label{Vint}
  {\cal V}_{\rm int} \sim \frac{{g_s}^{3/4} M N^{9/20} \sqrt{1-\frac{{r_0}^4}{r^4}} \left(-{N_f} \log \left(9 a^2 r^4+r^6\right)+\frac{16 \pi }{{g_s}}+\frac{\Omega}{g_s}\right){}^{4/3}}{ r \alpha _{\theta _1}^3 \alpha _{\theta _2}^2} \left(\frac{1}{2} \beta  ({\cal C}_{zz}^{\rm th}-2
   {\cal C}_{\theta_1z}^{\rm th})+1\right)\nonumber\\
& & \hskip -6.5in \times  \Biggl[{g_s} \log N  {N_f} \left(3 a^2-r^2\right) (2 \log (r)+1)+\log
   (r) \nonumber\\
   & & \hskip -6.4in \times \left(4 {g_s} {N_f} \left(r^2-3 a^2\right) \log \left(\frac{1}{4} \alpha _{\theta _1} \alpha _{\theta
   _2}\right)-24 \pi  a^2+r^2 (8 \pi -3 {g_s} {N_f})\right)\nonumber\\
& & \hskip -6.4in+2 {g_s} {N_f} \left(r^2-3 a^2\right) \log
   \left(\frac{1}{4} \alpha _{\theta _1} \alpha _{\theta _2}\right)+18 {g_s} {N_f} \left(r^2-3 a^2 (6
   r+1)\right) \log ^2(r)\Biggr],
\end{eqnarray}
(strictly speaking the periodicity of $x^3$, $\frac{2\pi}{M_{\rm KK}}$ where $M_{\rm KK} = \frac{2r_0}{L^2}\left[1 + {\cal O}\left(\frac{g_sM^2}{N}\right)\right]$, but this does not influence the computation of $l_{\rm max}$ and $l_{\rm crit}$) where \\ $\Omega\equiv \left({g_s} (2 \log N +3) {N_f}-4 {g_s} {N_f} \log
   \left(\frac{1}{4} \alpha _{\theta _1} \alpha _{\theta _2}\right)-8 \pi \right)$,
   $\alpha(r)$ and $\sigma(r)$ have been obtained from equation (\ref{type-IIA-dilaton}) and $h(r,\theta_{1,2})$ given in \cite{metrics, MQGP}, simplified form of the same are given below:
\begin{eqnarray}
\label{alpha}
\alpha(r) = \frac{3^{2/3} \left(-{N_f} \log \left(9 a^2 r^4+r^6\right)+\frac{16 \pi }{{g_s}}+\frac{\Omega}{g_s}\right){}^{2/3}}{8 \pi ^{7/6} \sqrt{\frac{{g_s} N}{r^4}}}\nonumber \\
   & & \hskip -3.5in  -\frac{27\ 3^{2/3} b^{10} \left(9 b^2+1\right)^4
   \beta  M \left(\frac{1}{N}\right)^{3/4} r \left(19683 \sqrt{6} \alpha _{\theta _1}^6+6642 \alpha _{\theta _2}^2
   \alpha _{\theta _1}^3-40 \sqrt{6} \alpha _{\theta _2}^4\right) \left(6 a^2+{r_0}^2\right) (r-2 {r_0})}{16 \pi ^{13/6} \left(3 b^2-1\right)^5 \left(6 b^2+1\right)^4 \log N ^4 \sqrt{N} {N_f}
   {r_0}^4 \alpha _{\theta _2}^3 \left(9 a^2+{r_0}^2\right) \sqrt{\frac{{g_s} N}{r^4}}}\nonumber \\
   & & \hskip -3.5in \times \left( \log
   ^3({r_0}) \left(-{N_f} \log \left(9 a^2 r^4+r^6\right)+\frac{16 \pi }{{g_s}}+\frac{\Omega}{g_s}\right){}^{2/3}\right)
   \end{eqnarray}
and
\begin{eqnarray}
\label{beta}
& &\hskip -0.8in \sigma(r) = \frac{4 \left(\frac{\pi }{3}\right)^{2/3} \left(\frac{32 \pi  a^2 {g_s} M^2 {N_f} ({c_1}+{c_2} \log
   ({r_0}))}{N \left(9 a^2+r^2\right) \left(-{N_f} \log \left(9 a^2 r^4+r^6\right)+\frac{16 \pi }{{g_s}}+\frac{\Omega}{g_s}\right)}+1\right)}{\left(1-\frac{{r_0}^4}{r^4}\right) \left(-{N_f} \log
   \left(9 a^2 r^4+r^6\right)+\frac{16 \pi }{{g_s}}+\frac{\Omega}{g_s}\right){}^{2/3}} \nonumber \\
& &  \hskip -0.8in -\frac{18
   \sqrt[3]{\frac{3}{\pi }} b^{10} \left(9 b^2+1\right)^4 \beta  M \left(\frac{1}{N}\right)^{5/4} r^5 \left(19683
   \sqrt{6} \alpha _{\theta _1}^6+6642 \alpha _{\theta _2}^2 \alpha _{\theta _1}^3-40 \sqrt{6} \alpha _{\theta
   _2}^4\right)}{\left(3 b^2-1\right)^5 \left(6
   b^2+1\right)^4 \log N ^4 {N_f} {r_0}^4 \alpha _{\theta _2}^3 \left(9 a^2+{r_0}^2\right)
   \left({r_0}^4-r^4\right)}\nonumber \\
     & & \hskip -0.8in \times \frac{\left(6 a^2+{r_0}^2\right) (r-2 {r_0}) \log ^3({r_0})}{ \left(-{N_f} \log \left(9 a^2 r^4+r^6\right)+\frac{16 \pi }{{g_s}}+\frac{\Omega}{g_s}\right){}^{2/3}},
\end{eqnarray}
$c_{1,2}$ appearing as the ${\cal O}\left(\frac{g_sM^2}{N}\right)$-correction $\left(\frac{g_sM^2}{N}\right)(c_1 + c_2 \log r_0)r_0$ in the $a-r_0$ relationship (inspired by \cite{EPJC-2,Bulk-Viscosity-McGill-IIT-Roorkee}). From equations (\ref{Vint}) and (\ref{alpha}) we obtained,
\begin{eqnarray}
\label{Hr}
& & H(r) \equiv {\cal V}_{\rm int}^2\alpha(r)^2 = h_0(r) + \beta h_1(r),
\end{eqnarray}
where,
\begin{eqnarray}
& & \hskip-0.7in h_0(r)\sim\frac{\sqrt{\text{gs}} M^2 r^2}{ \sqrt[10]{N} \alpha _{\theta _1}^6
   \alpha _{\theta _2}^4} \left(1-\frac{{r_0}^4}{r^4}\right)
 \lambda_1^2 \lambda_2^4,
   \end{eqnarray}
and
\begin{eqnarray}
& & \hskip-0.5in h_1(r)\sim\frac{{g_s}^{3/2} M^2 N^{9/10} \left(1-\frac{{r_0}^4}{r^4}\right)}{ r^2 \alpha
   _{\theta _1}^6 \alpha _{\theta _2}^4 \left(9 a^2+{r_0}^2\right)}\lambda_1^2  \lambda_2^{8/3}\left(\frac{\lambda_2^{4/3}}{64 \pi ^{10/3} \left(3 b^2-1\right)^5 \left(6 b^2+1\right)^4 {g_s} \log N^4 N^{3/2} {N_f} {r_0}^4 \alpha
   _{\theta _2}^3 }   \right)\nonumber\\
   & & 
 \hskip-0.5in\times  \left(\frac{3 \sqrt[3]{3} r^4 ({\cal C}_{zz}^{\rm th}-2 {\cal C}_{\theta_1 z}^{\rm th})}{64 \pi ^{7/3} {g_s} N} -81 \sqrt[3]{3} b^{10} \left(9 b^2+1\right)^4 M \left(\frac{1}{N}\right)^{3/4} r^5\Sigma_1 \left(6
   a^2+{r_0}^2\right) (r-2 {r_0}) \log ^3({r_0}) \right),\nonumber\\
   &&
\end{eqnarray}
where we have defined $\lambda_1$ and $\lambda_2$ as below:
\begin{eqnarray}
& & \lambda_1 =\Biggl[{g_s} \log N N_f \left(3 a^2-r^2\right) (2 \log (r)+1)\nonumber\\
 & &
 +\log (r) \left(4 {g_s} {N_f} \left(r^2-3
   a^2\right) \log \left(\frac{1}{4} \alpha _{\theta _1} \alpha _{\theta _2}\right)-24 \pi  a^2 + r^2 (8 \pi -3 {g_s}
   {N_f})\right)\nonumber\\
   & & +2 {g_s}
   {N_f} \left(r^2-3 a^2\right) \log \left(\frac{1}{4} \alpha _{\theta _1} \alpha _{\theta
   _2}\right)+18 {g_s}
   {N_f} \left(r^2-3 a^2 (6 r+1)\right) \log ^2(r)\Biggr],\\
   & & 
   \lambda_2=\left(-{N_f} \log \left(9 a^2 r^4+r^6\right)+\frac{8 \pi }{{g_s}}-4 {N_f} \log \left(\frac{\alpha_{\theta_1}\alpha _{\theta _2}}{4\sqrt{N}}\right)\right).\nonumber\\
\end{eqnarray}
Similarly, by replacing $r$ with $r^*$, we obtained $h_0(r^*)$ and $h_1(r^*)$. Therefore,
\begin{eqnarray}
\label{Hrstarbeta}
& & H(r^*) = h_0(r^*) + \beta h_1(r^*).
\end{eqnarray}
Now rewritting $\sigma(r)$ as,
\begin{eqnarray}
\label{sigmauptobeta}
& & \sigma(r) = \sigma_0(r) + \beta \sigma_1(r).
\end{eqnarray}
From equations (\ref{Hr}), (\ref{Hrstarbeta}) and (\ref{sigmauptobeta}), the integrand appearing in equation (\ref{lrstar}) can be written as,
\begin{eqnarray}
& &\sqrt{\frac{\sigma(r)}{H(r) - H(r^*)}} = \sqrt{\frac{\sigma_0(r)}{h_0(r) - h_0(r^*)}} + \beta\frac{\left(\sigma_1(r)(h_0(r) - h_0(r^*)) - \sigma_0(r)(h_1(r) - h_1(r^*))\right)}{2\sqrt{\sigma_0(r)}(h_0(r) - h_0(r^*))^{\frac{3}{2}}};
\nonumber\\
& & 
\end{eqnarray}
where,
{\footnotesize
\begin{eqnarray}
\label{h_0(r)}
 h_0(r^*) \sim \frac{M^2 \left({r^*}^4-{r_0}^4\right) \left({g_s} {N_f} \log \left(9 a^2
   {r^*}^4+{r^*}^6\right)-\Omega-16 \pi
   \right){}^4}{{g_s}^{7/2} \sqrt[10]{N} {r^*}^2 \alpha _{\theta _1}^6
   \alpha _{\theta _2}^4}\nonumber\\
& & \hskip -4.5in \times \Biggl[\log ({r^*}) \left(6 a^2 ({g_s} \log N  {N_f}-4 \pi )+4 {g_s} {N_f}
   \left({r^*}^2-3 a^2\right) \log \left(\frac{1}{4} \alpha _{\theta _1} \alpha _{\theta
   _2}\right)+{r^*}^2 (8 \pi -{g_s} (2 \log N +3) {N_f})\right)\nonumber\\
& & \hskip -4.5in+{g_s} {N_f} \left(3
   a^2-{r^*}^2\right) \left(\log N -2 \log \left(\frac{1}{4} \alpha _{\theta _1} \alpha _{\theta
   _2}\right)\right)+18 {g_s} {N_f} \left({r^*}^2-3 a^2 (6 {r^*}+1)\right) \log
   ^2({r^*})\Biggr].
\end{eqnarray}
}
\begin{itemize}
\item In $H(r^*)$ if
{
\begin{eqnarray}
\label{logr^*-1}
 \log r^*\to  \frac{\sqrt{72 {g_s}^2 {N_f}^2 \left(\log N -2 \log \left(\frac{1}{4} \alpha _{\theta _1}
   \alpha _{\theta _2}\right)\right)+\Omega{}^2}+\Omega }{36 {g_s} {N_f}}\nonumber\\
& & \hskip -3.5in = \frac{{g_s} {N_f} \left(\log N -2 \log \left(\frac{1}{4} \alpha _{\theta _1} \alpha _{\theta
   _2}\right)\right)}{8 \pi }+\frac{{g_s}^2 {N_f}^2 \left(-4 \log \left(\frac{1}{4} \alpha _{\theta _1} \alpha
   _{\theta _2}\right)+2 \log N +3\right)}{64 \pi ^2} \nonumber\\
   & & \hskip -3.5in \times \left(\log N -2 \log \left(\frac{1}{4} \alpha _{\theta _1} \alpha
   _{\theta _2}\right)\right)+\frac{{g_s}^3 {N_f}^3 \left(\log N -2 \log \left(\frac{1}{4}
   \alpha _{\theta _1} \alpha _{\theta _2}\right)\right)}{512 \pi ^3}\nonumber\\
& & \hskip -3.5in \times  \left(16 \log ^2\left(\frac{1}{4} \alpha _{\theta _1} \alpha
   _{\theta _2}\right)+4 \log N ^2-4 (4 \log N -3) \log \left(\frac{1}{4} \alpha _{\theta _1} \alpha _{\theta
   _2}\right)-6 \log N +9\right)
   \nonumber\\
   & &\hskip -3.5in +O\left({N_f}^4\right),
\end{eqnarray}
}
then,
\begin{eqnarray}
\label{sqrth0}
& & \sqrt{h_0}(r^*) = \frac{a M_{\rm UV} {N_f^{\rm UV}}  {r^*} \sqrt{\left| \log N -2 \log \left(\alpha _{\theta _1} \alpha _{\theta_2}\right)+\log (16)\right| }}{24 \sqrt{6} \pi ^{9/4} {g_s}^{3/4} \sqrt[20]{N} \alpha _{\theta _1}^3 \alpha
   _{\theta _2}^2}+O\left({N_f^{\rm UV}} ^2\right).\nonumber\\
& & 
\end{eqnarray}
Now, the integral in $l(r^*)$ is proportional to $\frac{\sqrt[20]{N} \alpha _{\theta _1}^3 \alpha _{\theta _2}^2}{{g_s}^{5/4} M {N_f}^{10/3} r^3 \log (r)
   (\log N -3 \log (r))^{7/3} | \log N -9 \log (r)| }\sim \frac{\alpha _{\theta _1}^3 \alpha _{\theta _2}^2 \left(\frac{1}{\log (N)}\right)^{10/3}}{{g_s}^{5/4} M
   {N_f}^{10/3} r^3 \log (r)}+O\left(\left(\frac{1}{\log N }\right)^{13/3}\right)$. Given that $\int \frac{dr}{r^3\log r} = {Ei}(-2 \log (r)) \\ = \frac{-2 \log ^2(r)+\log (r)-1}{4 r^2 \log ^3(r)}+O\left(\left(\frac{1}{r}\right)^3\right) \approx - \frac{1}{2r^2\log r}$. One hence obtains:
\begin{eqnarray}
\label{lr^*-ii}
& & \hskip -0.3in l(r^*) \sim \frac{a M_{\rm UV} {N_f^{\rm UV}}  \left(\frac{1}{\log (N)}\right)^{10/3} \sqrt{-2 \log \left(\alpha _{\theta _1} \alpha
   _{\theta _2}\right)+\log (N)+\log (16)}}{{g_s}^2 M {N_f}^{10/3} {r^*} \log ({r^*})}\stackrel{r^*\rightarrow\infty}{\longrightarrow}0,
\end{eqnarray}
as in \cite{Tc-EE}. As $r^*$ was assumed to be in the UV and as $l(r^*)$ is a monotonically dereasing function  in the UV, one sees that $l(r^*)$ attains a maximum at $r^*={\cal R}_{D5/\overline{D5}}^{\rm th}$. One can invert (\ref{lr^*-ii}) by solving: $l(r^*) = \frac{C}{r^*\log r^*}$ where $C\sim \frac{a \left(\frac{1}{\log (N)}\right)^{10/3} \sqrt{-2 \log \left(\alpha _{\theta _1} \alpha _{\theta
   _2}\right)+{\log N}+\log (16)}}{{g_s}^2 {N_f^{\rm UV}}^{7/3}}$. Hence,
\begin{equation}
\label{r^*=r^*(l)}
r^* = \frac{C}{l(r^*) W\left(\frac{C}{l(r^*)}\right)}.
\end{equation}

\item
Even though we will not use this in the rest of the paper, but as an academic curiosity, given that $r^*<{\cal R}_{\rm UV}<\left(4\pi g_s N\right)^{\frac{1}{4}}$, if we do not require therefore $l(r^*\rightarrow{\cal R}_{\rm UV})\rightarrow0$, and expand first w.r.t. $\log N > \log(r^*)$ (continuing to assume: $r^* > \sqrt{3} a$), then :
\begin{eqnarray}
\label{lr^*-iii}
& & \sqrt{h_0} \sim \frac{M_{\rm UV} {N_f^{\rm UV}}  {r^*}^3 \log ({r^*}) | 2 \log N -18 \log ({r^*})|  ({g_s}
   \log N  {N_f^{\rm UV}} -3 {g_s} {N_f^{\rm UV}}  \log ({r^*})+4 \pi )^2}{
   {g_s}^{3/4} \sqrt[20]{N} \alpha _{\theta _1}^3 \alpha _{\theta _2}^2}.\nonumber\\
& & 
\end{eqnarray}
One can similarly show that $l(r^*) \sim {r^*} (\log (N)-9 \log ({r^*})) (\log (N)-3 \log ({r^*}))^2$ and
that $l_{\rm max}$ corresponds to:
\begin{eqnarray}
\label{lr^*-iv}
& & r^*_{l_{\rm max}} = \exp \left(\frac{1}{18} \left(-\sqrt{4 \log ^2(N)-36 \log (N)+729}+4 \log (N)-27\right)\right)
\nonumber\\
& &  = \exp\left(\frac{\log (N)}{9}-1-\frac{9}{\log (N)}-\frac{81}{2 \log ^2(N)}+O\left(\left(\frac{1}{\log (N)}\right)^3\right)\right)
\nonumber\\
& & \sim \sqrt[9]{N} e^{-\frac{81}{2 \log ^2(N)}-\frac{9}{\log (N)}-1}.
\end{eqnarray}
As a numerical example, for $N=100$, the Fig. 1 has been obtained.
\begin{figure}
\begin{center}
\includegraphics[width=0.73\textwidth]{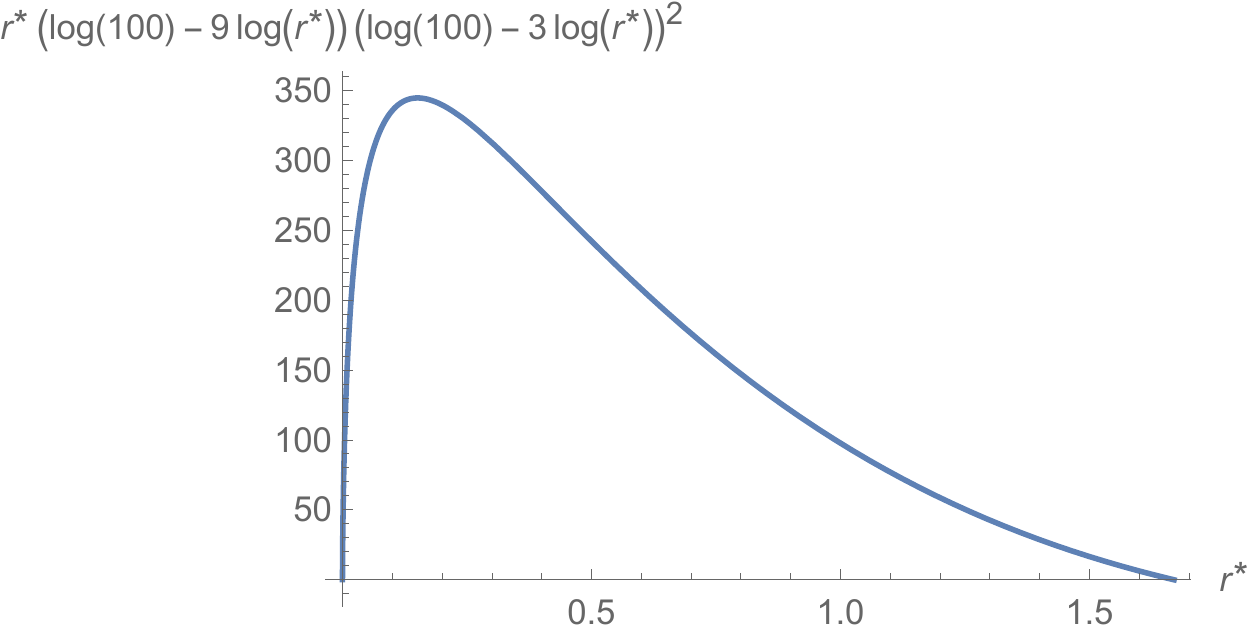}
\end{center}
\caption{Plot of ${r^*} (\log (N)-9 \log ({r^*})) (\log (N)-3 \log ({r^*}))^2$-versus-$r^*$}
\label{l-vs-rstar-lrstarlargenon0}
\end{figure}
\end{itemize}
\newpage
{\bf Entropy}

The connected region entropy catering to small $l$ and large $r$ is:
\begin{eqnarray}
\label{S-connected-large-r}
& & \frac{S}{{\cal V}_1} = \left\{\frac{\int_{{r^*}}^{{\cal R}_{\rm UV}} \frac{\sqrt{\sigma (r)} H(r)}{\sqrt{H(r)-H({r^*})}} \, dr}{2
   {G_N}^{(11)}}\right\}
\end{eqnarray}

\begin{eqnarray}
\label{S-connected-UV-finite}
& & \frac{S_{\rm connected}^{{\rm UV-finite},\ \beta^0}}{2{\cal V}_1} \sim \frac{\log N ^{4/3} M_{\rm UV} {N_f^{\rm UV}} ^{4/3} \alpha _{\theta _2}^2 \left({r_0}^4 {r^*}^2 \log
   ^3({r^*})+{r^*}^6 \log ^2({r^*})\right)}{\sqrt[12]{{g_s}} \sqrt[20]{N} {r^*}^2 \alpha
   _{\theta _1}^6 \log ({r^*})}\nonumber\\
& & \approx \frac{\log N ^{4/3} M_{\rm UV} {N_f^{\rm UV}} ^{4/3} {r^*}^4 \alpha _{\theta _2}^2 \log
   ({r^*})}{\sqrt[12]{{g_s}} \sqrt[20]{N} \alpha _{\theta _1}^6};\nonumber\\
& & \frac{S_{\rm connected}^{{\rm UV-finite}, \beta}}{2{\cal V}_1} \sim \frac{\beta  \sqrt[3]{\frac{1}{{g_s}}} M_{\rm UV} {r^*}^4 ({\cal C}_{zz}^{\rm th}-2 {\cal C}_{\theta_1z}^{\rm th})
   \left(\frac{1}{\log (N)}\right)^{17/3}}{{g_s}^{7/4} \sqrt[20]{N} {N_f^{\rm UV}} ^{2/3} \alpha _{\theta _1}^3 \alpha
   _{\theta _2}^2}.
\end{eqnarray}
{\it From (\ref{f_zz-f_theta1x-f_theta1z}), noting the extremely non-trivial cancelation of ${\cal O}(R^4)$ corrections ${\cal C}_{zz}^{\rm th}-2 {\cal C}_{\theta_1z}^{\rm th} = 0$, we note that there are no ${\cal O}\left({\cal R}^4\right)$-corrections to the connected region (and similarly disconnected region) entanglement entropy.}

The disconnected region entropy will be given by the following expression:
\begin{eqnarray}
\label{S-disconnected}
& & \frac{S_{\rm disconnected}}{2{\cal V}_1} = \left\{\frac{\int_{{r_0}}^{{\cal R}_{\rm UV}} \sqrt{\sigma (r) H(r)} \, dr}{2 {G_N}{}^{(11)}}\right\}\nonumber\\
& & = \left\{\frac{\left(\int_{{r_0}}^{{\cal R}_{D5/\overline{D5}}} + \int_{D5/\overline{D5}}^{{\cal R}_{\rm UV}}\right) \sqrt{\sigma (r) H(r)} \, dr}{2 {G_N}{}^{11}}\right\}\nonumber\\
& & = \frac{1}{\alpha_{\theta_1}^3\alpha_{\theta_2}^2}g_s^{\frac{5}{4}}N_f^{\frac{8}{3}}
\Biggl[\log^2 {\cal R}_{D5/\overline{D5}}{\cal R}_{D5/\overline{D5}}\left(-6a^2+{\cal R}_{D5/\overline{D5}}^2\right)\left(\log N - 3 \log {\cal R}_{D5/\overline{D5}}\right)^{\frac{5}{3}}\nonumber\\
& &  - \log^2r_0r_0^2\left(-6 a^2 + r_0^2\right)\left(\log N - 3\log r_0\right)^{\frac{5}{3}}\Biggr]\left(1 + \frac{{\cal C}_{zz}^{\rm th} - 2 {\cal C}_{\theta_1z}^{\rm th}}{2}\beta\right) + \frac{M_{\rm UV} \sqrt[20]{\frac{1}{N}} {N_f^{\rm UV}} ^{4/3} }{\sqrt[12]{{g_s}} \alpha _{\theta _1}^3 \alpha _{\theta _2}^2}\nonumber\\
& & \times \Biggl[{{\cal R}_{\rm UV}}^4 \log ({{\cal R}_{\rm UV}}) (\log N -3 \log
   ({{\cal R}_{\rm UV}}))^{4/3} \nonumber\\
   & & \hskip 0.3in - {\cal R}_{D5/\overline{D5}}^4 \log ({\cal R}_{D5/\overline{D5}})(\log N -3 \log
   ({\cal R}_{D5/\overline{D5}}))^{4/3}\Biggr].\nonumber\\
& & 
\end{eqnarray} 
Writing $a = \left(\frac{1}{\sqrt{3}} + \epsilon\right)r_0$, the contribution in $r\in\left[r_0, {\cal R}_{D5/\overline{D5}}\right]$ is given by:
\begin{eqnarray}
\label{S-disconnected-IR}
-\frac{\epsilon  {g_s}^{5/4} M \sqrt[20]{\frac{1}{N}} {N_f}^{8/3} {r_0}^4 (\beta  ({\cal C}_{zz}^{\rm th}-2
   {\cal C}_{\theta_1z}^{\rm th})+2)}{\alpha _{\theta _1}^3 \alpha _{\theta _2}^2}.
\end{eqnarray}

Noting that all radial distances are measured in units of ${\cal R}_{D5/\overline{D5}}^{\rm th}$, i.e., $ \log {\cal R}_{D5/\overline{D5}} \equiv 0$, one obtains:
\begin{eqnarray}
\label{final-S-connected+S-disconnected}
& & \frac{S_{\rm disconnected} - S_{\rm UV}^{\rm disconnected}}{2{\cal V}_1} \sim  -\frac{{g_s}^{5/4} M \sqrt[20]{\frac{1}{N}} {N_f}^{8/3} r_0^4 \log ^2(r_0) ({\log N}-3 \log
   (r_0))^{5/3}}{64 \sqrt[3]{2} 3^{5/6} \pi ^{41/12} \alpha _{\theta _1}^3 \alpha _{\theta _2}^2};\nonumber\\
& & \frac{S_{\rm connected} - S_{\rm UV}^{\rm connected}}{2{\cal V}_1} \sim \frac{{M_{\rm UV}} {N_f^{\rm UV}}^{4/3} \alpha _{\theta _2}^2 \log ^{\frac{4}{3}}(N) \left(r_0^4
   {r^*}^2 \log ^3({r^*})-{r^*}^6 \log ^2({r^*})\right)}{72\ 2^{2/3} 3^{5/6}
   \pi ^{25/12} \sqrt[12]{{g_s}} \sqrt[20]{N} {r^*}^2 \alpha _{\theta _1}^6 \log
   ({r^*})}\nonumber\\
& & \approx -\frac{{M_{\rm UV}} {N_f^{\rm UV}}^{4/3} {r^*}^4 \log ^{\frac{4}{3}}(N) \log ({r^*})}{72\ 2^{2/3} 3^{5/6} \pi
   ^{25/12} \sqrt[12]{{g_s}} \sqrt[20]{N}},
\end{eqnarray}
where,
\begin{eqnarray}
\label{SdisconnectedUV+SconnectedUV}
& & S_{\rm UV}^{\rm disconnected} \sim \frac{M_{\rm UV}N_f^{\rm UV}\ ^{\frac{4}{3}} {\cal R}_{\rm UV}^4\left(\log N - 3 \log {\cal R}_{\rm UV}\right)^{\frac{4}{3}}\log{\cal R}_{\rm UV} }{N^{\frac{1}{20}}\alpha_{\theta_1}^3\alpha_{\theta_2}^3},
\nonumber\\
& & S_{\rm UV}^{\rm connected} \sim \frac{M_{\rm UV}N_f^{\rm UV}\ ^{\frac{4}{3}} {\cal R}_{\rm UV}^4\left(\log N \right)^{\frac{4}{3}}\log{\cal R}_{\rm UV} \alpha_{\theta_2}^2}{N^{\frac{1}{20}}\alpha_{\theta_1}^6}.
\end{eqnarray}
One notes that in the large-$N$ limit, recalling ${\cal R}_{\rm UV}<\left(4\pi g_s N\right)^{\frac{1}{4}}$ and setting $\alpha_{\theta_1}^3 =  \alpha_{\theta_2}^4$, one sees that $S_{\rm UV}^{\rm disconnected} = S_{\rm UV}^{\rm connected}$.

As $l$ increases, i.e., $r^*$ decreases and reaches ${\cal R}_{D5/\overline{D5}}^{\rm th}$, $S_{\rm connected}$ changes from being negative to vanishing and $S_{\rm disconnected}$ stays negative implying disconnected region has lesser entropy.
At $r^* = r_{\rm criticial}$, $S_{\rm connected} = S_{\rm disconnected}$ and one sees that $r_{\rm critical}$ is the solution of the equation: $\left(r^*\right)^4\log r^* = \gamma$, where $\gamma \sim \frac{{g_s}^{4/3} M {N_f}^{8/3} \alpha _{\theta _1}^3 \log ^2(r_0) (\log (N)-3 \log
   (r_0))^{5/3}}{\pi ^{4/3} {M_{\rm UV}} {N_f^{\rm UV}}^{4/3} \alpha _{\theta _2}^4 \log
   ^{\frac{4}{3}}(N)}$. The solution is ${r^*}=e^{\frac{1}{4} W(4 \gamma )}\approx \sqrt{2} \sqrt[4]{\gamma}$. From Sec. {\bf 3}, setting $M_{\rm UV},N_f^{\rm} \sim \frac{1}{\log N}$ and $\log r_0 = -\frac{f_{r_0}}{3}\log N$ (and setting $f_{r_0}\approx 1$ \cite{MChPT}), one obtains:
\begin{equation}
\label{r-critical}
r_{\rm critical} = \frac{\sqrt[3]{{g_s}} M {N_f}^{2/3} \alpha _{\theta
   _1}^{3/4} \log ^{\frac{7}{6}}(N)}{\sqrt[4]{2} \sqrt[3]{\pi }
   \alpha _{\theta _2}} r_0.
\end{equation}
For $M=N_f=3, g_s=0.1-1$ (\ref{r-critical}) yields ${\cal O}(1)\frac{\alpha_{\theta_1}^{\frac{3}{4}}}{\alpha_{\theta_2}}$. From (\ref{r_h-r_0-relation}), one thus obtains by taking the 4D-limit effected via $M_{\rm KK}\rightarrow0$, or equivalently $r_0\rightarrow0$:
\begin{eqnarray}
\label{consistency}
& & {\cal O}(1)\frac{\alpha_{\theta_1}^{\frac{3}{4}}}{\alpha_{\theta_2}} \stackrel{\scriptsize 4D-{\rm limit}}{\longrightarrow} \frac{\sqrt[4]{\frac{\kappa_{\rm GHY}^{{\rm th},\ \beta^0}}{\kappa_{\rm GHY}^{{\rm bh},\ \beta^0}}}  {{\cal R}_{D5/\overline{D5}}^{\rm bh}} \sqrt[4]{\frac{\log
   \left(\frac{{{\cal R}_{\rm UV}}}{{{\cal R}_{D5/\overline{D5}}^{\rm th}}}\right)}{\log
   \left(\frac{{{\cal R}_{\rm UV}}}{{{\cal R}_{D5/\overline{D5}}^{\rm bh}}}\right)}}}{{{\cal R}_{D5/\overline{D5}}^{\rm th}}},
\end{eqnarray}
corresponding to the deconfinement temperature. As mentioned below, one would require $\alpha_{\theta_1}^3\alpha_{\theta_2}^2=2$, which would for $N=100, M=N_f=3, g_s=1$ would yield $l(r_{\rm critical})=3850$ (as obtained numerically below) for 
$\alpha_{\theta_2}\sim{\cal O}(10)$. 

Numerically, choosing $N=100, M_{\rm UV} = N_f^{\rm UV} = 0.01, M=N_f=3, \alpha_{\theta_{1,2}}: \alpha_{\theta_1}^3\alpha_{\theta_2}^2=2$, and $r_0=N^{-\frac{f_{r_0}}{3}}, f_{r_0}\approx 1$ \cite{MChPT}, one obtains the plots in Fig. 2.
\begin{figure}
\begin{center}
\includegraphics[width=0.73\textwidth]{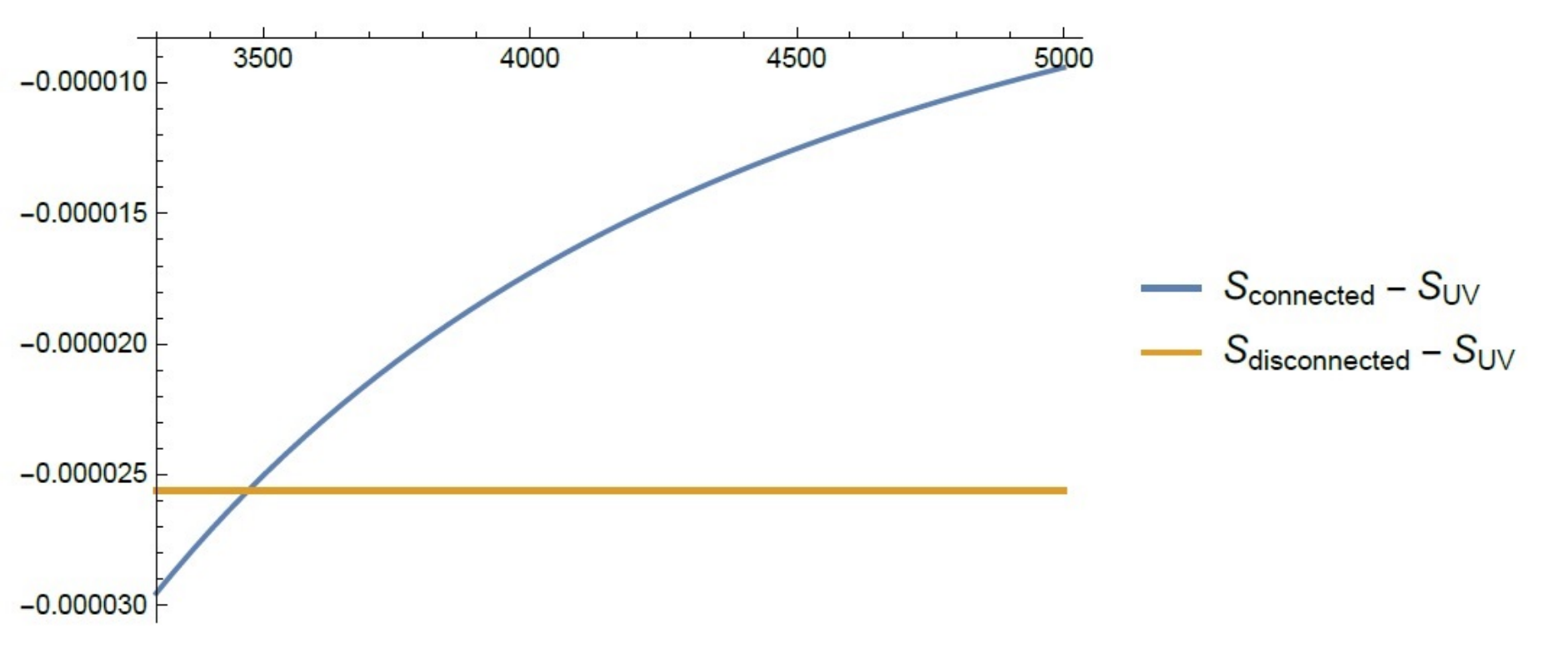}
\end{center}
\caption{Plot $S_{\rm connected}$ (blue) and $S_{\rm disconnected}$(orange) versus $l(r^*)$}
\label{S-cs-l}
\end{figure}

Let us also discuss the possibility of $r^*\in[r_0,{\cal R}_{D5/\overline{D5}}]$, i.e., IR-valued $r^*$. and splitting $\int_{r^*}^{{\cal R}_{\rm UV}}$ into $\left(\int_{r^*}^{{\cal R}_{D5/\overline{D5}}} + \int_{{\cal R}_{D5/\overline{D5}}}^{{\cal R}_{\rm UV}}\right)$, one obtains:
\begin{eqnarray}
\label{rstar-IR-i}
& & \int_{r^*}^{{\cal R}_{D5/\overline{D5}}}dr\frac{\sqrt{\sigma(r)} H(r)}{\sqrt{H(r) - H(r^*)}} \nonumber\\
& & \sim 2 \sqrt{{g_s}} M^2 {N_f}^{13/3} \int_{r^*}^{{\cal R}_{D5/\overline{D5}}} r^6 \sqrt{1-\frac{r_0^4}{r^4}} \log ^2(r) (\log N -3
   \log (r))^{11/3} \, dr\nonumber\\
& & = \int_{r^*}^{{\cal R}_{D5/\overline{D5}}} \left(2 \sqrt{{g_s}} \log N ^{11/3} M^2 {N_f}^{13/3} r^6 \sqrt{1-\frac{r_0^4}{r^4}} \log
   ^2(r)+O\left(\left(\log N\right)^{8/3}\right)\right) dr\nonumber\\
& & = \Biggl[\frac{2}{343} \sqrt{{g_s}} \log N ^{11/3} M^2 {N_f}^{13/3} r^7 \Biggl(7 \log (r) \Biggl(7 \log (r)
   \, _2F_1\left(-\frac{7}{4},-\frac{1}{2};-\frac{3}{4};\frac{r_0^4}{r^4}\right)\nonumber\\
& & -2 \,
   _3F_2\left(-\frac{7}{4},-\frac{7}{4},-\frac{1}{2};-\frac{3}{4},-\frac{3}{4};\frac{r_0^4}{r^4}\right
   )\Biggr)+2 \,
   _4F_3\left(-\frac{7}{4},-\frac{7}{4},-\frac{7}{4},-\frac{1}{2};-\frac{3}{4},-\frac{3}{4},-\frac{3}{4};\frac{r_0^4}{r^4}\right)\Biggr)\Biggr]_{r^*}^{{\cal R}_{D5/\overline{D5}}}\nonumber\\
& & = \frac{4}{343} \sqrt{{g_s}} \log N ^{11/3} M^2 {N_f}^{13/3}-\frac{2 r_0^4
   \left(\sqrt{{g_s}} \log N ^{11/3} M^2 {N_f}^{13/3}\right)}{27
   {{\cal R}_{D5/\overline{D5}}}^4}+O\left(r_0^5\right) \nonumber\\
& & -\frac{2}{7} \left(\sqrt{{g_s}} \log N ^{11/3} M^2 {N_f}^{13/3}  r^*\ ^7 \log
   ^2( r^*)\right)+\frac{1}{3} \sqrt{{g_s}} \log N ^{11/3} M^2 {N_f}^{13/3} r_0^4
    r^*\ ^3 \log ^2( r^*)+O\left(r_0^5\right),\nonumber\\
& & 
\end{eqnarray}
and similarly,
\begin{eqnarray}
\label{rstar-IR-ii}
& & \int_{{\cal R}_{D5/\overline{D5}}}^{{\cal R}_{\rm UV}}dr\frac{\sqrt{\sigma(r)} H(r)}{\sqrt{H(r) - H(r^*)}} \nonumber\\
& & =  \int_{{\cal R}_{D5/\overline{D5}}}^{{\cal R}_{\rm UV}}dr\frac{\sqrt[4]{\frac{1}{{g_s}}} \sqrt[6]{{g_s}} {M_{\rm UV}} {N_f^{\rm UV}}^{4/3} r^3 \alpha _{\theta
   _2}^2 \sqrt{1-\frac{r_0^4}{r^4}} \log (r) (\log N -3 \log (r))^{8/3}}{18\ 2^{2/3} 3^{5/6} \pi
   ^{25/12} \log N ^{4/3} \sqrt[20]{N} \alpha _{\theta _1}^3}+O\left(\left({N_f^{\rm UV}}\right)^2\right)\nonumber\\
& & =\int_{{\cal R}_{D5/\overline{D5}}}^{{\cal R}_{\rm UV}}dr \frac{\log N ^{4/3} {M_{\rm UV}} {N_f^{\rm UV}}^{4/3} r^3 \alpha _{\theta _2}^2
   \sqrt{1-\frac{r_0^4}{r^4}} \log (r)}{18\ 2^{2/3} 3^{5/6} \pi ^{25/12} \sqrt[12]{{g_s}}
   \sqrt[20]{N} \alpha _{\theta _1}^3} + {\cal O}\left(\sqrt[3]{\log N }\right)\nonumber\\
& & = \int_{{\cal R}_{D5/\overline{D5}}}^{{\cal R}_{\rm UV}}dr \frac{\log N ^{4/3} {M_{\rm UV}} {N_f^{\rm UV}}^{4/3} r^3 \alpha _{\theta _2}^2 \log (r)}{18\ 2^{2/3} 3^{5/6}
   \pi ^{25/12} \sqrt[12]{{g_s}} \sqrt[20]{N} \alpha _{\theta
   _1}^3}+O\left(\left(r_0^4\right)^1\right)\nonumber\\
& & = \Biggl[\frac{\log N ^{4/3} {M_{\rm UV}} {N_f^{\rm UV}}^{4/3} r^4 \alpha _{\theta _2}^2 (4 \log (r)-1)}{288\ 2^{2/3}
   3^{5/6} \pi ^{25/12} \sqrt[12]{{g_s}} \sqrt[20]{N} \alpha _{\theta _1}^3}\Biggr]_{{\cal R}_{D5/\overline{D5}}}^{{\cal R}_{\rm UV}}\nonumber\\
& & = \frac{\log N ^{4/3} {M_{\rm UV}} {N_f^{\rm UV}}^{4/3} {{\cal R}_{D5/\overline{D5}}}^4 \alpha _{\theta _2}^2}{288\
   2^{2/3} 3^{5/6} \pi ^{25/12} \sqrt[12]{{g_s}} \sqrt[20]{N} \alpha _{\theta
   _1}^3}+\frac{\log N ^{4/3} {M_{\rm UV}} {N_f^{\rm UV}}^{4/3} {\cal R}_{\rm UV}^4 \alpha _{\theta _2}^2 \log
   ({\cal R}_{\rm UV})}{72\ 2^{2/3} 3^{5/6} \pi ^{25/12} \sqrt[12]{{g_s}} \sqrt[20]{N} \alpha _{\theta _1}^3}.
\end{eqnarray}
\begin{itemize}
\item
One hence notes that the ``UV-divergent'' terms in $S_{\rm connected}^{r^*\in[{\cal R}_{D5/\overline{D5}},{\cal R}_{\rm UV}]}, S_{\rm disconnected}^{r^*=r_0}, S_{\rm connected}^{r^*\in[r_0,{\cal R}_{D5/\overline{D5}}]}$ are the same, denoted by $S_{\rm UV}\sim \frac{\log N ^{4/3} {M_{\rm UV}} {N_f^{\rm UV}}^{4/3} {\cal R}_{\rm UV}^4 \log ({\cal R}_{\rm UV})}{\sqrt[12]{{g_s}}
   \sqrt[20]{N}}$. 

\item
Further,
$S_{\rm connected}^{r^*\in[r_0,{\cal R}_{D5/\overline{D5}}]} - S_{\rm UV}\sim\sqrt{{g_s}} \log N ^{11/3} M^2 {N_f}^{13/3}$.

\item
One therefore obtains (similar to some of the examples in \cite{Tc-EE}):
\begin{eqnarray*}
\label{comparison}
& & S_{\rm connected}^{r^*\in[r_0,{\cal R}_{D5/\overline{D5}}]} - S_{\rm UV} > S_{\rm connected}^{r^*\in[{\cal R}_{D5/\overline{D5}},{\cal R}_{\rm UV}]} > S_{\rm disconnected}^{r^*=r_0}  - S_{\rm UV},\ l(r^*)>l(r_{\rm critical}),\nonumber\\
& & S_{\rm connected}^{r^*\in[r_0,{\cal R}_{D5/\overline{D5}}]} - S_{\rm UV} > S_{\rm connected}^{r^*\in[{\cal R}_{D5/\overline{D5}},{\cal R}_{\rm UV}]} - S_{\rm UV} < S_{\rm disconnected}^{r^*=r_0}  - S_{\rm UV},\ l(r^*)<l(r_{\rm critical}).
\end{eqnarray*}

\end{itemize}

\section{${\mathscr {M}}\chi$PT Compatibility}

In \cite{MChPT}, we worked out one-loop renormalised LECs of $SU(3)$ chiral perturbation theory Lagrangian up to ${\cal O}(p^4)$ from type IIA string dual of large-$N$ thermal QCD like-theory inclusive of ${\cal O}(R^4)$ corrections. We constrained ourselves to the chiral limit due to which some terms were absent in $SU(3)$ chiral perturbation theory Lagrangian of Gasser and Leutwyler \cite{GL}. Therefore we were able to calculate $L_{1,2,3,9,10}^r,F_\pi$ and $g_{YM}$. We match our results with phenomenological values of one-loop renormalised LECs as given in \cite{Ecker-2015}. Let us discuss this in some detail.

The meson sector in the type IIA dual background of top-down holographic type IIB setup is given by the flavor $D6$-branes action. Restricting to the Ouyang embedding (\ref{Ouyang-definition}) for a vanishingly small $|\mu_{\rm Ouyang}|$,
 one will assume that the embedding of the flavor $D6$-branes will be given by $\iota :\Sigma ^{1,6}\Bigg( R^{1,3},r,\theta_{2}\sim\frac{\alpha_{\theta_{2}}}{N^{\frac{3}{10}}},y\Bigg)\hookrightarrow M^{1,9}$, effected by: $z=z(r)$.  As obtained in \cite{Yadav+Misra+Sil-Mesons} one sees that $z$=constant is a solution and by choosing $z=\pm {\cal C}\frac{\pi}{2}$, one can choose the $D6/\overline{D6}$-branes to be at ``antipodal" points along the z coordinate. As in \cite{Yadav+Misra+Sil-Mesons}, we will be working with redefined $(r,z)$ in terms of new variables $(Y,Z)$:
\[r=r_{0}e^{\sqrt{Y^{2}+Z^{2}}}\]
\[z={\cal C}\arctan\frac{Z}{Y}.\]
 Vector mesons are obtained by considering gauge fluctuations of a background gauge field along the world volume of the embedded flavor $D6$-branes (with world volume ${\Sigma}_7(x^{0,1,2,3},Z,\theta_2,\tilde{y}) = {\Sigma}_2(\theta_2,\tilde{y})\times{\Sigma}_5(x^{0,1,2,3},Z)$). Turning on a gauge field fluctuation $\tilde{F}$ about a small background gauge field $F_0$ and the backround $i^*(g+B) [i:\Sigma_7\hookrightarrow M_{10}$, $M_{10}$ being the ten-dimensional ambient space-time]. $Y=0$ is the SYZ mirror of the Ouyang embedding \cite{Yadav+Misra+Sil-Mesons}. Picking up terms quadratic in $\tilde{F}$ in the DBI action:
 {\footnotesize
\begin{equation}
\label{DBI action}
{\rm S}^{IIA}_{D6}=\frac{T_{D_6}(2\pi\alpha^\prime)^{2}}{4} \left(\frac{\pi L^2}{r_0}\right){\rm Str}\int \prod_{i=0}^3dx^i dZd\theta_{2}dy \delta\Bigg(\theta_{2}-\frac{\alpha_{\theta_{2}}}{N^{\frac{3}{10}}}\Bigg) e^{-\Phi} \sqrt{-{\rm det}_{\theta_{2}y}(\iota^*(g+B))}\sqrt{{\rm det}_{{\mathbb  R}^{1,3},Z}(\iota^*g)}g^{\tilde{\mu}\tilde{\nu}}\tilde{F}_{\tilde{\nu}\tilde{\rho}}g^{\tilde{\rho}\tilde{\sigma}}\tilde{F}_{\tilde{\sigma}\tilde{\mu}},
\end{equation}
}
where $\tilde{\mu}=i(=0,1,2,3),Z$. Expanding five dimensional gauge fields as: \\ $A_\mu(x^\nu,Z) = \sum_{n=1}^\infty \rho^{(n)}_\mu(x^\nu)\psi_n(Z)$ and $A_Z(x^\nu,Z) = \sum_{n=0}^\infty\pi^{(n)}(x^\nu)\phi_n(Z) $. After a gauge transformation given below:
$$\rho_{\mu}^{(n)}\rightarrow\rho_{\mu}^{(n)}+{\cal M}_{(n)}^{-1}\partial_{\mu}\pi^{(n)}.$$ Action simplified as given below:
\begin{eqnarray}
& & -\int d^3x\ \ {\rm tr} \left[\frac{1}{2}\partial_{\mu}\pi^{(0)}\partial^{\mu}\pi^{(0)} + \sum_{n\ge 1}\left(\frac{1}{4}\tilde{F}^{(n)}_{\mu\nu}\tilde{F}^{(n)\mu\nu}+\frac{m_{n}^2}{2}\rho^{(n)}_\mu \rho^{(n)\mu }\right)\right].
\end{eqnarray}

Working in the $A_Z(x^\mu,Z)=0$-gauge, integrating out all higher order vector and axial vector meson fields except keeping only the lowest vector meson field \cite{HARADA} \\ $V_\mu^{(1)}(x^\mu) = g \rho_\mu(x^\mu) = \left(\begin{array}{ccc}
\frac{1}{\sqrt{2}}\left(\rho_\mu^0 + \omega_\mu\right) & \rho_\mu^+ & K_\mu^{*+} \\
\rho_\mu^- & -\frac{1}{\sqrt{2}}\left(\rho_\mu^0 - \omega_\mu\right) & K_\mu^{*0} \\
K_\mu^{*-} & {\bar K}_\mu^{*0}  & \phi_\mu
\end{array} \right)$ and lightest pseudo-scalar meson field i.e. $\pi = \frac{1}{\sqrt{2}}\left(\begin{array}{ccc}
\frac{1}{\sqrt{2}}\pi^0 + \frac{1}{\sqrt{6}}\eta_8 + \frac{1}{\sqrt{3}}\eta_0 & \pi^+ & K^+ \\
\pi^- & -\frac{1}{\sqrt{2}}\pi^0 + \frac{1}{\sqrt{6}}\eta_8 + \frac{1}{\sqrt{3}}\eta_0 & K^0 \\
K^- & {\bar K}^0 & -\frac{2}{\sqrt{6}}\eta_8 + \frac{1}{\sqrt{3}}\eta_0
\end{array}\right)$ meson, the gauge field $A_\mu(x^\nu,Z)$ up to ${\cal O}(\pi)$ is given by:
\begin{eqnarray}
\label{Amu-exp}
& & A_\mu(x^\nu,Z) = \frac{\partial_\mu\pi}{F_\pi}\psi_0(Z) - V_\mu(x^\nu)\psi_1(Z),
\end{eqnarray}
where $\psi_0(z) = \int^Z_0 dZ^\prime \phi_0(Z^\prime), V_\mu^{(1)}(x^\nu) = \rho^{(1)}_\mu - \frac{1}{{\cal M}_{(1)}}\partial_\mu\pi^{(1)}$.

To introduce external vector ${\cal V}_\mu$ and axial vector fields ${\cal A}_\mu$, one could use the Hidden Local Symmetry (HLS) formalism of \cite{HARADA} and references therein, wherein  $\frac{1}{F_\pi} \partial_\mu \pi\rightarrow\hat{\alpha}_{\mu \perp}= \frac{1}{F_\pi} \partial_\mu \pi + {\cal A}_\mu - \frac{i}{F_\pi}[{\cal V}_\mu,\pi] + \cdots$ , and one also works with $\hat{\alpha}_{\mu ||} \equiv -V_\mu + {\cal V}_\mu - \frac{i}{2F_\pi^2}[\partial_\mu\pi,\pi] + \cdots$.  Mode expansion of the gauge field contains infinite number of vector meson fields $V_\mu^{(n)}(x^\mu)$ and axial vector meson fields $A_\mu^{(n)}(x^\mu)$, therefore to obtain the low energy effective theory of QCD, we need to truncate the KK spectrum in such a way that we are left with lowest vector meson field $(V_\mu^{(1)} = g \rho_\mu)$ and lightest psuedo-scalar meson field ($\pi$ meson)\cite{HARADA},
\begin{equation}
A_\mu(x^\mu,z) = \hat{\alpha}_{\mu \perp}(x^\mu) \psi_0(z)
  + (\hat{\alpha}_{\mu ||}(x^\mu) + V_\mu^{(1)}(x^\mu)  )
  + \hat{\alpha}_{\mu ||}(x^\mu)  \psi_1(z),
  \end{equation}
therefore,
  \begin{eqnarray}
\label{F mu nu}
& &   F_{\mu\nu} = -V_{\mu\nu} \psi_1 +v_{\mu\nu}(1+\psi_1) + a_{\mu\nu} \psi_0 - i[\hat{\alpha}_{\mu ||},\hat{\alpha}_{\nu ||}]\psi_1(1+\psi_1)  + i[\hat{\alpha}_{\mu \perp},\hat{\alpha}_{\nu \perp}](1+\psi_1-\psi_0^{2}) \nonumber\\
  & &
  -i([\hat{\alpha}_{\mu \perp},\hat{\alpha}_{\nu ||}]+[\hat{\alpha}_{\mu ||},\hat{\alpha}_{\nu \perp}])\psi_1 \psi_0.
 \end{eqnarray}
Based on \cite{HLS-Physics-Reports}, as regards a chiral power counting, one notes that $M_\rho\equiv{\cal O}(p)$ implying $\hat{\alpha}_{\nu ||}\equiv \frac{{\cal O}(p^3)}{M_\rho^2}\equiv{\cal O}(p),
\hat{\alpha}_{\nu\perp}\equiv {\cal O}(p)$. Further, $V_{\mu\nu}, a_{\mu\nu}$ and $v_{\mu\nu}$ are of ${\cal O}(p^2)$. Hence, using (\ref{F mu nu}), $\left(F_{\mu\nu}F^{\mu\nu}\right)^m$ is of ${\cal O}(p^{4m}), m\in\mathbb{Z}^+$. Defining  parity as $x^i\rightarrow-x^i$, $i$ indexing the conformally Minkowskian spatial directions and $Z\rightarrow-Z$, given that $A_\mu(x, Z)$ will be odd, $\alpha_{\mu \perp}$ will be even, $\alpha_{\mu ||}$ will be odd and $V_\mu$ will be odd implies $\psi_0(Z)$  will be odd and $\psi_1(Z)$ will be even. As coupling constants are assumed to be scalars and they are given by $Z$-integrals, the $Z$-dependent terms in the action must be separately of even-$Z$ parity. As  $\psi_0$ has odd $Z$-parity and $\psi_1$ has even $Z$-parity, therefore at ${\cal O}(p^4)$, terms with $(3\hat{\alpha} _{\mu ||}s\ ,\ 1\hat{\alpha}_{\mu \perp}\ {\rm or}\ 3\hat{\alpha}_{\mu \perp} s\ ,\ 1\hat{\alpha} _{\mu ||} )$, are dropped as they involve coefficients of the type $\psi_0^{2m+1}\psi_1^{2n}(Z)$ for appropriate postive integral values of $n, m$. Similarly, at ${\cal O}(p^2)$, $tr(\hat{\alpha}_{\mu \perp}\hat{\alpha}_{||}^{\mu})$  accompanied by $\dot{\psi_0}\dot{\psi_1}(\ ^. \equiv \frac{d}{dZ})$ of odd-$Z$ parity, is dropped.
At ${\cal O}(p^4)$, one hence obtains \cite{HARADA}:
\begin{eqnarray}
\label{Lagrangian-Op4}
 & & {\mathcal{L}}_{(4)}
\ni
y_1 \,
{\rm tr}[{\hat{\alpha}}_{\mu\perp}{\hat{\alpha}}^{\mu}_{\perp}
{\hat{\alpha}}_{\nu\perp}{\hat{\alpha}}^{\nu}_{\perp}]
+
y_2 \,
{\rm tr}[{\hat{\alpha}}_{\mu\perp}{\hat{\alpha}}_{\nu\perp}
{\hat{\alpha}}^{\mu}_{\perp}{\hat{\alpha}}^{\nu}_{\perp}]
+y_3 \,
{\rm tr}[{\hat{\alpha}}_{\mu||}{\hat{\alpha}}^{\mu}_{||}
{\hat{\alpha}}_{\nu||}{\hat{\alpha}}^{\nu}_{||}]
+y_4 \,
{\rm tr}[{\hat{\alpha}}_{\mu||}{\hat{\alpha}}_{\nu||}
{\hat{\alpha}}^{\mu}_{||}{\hat{\alpha}}^{\nu}_{||}] \nonumber\\
& &
+y_5 \,
{\rm tr}[{\hat{\alpha}}_{\mu\perp}{\hat{\alpha}}^{\mu}_{\perp}
{\hat{\alpha}}_{\nu||}{\hat{\alpha}}^{\nu}_{||}]
+y_6 \,
{\rm tr}[{\hat{\alpha}}_{\mu\perp}{\hat{\alpha}}_{\nu\perp}
{\hat{\alpha}}^{\mu}_{||}{\hat{\alpha}}^{\nu}_{||}]
+y_7 \,
{\rm tr}[{\hat{\alpha}}_{\mu\perp}{\hat{\alpha}}_{\nu\perp}
{\hat{\alpha}}^{\nu}_{||}{\hat{\alpha}}^{\mu}_{||}] \nonumber\\
& &
+y_8 \,
\left\{ {\rm tr}[{\hat{\alpha}}_{\mu\perp}{\hat{\alpha}}^{\nu}_{||}
{\hat{\alpha}}_{\nu\perp}{\hat{\alpha}}^{\mu}_{||}]
+
{\rm tr}[{\hat{\alpha}}_{\mu\perp}{\hat{\alpha}}^{\mu}_{||}
{\hat{\alpha}}_{\nu\perp}{\hat{\alpha}}^{\nu}_{||}\right\}
+y_9 \,
{\rm tr}[{\hat{\alpha}}_{\mu\perp}{\hat{\alpha}}_{\nu ||}
{\hat{\alpha}}^{\mu}_{\perp}{\hat{\alpha}}^{\nu}_{||}]\nonumber\\
& &
+z_1 \,
{\rm tr}[v_{\mu\nu}v^{\mu\nu}]
+z_2 \,
{\rm tr}[a_{\mu\nu}a^{\mu\nu}]
+z_3 \,
{\rm tr}[v_{\mu\nu}V^{\mu\nu}]
+iz_4 \,
{\rm tr}[V_{\mu\nu}{\hat{\alpha}}^{\mu}_{\perp}
{\hat{\alpha}}^{\nu}_{\perp}]\nonumber\\
& &
+iz_5 \,
{\rm tr}[V_{\mu\nu}{\hat{\alpha}}^{\mu}_{||}
{\hat{\alpha}}^{\nu}_{||}]
+iz_6 \,
{\rm tr}[v_{\mu\nu}
{\hat{\alpha}}^{\mu}_{\perp}{\hat{\alpha}}^{\nu}_{\perp}]
+iz_7 \,
{\rm tr}[v_{\mu\nu}
{\hat{\alpha}}^{\mu}_{||}{\hat{\alpha}}^{\nu}_{||}]
-iz_8 \,
{\rm tr}\left[a_{\mu\nu}
\left({\hat{\alpha}}^{\mu}_{\perp}{\hat{\alpha}}^{\nu}_{||}
+{\hat{\alpha}}^{\mu}_{||}{\hat{\alpha}}^{\nu}_{\perp}
\right)\right]
\end{eqnarray}
where:
\begin{equation}
 v_{\mu\nu}
= \frac{1}{2} \left(
\xi_R {\cal R}_{\mu\nu} \xi^\dag_R + \xi_L {\cal L}_{\mu\nu} \xi_L^\dag \right)  \hspace{0.5cm}
{\rm and} \hspace{0.5cm}
a_{\mu\nu}
=  \frac{1}{2} \left(
\xi_R {\cal R}_{\mu\nu} \xi^\dag_R - \xi_L {\cal L}_{\mu\nu} \xi_L^\dag
\right),
\end{equation}
${\cal L}_{\mu\nu} = \partial_{[\mu}{\cal L}_{\nu]} - i[{\cal L}_\mu,{\cal L}_\nu]$ and ${\cal R}_{\mu\nu} = \partial_{[\mu}{\cal R}_{\nu]} - i[{\cal R}_\mu,{\cal R}_\nu]$ and ${\cal L}_\mu = {\cal V}_\mu - {\cal A}_\mu$ where ${\cal R}_\mu =  {\cal V}_\mu + {\cal A}_\mu $, and $\xi_L^\dagger(x^\mu) = \xi_R(x^\mu) = e^{\frac{i \pi(x^\mu)}{F_\pi}}$; also, $V_{\mu\nu} = \partial_{[\mu}V_{\nu]}-i[V_\mu,V_\nu]$, and $y_i$s and $z_j$s are given as radial integrals involving $\psi_0(Z)$ and $\psi_1(Z)$ (as reviewed in \cite{MChPT}).

In the chiral limit, the ${\cal O}(p^4)$ $SU(3)\ \chi$PT Lagrangian is given by \cite{GL}:
\begin{eqnarray}
\label{ChPT-Op4}
& & L_1 \left({\rm Tr}(\nabla_\mu U^\dagger \nabla^\mu U)\right)^2 + L_2\left({\rm Tr}(\nabla_\mu U^\dagger \nabla_\nu U)\right)^2 + L_3{\rm Tr} \left(\nabla_\mu U^\dagger \nabla^\mu U\right)^2\nonumber\\
& & - i L_9 Tr\left({\cal L}_{\mu\nu}\nabla^\mu U \nabla^\nu U^\dagger + {\cal R}_{\mu\nu}\nabla^\mu U \nabla^\nu U^\dagger\right) + L_{10} Tr\left(U^\dagger {\cal L}_{\mu\nu}U{\cal R}^{\mu\nu}\right) + H_1 Tr\left({\cal L}_{\mu\nu}^2 + {\cal R}_{\mu\nu}^2\right),\nonumber\\
& &
\end{eqnarray}
where $\nabla_\mu U\equiv \partial_\mu U - i {\cal L}_\mu U + i U {\cal R}_\mu,\ U=e^{\frac{2i\pi}{F_\pi}}$. $SU(3)$ chiral perturbation theory Lagrangian (\ref{ChPT-Op4}) can be  obtained from the HLS Lagrangian (\ref{Lagrangian-Op4}) by integrating out the $\rho$ mesons as done in \cite{HLS-Physics-Reports}. One hence obtains, relationships between the LECs $y_i, z_i$ of (\ref{Lagrangian-Op4}) and the $L_i$s of (\ref{ChPT-Op4}). Using these relations we have calculated LECs of $SU(3)$ $\chi$PT Lagrangian via gauge-gravity duality in \cite{MChPT}.

The parameters $L_i$ and $H_i$ are renormalized at one-loop level with all vertices in one-loop diagrams arising from the ${\cal O}(p^2)$ terms. Using dimensional regularization and performing renormalizations of the parameters via \cite{GL}:
\begin{equation}
L_i = L_i^r(\mu) + \Gamma_i \lambda(\mu) \ , \qquad
H_i = H_i^r(\mu) + \Delta_i \lambda(\mu) \ ,
\end{equation}
where $\mu$ is the renormalization point,
and $\Gamma_i$ and $\Delta_i$ are certain numbers given later;
$\lambda(\mu)$ is the divergent part given by:
\begin{equation}
\lambda(\mu) = - \frac{1}{2\left(4\pi\right)^2}
\left[   \frac{1}{\bar{\epsilon}} - \ln \mu^2 + 1 \right]
\ ,
\end{equation}
where
\begin{equation}
\frac{1}{\bar{\epsilon}} = \frac{2}{4-d}
- \gamma_E + \ln 4\pi 
\ ,
\end{equation}
$d$ being the non-radial non-compact space-time dimensionality to be set to four.
The constants $\Gamma_i$ and $\Delta_i$ for $SU(3)$ $\chi$PT theory were worked out in \cite{GL}:
\begin{equation}
\begin{array}{ccccc}
\Gamma_1 = \frac{3}{32} \ ,
& \Gamma_2 = \frac{3}{16} \ ,
& \Gamma_3 = 0 \ , 
& \Gamma_4 = \frac{1}{8} \ ,
& \Gamma_5 = \frac{3}{8} \ ,
\\
\Gamma_6 = \frac{11}{144} \ ,
& \Gamma_7 = 0 \ ,
& \Gamma_8 = \frac{5}{48} \ ,
& \Gamma_9 = \frac{1}{4} \ , 
& \Gamma_{10} = - \frac{1}{4} \ ;
\\
\Delta_1 = - \frac{1}{8} \ ,
& \Delta_2 = \frac{5}{24} \ .
& & &
\end{array}
\end{equation}
The analog of the 1-loop renormalization in $\chi$PT can be understood on the gravity dual side by noting that the latter requires holographic renormalization as discussed in Section {\bf 3}.

In \cite{MChPT}, we had shown how, in five steps, it is possible to match the phenomenological values of the ${\cal O}(p^4)$  $SU(3)$ $\chi$PT Lagrangian \cite{GL} one-loop renormalized LECs $L_{1,9,10}^r$ as well as $F_\pi^2, g_{\rm YM}(\Lambda_{\rm QCD}=0.4 {\rm GeV}, \Lambda=1.1 {\rm Gev}, \mu=M_\rho)$ exactly, where $\Lambda$ is the ``HLS-QCD'' matching scale \cite{HLS-Physics-Reports} and $\mu$ is the renormalization scale, as well as the order of magnitude and signs of $L_{2,3}^r$. 

On matching LECs computed holographically with their phenomenological/experimental values, it turns out that a particular combination of integration constants, $({\cal C}_{zz}^{\rm th}-{\cal C}_{\theta_1 z}^{\rm th}+2 {\cal C}_{\theta_1 x}^{\rm th})$,  is appearing in all the LECs of $SU(3)$ chiral perturbation theory Lagrangian up to ${\cal O}(p^4)$. Here, ${\cal C}_{MN}^{\rm th}$ are integration constants appearing in solutions to ${\cal O}(R^4)$ corrections as worked out in \cite{MChPT} to the MQGP metric  for the thermal gravitational dual correpsonding to (\ref{TypeIIA-from-M-theory-Witten-prescription-T<Tc}). This combination, $({\cal C}_{zz}^{\rm th}-{\cal C}_{\theta_1 z}^{\rm th}+2 {\cal C}_{\theta_1 x}^{\rm th})$, encodes information about the compact part of non-compact four cycle around which flavor $D7$-branes are wrapping in type IIB setup.

Now, as we are working up to ${\cal O}(\beta)$ and further due to the smallness of $\epsilon$ assuming working up to ${\cal O}(\epsilon^2)$, where $\epsilon$ is defined via the relation $a=\left(\frac{1}{\sqrt{3}}+\epsilon\right)r_0$, simplified form of $L_{1,2,3}^r$, as worked out in \cite{MChPT}, are given below:
\begin{eqnarray}
\label{L1_c}
& & \hskip -0.4in  L_1^r = \frac{L_2^r}{2} = - \frac{L_3^r}{6} = \frac{1}{g_{\rm YM}^2} - z_4 + y_2 \nonumber\\
& & \hskip -0.4in \sim \frac{1}{ {f_{r_0}}
   {g_s}^8 \log N  M^4 N_f ^8 \alpha _{\theta _1}^3 }\Biggl\{3 \pi  ({f_{r_0}}+1) N^{7/5} \Biggl(\frac{3 {g_s}^9 \log r_0  M^4
   N_f ^8  }{({f_{r_0}}+1)^2} \nonumber\\
& & \hskip -0.4in  -\frac{  {f_{r_0}} {g_s}^9 \log N  M^4 N_f ^8
   \alpha _{\theta _1}^2   }{{f_{r_0}}+1} -\frac{576  {f_{r_0}} {g_s}^8 \log N  M^4 N^{\frac{1}{3}}
   N_f ^8 \alpha _{\theta _1}^5}{\pi}\left(\frac{\epsilon ^2}{ \Omega}\right)\Biggr)
\Biggr\},
\end{eqnarray}
where:
\begin{equation}
\label{Omega}
 \Omega \equiv 7 \beta  \left({\cal C}_{zz}^{\rm th} - 2 {\cal C}_{\theta_1z}^{\rm th} + 2 {\cal C}_{\theta_1x}^{\rm th}\right) {f_{r_0}}^2 \gamma ^2 {g_s}^2
   \left(\log N\right) ^2 M^4+3456 \epsilon ^2 ({f_{r_0}}+1) N^2.
\end{equation}
Further, as the $\log r_0$ in (\ref{L1_c}) is in fact $\log\left(\frac{r_0}{{\cal R}_{D5/\overline{D5}}}\right)$ - ${\cal R}_{D5/\overline{D5}}>r_0$ being the $D5-\overline{D5}$ separation - one sees from (\ref{L1_c}) that in order to obtain a positive value (as required from phenomenological value of $L_1^r$), $\Omega<0$. Note, as shown in \cite{MChPT}, matching with the experimental value of the pion decay constant $F_\pi$ requires an $N$-suppression in $\alpha_{\theta_1}$, implying the $N$ enhancement in the last term in (\ref{L1_c}) is artificial.  So, to ensure one does not pick up an ${\cal O}\left(\frac{1}{\beta}\right)$ contribution in $L_1$ from $\frac{\epsilon^2}{\Omega}$ in (\ref{L1_c}) and also to ensure that the third term in (\ref{L1_c}) required to partly compensate the first two negative terms in the same (as explained above) to produce a positive term, is not vanishingly small, from (\ref{Omega}), one needs to set: 
\begin{equation}
\label{epsilon-sqrtbeta-over-N-lambdaeps}
\epsilon = \lambda_{\epsilon}\frac{\sqrt{\beta}}{N}.
\end{equation}
Finally, combining the above observations with the requirement to match the experimental value $L_1^{\rm exp}=0.64\times10^{-3}$, as shown in \cite{MChPT}, one therefore requires to implement the following constraint (which follows from the discussion above):
\begin{eqnarray}
\label{CCsO4}
& & \left({\cal C}_{zz}^{\rm th} - 2 {\cal C}_{\theta_1z}^{\rm th} + 2 {\cal C}_{\theta_1x}^{\rm th}\right) = -\frac{493.7 (\delta +1) ({f_{r_0}}+1) \lambda_{\epsilon}^2}{{f_{r_0}}^2 \gamma ^2 {g_s}^2 \left(\log N\right) ^2 M^4},\nonumber\\
& & \delta = \frac{0.053 \alpha _{\theta _1}^3 N^{-\frac{2 {f_{r_0}}}{3}-1}}{{g_s}}.
\end{eqnarray}

{\it We hence see from (\ref{epsilon-sqrtbeta-over-N-lambdaeps}) that $\epsilon$ provides an expansion parameter connecting the $\frac{1}{N}$ and $\beta$ expansions. Also, together with (\ref{CCsO4}), as was noted in \cite{MChPT}, this was the first connection between large-$N$ and higher derivative corrections in the context of ${\cal M}$-theory dual of large-$N$ thermal QCD-like theories.}

Similarly, it was showin in \cite{MChPT} that:
\begin{eqnarray}
\label{L9}
& & L_9^r = \frac{1}{8}\left(\frac{2}{g_{\rm YM}^2} - 2z_3 - z_4 - z_6\right).
\nonumber\\
& & = -\frac{0.0031 ({f_{r_0}}-1) {g_s} N^{\frac{2 {f_{r_0}}}{3}+\frac{4}{5}}}{({f_{r_0}}+1) \alpha _{\theta _1}^3},
\end{eqnarray}
where,
\begin{equation}
\label{f_{r_0}}
{f_{r_0}} = 1 - \omega \alpha_{\theta_1}^3,
\end{equation}
and
\begin{equation}
\label{fr0-omega}
\omega = \frac{4.6 N^{-\frac{2 {f_{r_0}}}{3}-\frac{4}{5}}}{{g_s}}.
\end{equation}
One sees one gets a match with the phenomenological/experimental value $L_9^{\rm exp}=6.9\times10^{-3}$. Substituting (\ref{f_{r_0}}) - (\ref{fr0-omega}) into (\ref{CCsO4}), one obtains: 
\begin{equation}
\label{CCsO4-lambda-epsilon}
\left({\cal C}_{zz}^{\rm th} - 2 {\cal C}_{\theta_1z}^{\rm th} + 2 {\cal C}_{\theta_1x}^{\rm th}\right) \approx -\frac{987.4 \lambda_{\epsilon}^2}{ \gamma ^2 {g_s}^2 \left(\log N\right) ^2 M^4}<0.
\end{equation}

From equation (\ref{CCsO4-lambda-epsilon}), it is clear that combination of constants of integration required to have definite sign which is negative. In appendix {\bf A}, we have obtained values of these integration constants explicitly by taking the decompactification limit (i.e. $M_{\rm KK}\rightarrow0$ limit) of  the spatial circle $S^1(x^3)$ appearing as part of the  ${\mathscr{M}}$-theory metric used in \cite{MChPT}, i.e., in $S^1(x^0)\times R^2 \times S^1\left(\frac{1}{M_{KK}}\right)$ to recover 4D QCD-like theory. In fact, using those values we derived that the abovementioned combination of integration constants which  appears in all the LECs of $SU(3)$ $\chi$PT Lagrangian up to ${\cal O}(p^4)$, indeed can be made to have a negative sign as required by matching with the phenomenological/experimental values of the one-loop renormalised LECs of $SU(3)$ $\chi$PT Lagrangian  as discussed above. In particular from 
(\ref{f_zz-f_theta1x-f_theta1z}), we see that close to the Ouyang embedding of the flavor $D7$-branes in the parent type IIB dual, there occurs a delicate cancelation between the contributions arising  from the metric corrections at ${\cal O}(R^4)$ in the ${\mathscr{M}}$ theory uplift along the $S^1(\psi/z)$-fiber (considering the $S^3(\theta_1,\phi_1/x,\psi/z)$ as an $S^1$-fibration over the vanishing two-cycle $S^2(\theta_1,\phi_1/x)$) and  $S^2(\theta_1,\psi/z)$ resulting in a non-zero contribution only along  $S^2(\theta,\phi_1/x)$ surviving :
$ {\cal C}_{zz}^{\rm th} - 2 {\cal C}_{\theta_1z}^{\rm th} + 2 {\cal C}_{\theta_1x}^{\rm th} = 2 {\cal C}_{\theta_1x}^{\rm th} \sim \frac{\left(\frac{1}{N}\right)^{7/6} \Sigma _1}{ \epsilon ^{11} {g_s}^{9/4}
   \log N ^4 {N_f}^3 {r_0}^5 \alpha _{\theta _1}^7 \alpha _{\theta _2}^6}$ which is negative for $\Sigma_1<0$. Once again, the above discussion too is a manifestation of the "Flavor Memory" effect discussed in Section {\bf 4}. 

\section{Summary}

Let us  summarise our main results. In this paper, in the context of a top-down holographic evaluation of the deconfinement temperature $T_c$ in QCD at intermediate coupling, we obtain $T_c$ from the ${\mathscr {M}}$-theory dual of large-$N$ thermal QCD-like theories (belonging to the class of theories that display IR confinement, UV conformality with quarks in the fundamental representation of the flavor and color symmetry groups) inclusive of the ${\cal O}(R^4)$ corrections. In this process, there are the following conclusions that are arrived upon. 

\begin{itemize}
\item
{\bf UV-IR Mixing and Flavor Memory}: Performing a semiclassical computation as advocated in \cite{Witten-Hawking-Page-Tc}, by matching the actions at the deconfinement temperature of the ${\mathscr {M}}$-theory uplifts of the thermal and black-hole backgrounds at the UV cut-off, one sees that one obtains a relationship in the IR between the ${\cal O}(R^4)$ corrections to the ${\mathscr {M}}$-theory metric  along the ${\mathscr {M}}$-theory circle in the thermal background and the ${\cal O}(R^4)$ correction to a specific combination of the  ${\mathscr {M}}$-theory metric components along the compact part of the four-cycle "wrapped" by the flavor $D7$-branes of the parent type IIB (warped resolved deformed) conifold geometry - the latter referred to as "Flavor Memory" in the ${\cal M}$-theory uplift.

\item
{\bf Non-Renormalization of $T_c$}: 
\begin{itemize}
\item
{\bf Semiclassical computation}: We further show that the LO result for $T_c$ also holds even after inclusion of the ${\cal O}(R^4)$ corrections. The dominant contribution from the ${\cal O}(R^4)$ terms in the large-$N$ limit arises from the $t_8t_8R^4$ terms  (along with the sub-dominant $\epsilon_{11}\epsilon_{11}R^4$ term), which from a type IIB perspective in the zero-instanton sector, correspond to the tree-level contribution at ${\cal O}\left((\alpha^\prime)^3\right)$ as well as one-loop contribution to four-graviton scattering amplitude and  obtained from integration of the fermionic zero modes. As from the type IIB perspective, the $SL(2,\mathbb{Z})$ completion of these $R^4$ terms \cite{Green and Gutperle} suggests that they are not renormalized  perturbatively beyond one loop in the zero-instanton sector, this therefore suggests the non-renormalization of $T_c$ at all loops in ${\mathscr {M}}$-theory at ${\cal O}(R^4)$.
\item
{\bf $T_c$ from Entanglement entropy}: With an obvious generalization of \cite{Tc-EE} to ${\mathscr {M}}$-theory, we calculated the entanglement entropy between two regions by dividing one of the spatial coordinates of the thermal ${\mathscr {M}}$-theory background into a segment of finite length ${\it l}$ and its complement. Like \cite{Tc-EE}, there are two RT surfaces - connected and disconnected. There is a critical value of ${\it l}$ which is denoted by ${\it l_{crit}}$ such that if one is below the critical value ${\it l_{crit}}$ then it is the connected surface that dominates the entanglement entropy, and if one is above the critical value ${\it l_{crit}}$ then it is the disconnected surface that dominates the entanglement entropy; ${\it l}<{\it l_{crit}}$ corresponds to confining phase of large $N_c$ gauge theories whereas ${\it l}>{\it l_{crit}}$ corresponds to deconfining phase of the same. This is interpreted as confinement-deconfinement phase transition in large $N_c$ gauge theories.

Remarkably, when evaluating the deconfinement temperature from an entanglement entropy computation in the thermal gravity dual, due to an exact and delicate cancelation between the ${\cal O}(R^4)$ corrections from a subset of the abovementioned metric components, one sees that there are consequently no corrections to $T_c$ at quartic order in the curvature supporting the conjecture made in Section ${\bf 4}$ - and summarized above - on the basis of a semiclassical computation.

\end{itemize}

\item
{\bf Deriving ${\mathscr {M}}\chi$PT-Phenomenology compatibility}: As shown in \cite{MChPT}, matching the phenomenological value of the 1-loop renormalized coupling constant corresponding to the ${\cal O}(p^4)$ $SU(3)\ \chi$PT Lagrangian term ``$\left(\nabla_\mu U^\dagger\nabla^\mu U\right)^2$" with the value obtained from the type IIA dual of thermal QCD-like theories inclusive of the aforementioned ${\cal O}(R^4)$ corrections, required the ${\cal O}(R^4)$ corrections arising from  contributions due to the very same abovementioned combination of  ${\mathscr {M}}$-theory metric components evaluated at the IR cut-off, to have a definite sign (negative). The thermal supergravity background dual to type IIB (solitonic) $D3$-branes at low temperatures, includes $\mathbb{R}^2\times S^1(\frac{1}{M_{\rm KK}})$. By taking the $M_{\rm KK}\rightarrow0$ limit (to recover a boundary four-dimensional QCD-like theory after compactifying on the base of a $G_2$-structure cone),  remarkably, we obtain the values of the aforementioned ${\cal O}(R^4)$ corrections to the ${\mathscr {M}}$-theory uplift's metric and therefore {\it derive} the M$\chi$PT requirement of the sign of their specific relevant combination evaluated at the IR cut-off. Close to the Ouyang embedding of the flavor $D7$-branes in the parent type IIB dual, there occurs a delicate cancellation between the contributions arising  from the metric corrections at ${\cal O}(R^4)$ in the ${\mathscr{M}}$ theory uplift along the $S^1(\psi/z)$-fiber (considering the $S^3(\theta_1,\phi_1/x,\psi/z)$, the compact part of the four-cycle $S^3\times \mathbb{R}_{>0}$ "wrapped" by the flavor $D7$-branes in the parent type IIB dual of \cite{metrics}, as an $S^1(\psi/z)$-fibration over the vanishing two-cycle $S^2(\theta_1,\phi_1/x)$) and  $S^2(\theta_1,\psi/z)$ resulting in a non-zero contribution only along $S^2(\theta,\phi_1/x)$ surviving. This further reinforces the "Flavor Memory" discussed earlier. 

\item
{\bf Wald Entropy at ${\cal O}(R^4)$}: Equivalence with Wald entropy for the black hole in the high-temperature ${\mathscr {M}}$-theory dual at ${\cal O}(R^4)$ imposes a linear constraint on the same linear combination of the abovementioned metric corrections.

\end{itemize} 

\section*{Acknowledgement}

GY is supported by a Senior Research Fellowship (SRF) from the Council of Scientific and Industrial Research, Govt. of India.  AM was partially supported by a grant from the Council of Scientific and Industrial Research, Government of India, grant number CSR-1477-PHY. We thank A. Ravichandran for verifying some portion of the calculations of Sec. {\bf 3} as part of his UG research project.

\appendix
\section{${\cal O}(R^4)$ Corrections to the ${\mathscr {M}}$-theory metric of \cite{MQGP} in the MQGP limit near the $\psi=2n\pi, n=0, 1, 2$-branches}
\setcounter{equation}{0}\seceqaa

The ${\cal O}(\beta)$-corrected ${\mathscr {M}}$-theory metric of \cite{MQGP} in the MQGP limit near the $\psi=2n\pi, n=0, 1, 2$-branches  up to ${\cal O}((r-r_h)^2)$ [and up to ${\cal O}((r-r_h)^3)$ for some of the off-diagonal components along the delocalized $T^3(x,y,z)$] - the components which do not receive an ${\cal O}(\beta)$ corrections, are not listed in (\ref{ M-theory-metric-psi=2npi-patch}) - was worked out in \cite{OR4-Yadav+Misra} and is given below:
{\footnotesize
\begin{eqnarray}
\label{ M-theory-metric-psi=2npi-patch}
 & &   G_{tt} =  G^{\rm MQGP}_{tt}\Biggl[1 + \frac{1}{4}  \frac{4 b^8 \left(9 b^2+1\right)^3 \left(4374 b^6+1035 b^4+9 b^2-4\right) \beta  M \left(\frac{1}{N}\right)^{9/4} \Sigma_1
   \left(6 a^2+  {r_h}^2\right) \log (  {r_h})}{27 \pi  \left(18 b^4-3 b^2-1\right)^5  \log N ^2   {N_f}   {r_h}^2
   \alpha _{\theta _2}^3 \left(9 a^2+  {r_h}^2\right)} (r-  {r_h})^2\Biggr]
\nonumber\\
& & G_{x^{1,2,3}x^{1,2,3}}  =   G^{\rm MQGP}_{x^{1,2,3}x^{1,2,3}}
\Biggl[1 - \frac{1}{4} \frac{4 b^8 \left(9 b^2+1\right)^4 \left(39 b^2-4\right) M \left(\frac{1}{N}\right)^{9/4} \beta  \left(6 a^2+{r_h}^2\right) \log
   ({r_h})\Sigma_1}{9 \pi  \left(3 b^2-1\right)^5 \left(6 b^2+1\right)^4 \log N ^2 {N_f} {r_h}^2 \left(9 a^2+{r_h}^2\right) \alpha
   _{\theta _2}^3} (r - {r_h})^2\Biggr]\nonumber\\
& & G_{rr}  =  G^{\rm MQGP}_{rr}\Biggl[1 + \Biggl(- \frac{2 \left(9 b^2+1\right)^4 b^{10} M   \left(6 a^2+{r_h}^2\right) \left((r-{r_h})^2+{r_h}^2\right)\Sigma_1}{3 \pi
   \left(-18 b^4+3 b^2+1\right)^4 \log N  N^{8/15} {N_f} \left(-27 a^4+6 a^2 {r_h}^2+{r_h}^4\right) \alpha _{\theta
   _2}^3}\nonumber\\
& & +{\cal C}_{zz}^{\rm bh}-2 {\cal C}_{\theta_1z}^{\rm bh}+2 {\cal C}_{\theta_1x}^{\rm bh}\Biggr)\beta\Biggr]\nonumber\\
 & & G_{\theta_1x}  =  G^{\rm MQGP}_{\theta_1x}\Biggl[1 + \Biggl(
- \frac{\left(9 b^2+1\right)^4 b^{10} M  \left(6 a^2+{r_h}^2\right) \left((r-{r_h})^2+{r_h}^2\right)
   \Sigma_1}{3 \pi  \left(-18 b^4+3 b^2+1\right)^4 \log N  N^{8/15} {N_f} \left(-27 a^4+6 a^2
   {r_h}^2+{r_h}^4\right) \alpha _{\theta _2}^3}+{\cal C}_{\theta_1x}^{\rm bh}
\Biggr)\beta\Biggr]\nonumber\\
& & G_{\theta_1z}  =  G^{\rm MQGP}_{\theta_1z}\Biggl[1 + \Biggl(\frac{16 \left(9 b^2+1\right)^4 b^{12}  \beta  \left(\frac{(r-{r_h})^3}{{r_h}^3}+1\right) \left(19683
   \sqrt{3} \alpha _{\theta _1}^6+3321 \sqrt{2} \alpha _{\theta _2}^2 \alpha _{\theta _1}^3-40 \sqrt{3} \alpha _{\theta _2}^4\right)}{243
   \pi ^3 \left(1-3 b^2\right)^{10} \left(6 b^2+1\right)^8 {g_s}^{9/4} \log N ^4 N^{7/6} {N_f}^3 \left(-27 a^4 {r_h}+6 a^2
   {r_h}^3+{r_h}^5\right) \alpha _{\theta _1}^7 \alpha _{\theta _2}^6}+{\cal C}_{\theta_1z}^{\rm bh}\Biggr)\Biggr]\nonumber\\
 & &   G_{\theta_2x}  =  G^{\rm MQGP}_{\theta_2x}\Biggl[1 + \Biggl(
   \frac{16 \left(9 b^2+1\right)^4 b^{12} \left(\frac{(r-{r_h})^3}{{r_h}^3}+1\right) \left(19683 \sqrt{3}
   \alpha _{\theta _1}^6+3321 \sqrt{2} \alpha _{\theta _2}^2 \alpha _{\theta _1}^3-40 \sqrt{3} \alpha _{\theta _2}^4\right)}{243 \pi ^3 \left(1-3
   b^2\right)^{10} \left(6 b^2+1\right)^8 {g_s}^{9/4} \log N ^4 N^{7/6} {N_f}^3 \left(-27 a^4 {r_h}+6 a^2
   {r_h}^3+{r_h}^5\right) \alpha _{\theta _1}^7 \alpha _{\theta _2}^6}+{\cal C}_{\theta_2x}^{\rm bh}\Biggr)\beta\Biggr]\nonumber\\
& & G_{\theta_2y}  =  G^{\rm MQGP}_{\theta_2y}\Biggl[1 +  \frac{3 b^{10} \left(9 b^2+1\right)^4 M \beta \left(6 a^2+{r_h}^2\right) \left(1-\frac{(r-{r_h})^2}{{r_h}^2}\right) \log
   ({r_h}) \Sigma_1}{\pi  \left(3 b^2-1\right)^5 \left(6 b^2+1\right)^4 \log N ^2 N^{7/5} {N_f} \left(9 a^2+{r_h}^2\right) \alpha
   _{\theta _2}^3}\Biggr]\nonumber\\
& & G_{\theta_2z}  =  G^{\rm MQGP}_{\theta_2z}\Biggl[1 + \Biggl(\frac{3 \left(9 b^2+1\right)^4 b^{10} M  \left(6 a^2+{r_h}^2\right) \left(1-\frac{(r-{r_h})^2}{{r_h}^2}\right) \log
   ({r_h}) \left(19683 \sqrt{6} \alpha _{\theta _1}^6+6642 \alpha _{\theta _2}^2 \alpha _{\theta _1}^3-40 \sqrt{6} \alpha _{\theta
   _2}^4\right)}{\pi  \left(3 b^2-1\right)^5 \left(6 b^2+1\right)^4 {\log N}^2 N^{7/6} {N_f} \left(9 a^2+{r_h}^2\right) \alpha
   _{\theta _2}^3}\nonumber\\
& & +{\cal C}_{\theta_2 z}^{\rm bh}\Biggr)\beta\Biggr]\nonumber\\
& & G_{xy}  =  G^{\rm MQGP}_{xy}\Biggl[1 + \Biggl(\frac{3 \left(9 b^2+1\right)^4 b^{10} M  \left(6 a^2+{r_h}^2\right) \left(\frac{(r-{r_h})^2}{{r_h}^2}+1\right) \log
   ({r_h}) \alpha _{\theta _2}^3\Sigma_1}{\pi  \left(3 b^2-1\right)^5 \left(6 b^2+1\right)^4 \log N ^2 N^{21/20} {N_f} \left(9
   a^2+{r_h}^2\right) \alpha _{\theta _{2 l}}^6}+{\cal C}_{xy}^{\rm bh}\Biggr)\beta\Biggr]\nonumber\\
& & G_{xz}   =  G^{\rm MQGP}_{xz}\Biggl[1 + \frac{18 b^{10} \left(9 b^2+1\right)^4 M \beta  \left(6 a^2+{r_h}^2\right)
   \left(\frac{(r-{r_h})^2}{{r_h}^2}+1\right) \log ^3({r_h}) \Sigma_1}{\pi  \left(3b^2-1\right)^5 \left(6 b^2+1\right)^4 \log N ^4 N^{5/4} {N_f} \left(9 a^2+{r_h}^2\right) \alpha
   _{\theta _2}^3}\Biggr]\nonumber\\
& & G_{yy}  =  G^{\rm MQGP}_{yy}\Biggl[1  - \frac{3 b^{10} \left(9 b^2+1\right)^4 M \left(\frac{1}{N}\right)^{7/4} \beta  \left(6 a^2+{r_h}^2\right) \log \left(\frac{r_h}{{\cal R}_{D5/\overline{D5}}^{\rm bh}}\right)\Sigma_1
   \left(\frac{(r-{r_h})^2}{r_h^2}+1\right)}{\pi  \left(3 b^2-1\right)^5 \left(6 b^2+1\right)^4 \log N ^2 {N_f} {r_h}^2 \left(9
   a^2+{r_h}^2\right) \alpha _{\theta _2}^3}\Biggr]\nonumber\\
& &  G_{yz}  =  G^{\rm MQGP}_{yz}\Biggl[1 + \Biggl(\frac{64 \left(9 b^2+1\right)^8 b^{22} M \left(\frac{1}{N}\right)^{29/12}  \left(6 a^2+{r_h}^2\right)
   \left(\frac{(r-{r_h})^3}{{r_h}^3}+1\right) \log \left(\frac{r_h}{{\cal R}_{D5/\overline{D5}}^{\rm bh}}\right) }{27 \pi ^4 \left(3 b^2-1\right)^{15} \left(6 b^2+1\right)^{12}
   {g_s}^{9/4} \log N ^6  {N_f}^4 {r_h}^3 \left({r_h}^2-3 a^2\right) \left(9 a^2+{r_h}^2\right)^2 \alpha
   _{\theta _1}^7 \alpha _{\theta _2}^9}\nonumber\\
& & \times \left(387420489 \sqrt{2} \alpha _{\theta _1}^{12}+87156324 \sqrt{3}
   \alpha _{\theta _2}^2 \alpha _{\theta _1}^9+5778054 \sqrt{2} \alpha _{\theta _2}^4 \alpha _{\theta _1}^6-177120 \sqrt{3} \alpha _{\theta
   _2}^6 \alpha _{\theta _1}^3+1600 \sqrt{2} \alpha _{\theta _2}^8\right)+{\cal C}_{yz}^{\rm bh}\Biggr)\beta\Biggr]\nonumber\\
& & G_{zz}  =  G^{\rm MQGP}_{zz}\Biggl[1 + \Biggl({\cal C}_{zz}^{\rm bh}-\frac{b^{10} \left(9 b^2+1\right)^4 M \left({r_h}^2-\frac{(r-{r_h})^3}{{r_h}}\right) \log \left(\frac{r_h}{{\cal R}_{D5/\overline{D5}}^{\rm bh}}\right)
   \Sigma_1}{27 \pi ^{3/2} \left(3 b^2-1\right)^5 \left(6 b^2+1\right)^4 \sqrt{{g_s}} \log N ^2 N^{23/20} {N_f} \alpha
   _{\theta _2}^5}\Biggr)\beta\Biggr]\nonumber\\
& & G_{x^{10}x^{10}}  =  G^{\rm MQGP}_{x^{10}x^{10}}\Biggl[1 -\frac{27 b^{10} \left(9 b^2+1\right)^4 M \left(\frac{1}{N}\right)^{5/4} \beta  \left(6 a^2+{r_h}^2\right)
   \left(1-\frac{(r-{r_h})^2}{{r_h}^2}\right) \log ^3({r_h}) \Sigma_1}{\pi  \left(3 b^2-1\right)^5 \left(6 b^2+1\right)^4 \log N ^4
   {N_f} {r_h}^2 \left(9 a^2+{r_h}^2\right) \alpha _{\theta _2}^3}\Biggr],
\end{eqnarray}
}
where $\Sigma_1$ is defined as:
\begin{eqnarray}
\label{Sigma_1-def}
& & \hskip -0.8in\Sigma_1 \equiv 19683
   \sqrt{6} \alpha _{\theta _1}^6+6642 \alpha _{\theta _2}^2 \alpha _{\theta _1}^3-40 \sqrt{6} \alpha _{\theta _2}^4,
\end{eqnarray}
and $G^{\rm MQGP}_{MN}$ are the ${\mathscr {M}}$-theory metric components in the MQGP limit at ${\cal O}(\beta^0)$ 
\cite{VA-Glueball-decay}. The explicit dependence on $\theta_{10,20}$ of the ${\mathscr {M}}$-theory metric components up to ${\cal O}(\beta)$, using (\ref{small-theta_12}), is effected by the replacemements: 
$\alpha_{\theta_1}\rightarrow N^{\frac{1}{5}}\sin\theta_{10},\ \alpha_{\theta_2}\rightarrow N^{\frac{3}{10}}\sin\theta_{20}$ in (\ref{ M-theory-metric-psi=2npi-patch}).

\section{Thermal $f_{MN}$ EOMs, their Solutions in the IR, $4D$-Limit and  ${\mathscr {M}}\chi$PT Compatibility}
\setcounter{equation}{0} \seceqbb

In this appendix, we discuss the independent EOMs for the metric perturbations $f_{MN}$ of (\ref{TypeIIA-from-M-theory-Witten-prescription-T<Tc}) close to the IR cut-off $r_0$ up to LO in $N$, their solutions and constraints and values of the same in the decompactification-limit of a spatial direction (that plays a crucial role in providing evidence of an all-loop non-renormalization of $T_c$ at ${\cal O}(R^4)$). In this way, we are able to obtain the values of the metric perturbations (in the deep IR) along the three-cycle $S^3(\theta_1,x,z)$ - the delocalized version of $S^3(\theta_1,\phi_1,\psi)$ -strictly speaking along the fiber $S^1(z)$ (the $S^3(\theta_1,x,z)$ is an $S^1(z)$ fibration over the vanishing two-cycle $S^2(\theta_1,x)$) and along another two-cycle $S^2(\theta_1,z)$ (which is also an $S^1(z)$-fibration). A specific linear combination of the contributions from $\left.f_{MN}\right|_{S^3(\theta_1,x,z)}$, near the Ouyang embedding in the parent type IIB dual, appears ubiquitously in computations of $T_c$ in this work and the LECs of $SU(3)\ \chi$PT Lagrangian at ${\cal O}(p^4)$ from ${\mathscr {M}}\ \chi$PT in \cite{MChPT}. We will prove that the negative sign of this combination, as required by matching with the phenomenological values of the LECs in the $\chi$PT Lagrangian at ${\cal O}(p^4)$ in  \cite{MChPT}, can hence be derived; the bonus is the actual values of the contributions appearing in the aforementioned linear combination which further demonstrates a delicate cancellation with the contribution along the vanishing two-cycle $S^2(\theta_1,x)$ being the only one surviving. These are manifestations of the "Flavor Memory" as discussed in Sections {\bf 4, 6}.

The EOMs are given below:
{\footnotesize
\begin{eqnarray}
\label{EOM-fMN-thermal}
& & {\rm EOM}_{tt}:\nonumber\\
& &  -\frac{\beta  \left(\frac{1}{N}\right)^{9/4} \left(19683 \sqrt{6} \alpha _{\theta _1}^6+6642 \alpha _{\theta _2}^2
   \alpha _{\theta _1}^3-40 \sqrt{6} \alpha _{\theta _2}^4\right) \log ({r_0}) }{156728328192 \pi ^3 {g_s}^5 \log N ^4 M {N_f}^3
   \epsilon ^{10} \alpha _{\theta _2}^3}\nonumber\\
& & \times \left(-\frac{98 \pi ^3 (2
   {f_{zz}}({r_0})-3 {f_{x^{10}x^{10}}}({r_0})-4 {f_{\theta_1z}}({r_0})-5 {f_{\theta_2y}}({r_0}))}{\log
   ^2({r_0})}-\frac{1728 {g_s}^3 M^2 \left(\frac{1}{N}\right)^{2/5} {N_f}^2 ({f_{yy}}({r_0})-2
   {f_{yz}}({r_0}))}{\alpha _{\theta _2}^2}\right) = 0\nonumber\\
& & \nonumber\\
& &  {\rm EOM}_{\theta_1\theta_1}:\nonumber\\
& &  \frac{81 \alpha _{\theta _1}^2 \left(\frac{1152 {g_s}^3 M^2 \left(\frac{1}{N}\right)^{2/5} {N_f}^2 \log
   ^2({r_0}) \left(\alpha _{\theta _2} {f_{zz}}({r_0})-2 \alpha _{\theta _2} {f_{yz}}({r_0})+\alpha
   _{\theta _2} {f_{yy}}({r_0})\right)}{\pi ^3}-98 \alpha _{\theta _2}^3 (2
   {f_{x^{10}x^{10}}}({r_0})+{f_{\theta_2y}}({r_0}))\right)}{1024 \alpha _{\theta _2}^5}\nonumber\\
& & +\frac{\beta  M
   \left(\frac{1}{N}\right)^{19/10} \left(-19683 \sqrt{6} \alpha _{\theta _1}^6-6642 \alpha _{\theta _2}^2 \alpha
   _{\theta _1}^3+40 \sqrt{6} \alpha _{\theta _2}^4\right)}{4782969 \pi ^{5/4} \sqrt[4]{{g_s}} \log N ^2
   {N_f} {r_0}^2 \epsilon ^7 \alpha _{\theta _2}^5}=0\nonumber\\
& & \nonumber\\
& & {\rm EOM}_{\theta_1\theta_2}:\nonumber\\
& &  -\frac{441 \sqrt{3} {r_0} \alpha _{\theta _1}^6 (2 {f_{x^{10}x^{10}}}({r_0})+{f_{\theta_2y}}({r_0}))}{\alpha _{\theta
   _2}^3}-\frac{64 \sqrt{2} {g_s}^{3/2} M {N_f} {fr}({r_0})}{\pi ^{3/2} \sqrt[5]{N}}=0\nonumber\\
& & \nonumber\\
& &  {\rm EOM}_{\theta_2\theta_2}:\nonumber\\
& & \frac{{f_{zz}}({r_0})-{f_{x^{10}x^{10}}}({r_0})-2 {f_{\theta_1z}}({r_0})-{fr}({r_0})}{9 \log N ^2
   \alpha _{\theta _1}^4}+\frac{9 \sqrt{6} {g_s}^{3/2} M {N_f} {r_0} \log ^2({r_0})
   ({f_{\theta_1y}}({r_0})-{f_{yz}}({r_0}))}{\pi ^{3/2} \log N ^2 N^{2/5} \alpha _{\theta _2}^3}=0\nonumber\\
& & \nonumber\\
& &  {\rm EOM}_{\theta_2y}:\nonumber\\
& & \frac{-\frac{49 \sqrt{2} \pi ^3 \alpha _{\theta _2}^3 (7 {f_{zz}}({r_0})-15 {f_{x^{10}x^{10}}}({r_0})-14
   {f_{\theta_1z}}({r_0})+5 {f_{\theta_2y}}({r_0}))}{\log ^2({r_0})}-864 \sqrt{2} {g_s}^3 M^2
   \left(\frac{1}{N}\right)^{2/5} {N_f}^2 \left(2 \alpha _{\theta _2} {f_{yz}}({r_0})-\alpha _{\theta _2}
   {f_{yy}}({r_0})\right)}{10368 \sqrt[4]{\pi } {g_s}^{13/4} \log N ^2 M^2 {N_f}^2 \epsilon ^2 \alpha
   _{\theta _1}^2 \alpha _{\theta _2}^2}\nonumber\\
& & +\frac{32 \beta  \left(\frac{1}{N}\right)^{3/2} \left(19683 \sqrt{6} \alpha
   _{\theta _1}^6+6642 \alpha _{\theta _2}^2 \alpha _{\theta _1}^3-40 \sqrt{6} \alpha _{\theta _2}^4\right)}{3486784401
   {g_s}^2 \log N ^2 {N_f}^2 \epsilon ^8 \alpha _{\theta _1}^6 \alpha _{\theta _2}}=0\nonumber\\
& &  {\rm EOM}_{xx}: \nonumber\\
& &\frac{-\frac{1024 \pi ^{3/2} {g_s}^3 \alpha _{\theta _2}^3 ({f_{zz}}({r_0})-2 {f_{\theta_1z}}({r_0})+2
   {f68}({r_0})-{fr}({r_0}))}{\alpha _{\theta _1}^4}-\frac{11907 \pi ^{9/2}
   \left(\frac{1}{N}\right)^{2/5} \alpha _{\theta _2}^5 (3 {f_{x^{10}x^{10}}}({r_0})+5
   {f_{\theta_2y}}({r_0}))}{\log N ^2 M^2 {N_f}^2 \log ^2({r_0})}}{279936 \pi ^2 {g_s}^{7/2} \epsilon ^2
   \alpha _{\theta _2}^5}\nonumber\\
& & +\frac{64 \beta  \left(\frac{1}{N}\right)^{23/20}}{43046721 \sqrt[4]{\pi } {g_s}^{9/4}
   \log N ^3 {N_f}^2 \epsilon ^7 \alpha _{\theta _1}^4 \alpha _{\theta _2}^4}=0\nonumber\\
& &  {\rm EOM}_{yy}:\nonumber\\
& & \frac{49 \pi ^{5/2} \left(16 {f_{zz}}({r_0})+3 {f_{x^{10}x^{10}}}({r_0})+4 {f_{\theta_1z}}({r_0})+\frac{18 \alpha
   _{\theta _2}^2 {f_{\theta_1y}}({r_0})}{\sqrt[5]{N} \alpha _{\theta _1}^2}+5 {f_{\theta_2y}}({r_0})-36
   {f_{yz}}({r_0})+18 {f_{yy}}({r_0})\right)}{1152 {g_s}^{7/2} \log N ^2 M^2 {N_f}^2 \epsilon
   ^2 \log ^2({r_0})}\nonumber\\
& & +\frac{2 \beta  \left(19683 \alpha _{\theta _1}^6+1107 \sqrt{6} \alpha _{\theta _2}^2 \alpha
   _{\theta _1}^3-40 \alpha _{\theta _2}^4\right)}{387420489 \sqrt[4]{\pi } {g_s}^{9/4} \log N ^3 N^{3/2}
   {N_f}^2 \epsilon ^9 \alpha _{\theta _1}^4 \alpha _{\theta _2}^2}=0.\nonumber\\
& & \nonumber\\   
\end{eqnarray}
}
The solutions to (\ref{EOM-fMN-thermal}) are given as under:
\begin{eqnarray}
\label{solutions-fMN}
& & f_t(r) = f_t(r_0),\nonumber\\
& & f(r) = f(r_0),\nonumber\\
& & f_r(r) = -\frac{99 \sqrt{\frac{3}{2}} \beta  {g_s}^{3/2} M \sqrt[5]{\frac{1}{N}} {N_f} {r_0} \alpha _{\theta _1}^6
   {f_{x^{10}x^{10}}}({r_0}) \log ^2({r_0})}{2 \pi ^{3/2} \alpha _{\theta _2}^5},\nonumber\\
& & f_{\theta_1\theta_1}(r) = f_{\theta_1\theta_1}(r_0),\nonumber\\
& & f_{\theta_1\theta_2}(r) = f_{\theta_1\theta_2}(r_0),\nonumber\\
& & f_{\theta_1x}(r) = -\frac{99 \sqrt{\frac{3}{2}} {g_s}^{3/2} M \sqrt[5]{\frac{1}{N}} {N_f} {r_0} \alpha _{\theta _1}^6
   {f_{x^{10}x^{10}}}({r_0}) \log ^2({r_0})}{4 \pi ^{3/2} \alpha _{\theta _2}^5}-{f_{x^{10}x^{10}}}({r_0}),\nonumber\\
& & f_{\theta_1y}(r) =  f_{\theta_1y}(r_0),\nonumber\\
& &  f_{\theta_1z} = \frac{539 \pi ^3 N^{2/5} \alpha _{\theta _2}^2 {f_{x^{10}x^{10}}}({r_0})}{1728 {g_s}^3 M^2 {N_f}^2 \log
   ^2({r_0})}-\frac{185 {f_{x^{10}x^{10}}}({r_0})}{108},\nonumber\\
& & f_{\theta_2\theta_2}(r) = f_{\theta_2\theta_2}(r_0),\nonumber\\
& & f_{\theta_2x}(r) = f_{\theta_2x}(r_0),\nonumber\\
& & f_{\theta_2y}(r) = \frac{352 {g_s}^3 M^2 \left(\frac{1}{N}\right)^{2/5} {N_f}^2 {f_{x^{10}x^{10}}}({r_0}) \log ^2({r_0})}{49 \pi
   ^3 \alpha _{\theta _2}^2}-2 {f_{x^{10}x^{10}}}({r_0}),\nonumber\\
& & f_{\theta_2z}(r) = f_{\theta_2z}(r_0),\nonumber\\
& & f_{xx}(r) = f_{xx}(r_0),\nonumber\\
& & f_{xy}(r) = f_{xy}(r_0),\nonumber\\
& & f_{xz}(r) = f_{xz}(r_0),\nonumber\\
& & f_{yy}(r) = \frac{N^{2/5} {f_{x^{10}x^{10}}}({r_0}) \left(32 \sqrt{6} \pi ^{3/2} {g_s}^{3/2} M {N_f} \alpha _{\theta
   _2}^3-4851 \pi ^3 {r_0} \alpha _{\theta _1}^4 \alpha _{\theta _2}^2\right)}{7776 {g_s}^3 M^2 {N_f}^2
   {r_0} \alpha _{\theta _1}^4 \log ^2({r_0})}+\frac{55 {f_{x^{10}x^{10}}}({r_0})}{27} \nonumber\\
   & & \hskip 0.6in +{f_{\theta_1y}}({r_0}),\nonumber\\
& &  f_{yz}(r) = \frac{\pi ^{3/2} N^{2/5} \alpha _{\theta _2}^3 {f_{x^{10}x^{10}}}({r_0})}{81 \sqrt{6} {g_s}^{3/2} M {N_f}
   {r_0} \alpha _{\theta _1}^4 \log ^2({r_0})}+\frac{{f_{\theta_1y}}({r_0})}{2},\nonumber\\
& & f_{zz}(r) = \frac{539 \pi ^3 N^{2/5} \alpha _{\theta _2}^2 {f_{x^{10}x^{10}}}({r_0})}{864 {g_s}^3 M^2 {N_f}^2 \log
   ^2({r_0})}-\frac{77 {f_{x^{10}x^{10}}}({r_0})}{54}(r_0),\nonumber\\
& & f_{x^{10}x^{10}}(r) = f_{x^{10}x^{10}}(r_0).
\end{eqnarray}

We will now show consistency between the solutions to the EOMs for ${\cal O}(R^4)$ metric perturbations as given in (\ref{solutions-fMN}) and the ones obtained by taking $\tilde{g}(r)\rightarrow1$-limit of the ${\cal O}(R^4)$ corrections to (\ref{TypeIIA-from-M-theory-Witten-prescription-T<Tc}) as obtained in \cite{MChPT}. In the $M_{\rm KK}\rightarrow0$ limit yielding 4D gauge theory after dimensional reduction to $M_5(r,t,x^{1,2,3})$, consistency of the aforementioned pair of solutions $f_{MN}$-wise requires:
\begin{eqnarray*}
& & f_t(r) = f_t(r_0) = \left.- \frac{1}{4} \frac{4 b^8 \left(9 b^2+1\right)^4 \left(39 b^2-4\right) M \left(\frac{1}{N}\right)^{9/4} \beta  \left(6 a^2+{r_0}^2\right) \log
   ({r_0})\Sigma_1}{9 \pi  \left(3 b^2-1\right)^5 \left(6 b^2+1\right)^4 \log N ^2 {N_f} {r_0}^2 \left(9 a^2+{r_0}^2\right) \alpha
   _{\theta _2}^3} (r - {r_0})^2\right|_{r=r_0} \nonumber\\
& & = f(r) = f(r_0) = f_{x^3x^3}\nonumber\\
& &  = \left. \frac{1}{4}  \frac{4 b^8 \left(9 b^2+1\right)^3 \left(4374 b^6+1035 b^4+9 b^2-4\right) \beta  M \left(\frac{1}{N}\right)^{9/4} \Sigma_1
   \left(6 a^2+  {r_0}^2\right) \log (  {r_0})}{27 \pi  \left(18 b^4-3 b^2-1\right)^5  \log N ^2   {N_f}   {r_0}^2
   \alpha _{\theta _2}^3 \left(9 a^2+  {r_0}^2\right)} (r-  {r_0})^2\right|_{r=r_0},\nonumber\\
& & f_r(r) = -\frac{99 \sqrt{\frac{3}{2}} \beta  {g_s}^{3/2} M \sqrt[5]{\frac{1}{N}} {N_f} {r_0} \alpha _{\theta _1}^6
   {f_{x^{10}x^{10}}}({r_0}) \log ^2({r_0})}{2 \pi ^{3/2} \alpha _{\theta _2}^5} \nonumber\\
& &  = - \frac{2 \left(9 b^2+1\right)^4 b^{10} M   \left(6 a^2+{r_0}^2\right){r_0}^2\Sigma_1}{3 \pi
   \left(-18 b^4+3 b^2+1\right)^4 \log N  N^{8/15} {N_f} \left(-27 a^4+6 a^2 {r_0}^2+{r_0}^4\right) \alpha _{\theta
   _2}^3}\nonumber\\
& & +{\cal C}_{zz}^{\rm th}-2 {\cal C}_{\theta_1z}^{\rm th}+2 {\cal C}_{\theta_1x}^{\rm th},\nonumber\\
& & f_{\theta_1\theta_1}(r) = f_{\theta_1\theta_1}(r_0) = 0,\nonumber\\
& & f_{\theta_1\theta_2}(r) = f_{\theta_1\theta_2}(r_0) = 0,\nonumber\\
& & f_{\theta_1x}(r) = -\frac{99 \sqrt{\frac{3}{2}} {g_s}^{3/2} M \sqrt[5]{\frac{1}{N}} {N_f} {r_0} \alpha _{\theta _1}^6
   {f_{x^{10}x^{10}}}({r_0}) \log ^2({r_0})}{4 \pi ^{3/2} \alpha _{\theta _2}^5}-{f_{x^{10}x^{10}}}({r_0}) \nonumber\\
& &  = - \frac{\left(9 b^2+1\right)^4 b^{10} M  \left(6 a^2+{r_0}^2\right){r_0}^2
   \Sigma_1}{3 \pi  \left(-18 b^4+3 b^2+1\right)^4 \log N  N^{8/15} {N_f} \left(-27 a^4+6 a^2
   {r_0}^2+{r_0}^4\right) \alpha _{\theta _2}^3}+{\cal C}_{\theta_1x}^{\rm th},\nonumber\\
& & f_{\theta_1y}(r) =  f_{\theta_1y}(r_0) = 0,\nonumber\\
& &  f_{\theta_1z} = \frac{539 \pi ^3 N^{2/5} \alpha _{\theta _2}^2 {f_{x^{10}x^{10}}}({r_0})}{1728 {g_s}^3 M^2 {N_f}^2 \log
   ^2({r_0})}-\frac{185 {f_{x^{10}x^{10}}}({r_0})}{108} \nonumber\\
& &  = \frac{16 \left(9 b^2+1\right)^4 b^{12}  \beta  \left(19683
   \sqrt{3} \alpha _{\theta _1}^6+3321 \sqrt{2} \alpha _{\theta _2}^2 \alpha _{\theta _1}^3-40 \sqrt{3} \alpha _{\theta _2}^4\right)}{243
   \pi ^3 \left(1-3 b^2\right)^{10} \left(6 b^2+1\right)^8 {g_s}^{9/4} \log N ^4 N^{7/6} {N_f}^3 \left(-27 a^4 {r_0}+6 a^2
   {r_0}^3+{r_0}^5\right) \alpha _{\theta _1}^7 \alpha _{\theta _2}^6}+{\cal C}_{\theta_1z}^{\rm th},\nonumber\\
& & f_{\theta_2\theta_2}(r) = f_{\theta_2\theta_2}(r_0) = 0,\nonumber\\
& & f_{\theta_2x}(r) = f_{\theta_2x}(r_0) \nonumber\\
& & =  \frac{16 \left(9 b^2+1\right)^4 b^{12}\left(19683 \sqrt{3}
   \alpha _{\theta _1}^6+3321 \sqrt{2} \alpha _{\theta _2}^2 \alpha _{\theta _1}^3-40 \sqrt{3} \alpha _{\theta _2}^4\right)}{243 \pi ^3 \left(1-3
   b^2\right)^{10} \left(6 b^2+1\right)^8 {g_s}^{9/4} \log N ^4 N^{7/6} {N_f}^3 \left(-27 a^4 {r_0}+6 a^2
   {r_0}^3+{r_0}^5\right) \alpha _{\theta _1}^7 \alpha _{\theta _2}^6}+{\cal C}_{\theta_2x}^{\rm th},\nonumber\\
& & f_{\theta_2y}(r) = \frac{352 {g_s}^3 M^2 \left(\frac{1}{N}\right)^{2/5} {N_f}^2 {f_{x^{10}x^{10}}}({r_0}) \log ^2({r_0})}{49 \pi
   ^3 \alpha _{\theta _2}^2}-2 {f_{x^{10}x^{10}}}({r_0}) \nonumber\\
& & = \frac{3 b^{10} \left(9 b^2+1\right)^4 M \beta \left(6 a^2+{r_0}^2\right)  \log
   ({r_0}) \Sigma_1}{\pi  \left(3 b^2-1\right)^5 \left(6 b^2+1\right)^4 \log N ^2 N^{7/5} {N_f} \left(9 a^2+{r_0}^2\right) \alpha
   _{\theta _2}^3} ,\nonumber\\
& & f_{\theta_2z}(r) = f_{\theta_2z}(r_0) = \frac{3 \left(9 b^2+1\right)^4 b^{10} M  \left(6 a^2+{r_0}^2\right)  \log
   ({r_0})}{\pi  \left(3 b^2-1\right)^5 \left(6 b^2+1\right)^4 {\log N}^2 N^{7/6} {N_f} \left(9 a^2+{r_0}^2\right) \alpha
   _{\theta _2}^3}\nonumber\\
& & \hskip 1.5in \times \left(19683 \sqrt{6} \alpha _{\theta _1}^6+6642 \alpha _{\theta _2}^2 \alpha _{\theta _1}^3-40 \sqrt{6} \alpha _{\theta
   _2}^4\right) +{\cal C}_{\theta_2 z}^{\rm th},\nonumber\\
\end{eqnarray*}
\begin{eqnarray}
\label{consistency-f_MN's}
& & f_{xx}(r) = f_{xx}(r_0) = 0,\nonumber\\
& & f_{xy}(r) = f_{xy}(r_0) = \frac{3 \left(9 b^2+1\right)^4 b^{10} M  \left(6 a^2+{r_0}^2\right)  \log
   ({r_0}) \alpha _{\theta _2}^3\Sigma_1}{\pi  \left(3 b^2-1\right)^5 \left(6 b^2+1\right)^4 \log N ^2 N^{21/20} {N_f} \left(9
   a^2+{r_0}^2\right) \alpha _{\theta _{2 l}}^6}+{\cal C}_{xy}^{\rm th},\nonumber\\
& & f_{xz}(r) = f_{xz}(r_0) = \frac{18 b^{10} \left(9 b^2+1\right)^4 M \beta  \left(6 a^2+{r_0}^2\right)
   \log ^3({r_0}) \Sigma_1}{\pi  \left(3b^2-1\right)^5 \left(6 b^2+1\right)^4 \log N ^4 N^{5/4} {N_f} \left(9 a^2+{r_0}^2\right) \alpha
   _{\theta _2}^3},\nonumber\\
& & f_{yy}(r) = \frac{N^{2/5} {f_{x^{10}x^{10}}}({r_0}) \left(32 \sqrt{6} \pi ^{3/2} {g_s}^{3/2} M {N_f} \alpha _{\theta
   _2}^3-4851 \pi ^3 {r_0} \alpha _{\theta _1}^4 \alpha _{\theta _2}^2\right)}{7776 {g_s}^3 M^2 {N_f}^2
   {r_0} \alpha _{\theta _1}^4 \log ^2({r_0})}+\frac{55 {f_{x^{10}x^{10}}}({r_0})}{27}+{f_{\theta_1y}}({r_0})\nonumber\\
& &  =  - \frac{3 b^{10} \left(9 b^2+1\right)^4 M \left(\frac{1}{N}\right)^{7/4} \beta  \left(6 a^2+{r_0}^2\right) \log ({r_0})\Sigma_1
   }{\pi  \left(3 b^2-1\right)^5 \left(6 b^2+1\right)^4 \log N ^2 {N_f} {r_0}^2 \left(9
   a^2+{r_0}^2\right) \alpha _{\theta _2}^3},\nonumber\\
& &  f_{yz}(r) = \frac{\pi ^{3/2} N^{2/5} \alpha _{\theta _2}^3 {f_{x^{10}x^{10}}}({r_0})}{81 \sqrt{6} {g_s}^{3/2} M {N_f}
   {r_0} \alpha _{\theta _1}^4 \log ^2({r_0})}+\frac{{f_{\theta_1y}}({r_0})}{2}\nonumber\\
& &  = \frac{64 \left(9 b^2+1\right)^8 b^{22} M \left(\frac{1}{N}\right)^{29/12}  \left(6 a^2+{r_0}^2\right)
   \log ({r_0}) }{27 \pi ^4 \left(3 b^2-1\right)^{15} \left(6 b^2+1\right)^{12}
   {g_s}^{9/4} \log N ^6  {N_f}^4 {r_0}^3 \left({r_0}^2-3 a^2\right) \left(9 a^2+{r_0}^2\right)^2 \alpha
   _{\theta _1}^7 \alpha _{\theta _2}^9}\nonumber\\
& & \hskip -0.3in \times \left(387420489 \sqrt{2} \alpha _{\theta _1}^{12}+87156324 \sqrt{3}
   \alpha _{\theta _2}^2 \alpha _{\theta _1}^9+5778054 \sqrt{2} \alpha _{\theta _2}^4 \alpha _{\theta _1}^6-177120 \sqrt{3} \alpha _{\theta
   _2}^6 \alpha _{\theta _1}^3+1600 \sqrt{2} \alpha _{\theta _2}^8\right) \nonumber\\
   & & +{\cal C}_{yz}^{\rm th},\nonumber\\
& & f_{zz}(r) = \frac{539 \pi ^3 N^{2/5} \alpha _{\theta _2}^2 {f_{x^{10}x^{10}}}({r_0})}{864 {g_s}^3 M^2 {N_f}^2 \log
   ^2({r_0})} = -\frac{b^{10} \left(9 b^2+1\right)^4 M{r_0}^2 \log ({r_0})
   \Sigma_1}{27 \pi ^{3/2} \left(3 b^2-1\right)^5 \left(6 b^2+1\right)^4 \sqrt{{g_s}} \log N ^2 N^{23/20} {N_f} \alpha
   _{\theta _2}^5} \nonumber\\
   & & \hskip 2.5in + {\cal C}_{zz}^{\rm th},\nonumber\\
& & f_{x^{10}x^{10}}(r) = f_{x^{10}x^{10}}(r_0) \nonumber\\
& & = -\frac{27 b^{10} \left(9 b^2+1\right)^4 M \left(\frac{1}{N}\right)^{5/4} \beta  \left(6 a^2+{r_0}^2\right)
    \log ^3({r_0}) \Sigma_1}{\pi  \left(3 b^2-1\right)^5 \left(6 b^2+1\right)^4 \log N ^4
   {N_f} {r_0}^2 \left(9 a^2+{r_0}^2\right) \alpha _{\theta _2}^3}.
\end{eqnarray}
From (\ref{consistency-f_MN's}), near $r=r_0$, one sees one can set:
\begin{equation}
\label{f_MN's-zero}
f_t = f = f_{\theta_i\theta_j} = f_{\theta_1y} = f_{xx} = 0,
\end{equation}
and for $r_0$-dependent values of ${\cal C}_{\theta_2x}^{\rm th}, {\cal C}_{\theta_2z}^{\rm th}, {\cal C}_{xy}^{\rm th}$ determined by (\ref{consistency-f_MN's}),
\begin{equation}
\label{f_MN's-zero-ii}
f_{\theta_2x} = f_{\theta_2z} = f_{xy} = 0.
\end{equation}
which is what has been used in our calculations. By matching near $r=r_0$ of $f_{r}(r), f_{\theta_1x}, f_{\theta_1z}, f_{zz}$, one can solve for ${\cal C}_{\theta_1x}^{\rm th}, {\cal C}_{\theta_1z}^{\rm th}, {\cal C}_{zz}^{\rm th}, f_{x^{10}x^{10}}(r_0)$. 

The function $f_{x^{10}x^{10}}$ can independently be determined from:
\begin{itemize}
\item
{\it matching $f_{\theta_2y}$}: Substituting $b = \frac{1}{\sqrt{3}} + \epsilon$ \cite{OR4-Yadav+Misra}, \cite{MChPT} this yields
\begin{equation}
\label{f1111-theta2y}
 f_{x^{10}x^{10}}(r_0) \sim - \frac{M \Sigma_1(\alpha_{\theta_1}, \alpha_{\theta_2}) \log r_0}{N^{\frac{7}{5}}\log^2N N_f\epsilon^5\alpha_{\theta_2}^3}.
\end{equation}

\item
{\it matching $f_{yy}$}: This yields
\begin{equation}
\label{f1111-yy}
 f_{x^{10}x^{10}}(r_0) \sim -\frac{g_s^{\frac{3}{2}}M^2\log^3r_0\alpha_{\theta_1}^4\Sigma_1(\alpha_{\theta_1}, \alpha_{\theta_2})}{\epsilon^5\log^Nr_0\alpha_{\theta_2}^6N^{\frac{43}{20}}}.
\end{equation}
One can show that (\ref{f1111-theta2y}) and (\ref{f1111-yy}) are mutually consistent provided the IR cut-off
$r_0$:
\begin{equation}
\label{r_0}
r_0 = W\left(\frac{1}{2 \sqrt{a}}\right)\approx a e^{-\frac{4 \log ^2(-\log (a))}{\log ^2(a)}} \log
   ^2(a),
\end{equation}
$W$ being the Lambert's product log function and $a\equiv \frac{81 \sqrt{6} {g_s}^{3/2} M
   \left(\frac{1}{N}\right)^{3/4} {N_f} \alpha
   _{\theta _1}^4}{\pi ^{3/2} \alpha _{\theta
   _2}^3}$. 

\item
{\it matching $f_{x^{10}x^{10}}$}: This yields
\begin{equation}
\label{f1111-1111}
f_{x^{10}x^{10}}(r_0) \sim -\frac{M \log^3r_0\Sigma_1(\alpha_{\theta_1}, \alpha_{\theta_2})}{\epsilon^5 r_0^2\log^N N_f \alpha_{\theta_2}^3 N^{\frac{5}{4}}}.
\end{equation}
Numerically, e.g., for $M=N_f=3, N=10^2,g_s=0.1$ and ${\cal O}(1)$ $\alpha_{\theta_{1,2}}$, one can show that (\ref{f1111-1111}) is consistent with (\ref{f1111-theta2y}) and (\ref{f1111-yy}). 

\item
{\it matching $f_t(r)$}
\begin{eqnarray}
\label{f_r}
& & -\frac{2 \left(9 b^2+1\right)^4 b^{10} M {r_0}^2 \Sigma _1 \left(6 a^2+{r_0}^2\right)}{3 \pi  \left(-18 b^4+3
   b^2+1\right)^4 \log N  N^{8/15} {N_f} \alpha _{\theta _2}^3 \left(6 a^2 {r_0}^2-27
   a^4+{r_0}^4\right)}+{\cal C}_{zz}^{\rm th}-2 {\cal C}_{\theta_1z}^{\rm th}+2 {\cal C}_{\theta_1x}^{\rm th}  \nonumber \\
 & &
 =\frac{99 \sqrt{\frac{3}{2}} {g_s}^{3/2}
   M \sqrt[5]{\frac{1}{N}} {N_f} {r_0} \alpha _{\theta _1}^6 {f_{x^{10}x^{10}}}({r_0}) \log ^2({r_0})}{2 \pi
   ^{3/2} \alpha _{\theta _2}^5}.
\end{eqnarray}

\item
{\it matching $f_{\theta_1x}$}
\begin{eqnarray}
\label{f_theta1x}
& & {{\cal C}^{\rm th}_{\theta_{1} x}}-\frac{b^{10} \left(9 b^2+1\right)^4 M {r_0}^2 \Sigma _1 \left(6 a^2+{r_0}^2\right)}{3 \pi 
   \left(-18 b^4+3 b^2+1\right)^4 \log N  N^{8/15} {N_f} \alpha _{\theta _2}^3 \left(6 a^2 {r_0}^2-27
   a^4+{r_0}^4\right)} \nonumber \\
 & &
   =-\frac{99 \sqrt{\frac{3}{2}} {g_s}^{3/2} M \sqrt[5]{\frac{1}{N}} {N_f} {r_0}
   \alpha _{\theta _1}^6 {f_{x^{10}x^{10}}}({r_0}) \log ^2({r_0})}{4 \pi ^{3/2} \alpha _{\theta
   _2}^5}-{f_{x^{10}x^{10}}}({r_0}).
\end{eqnarray}

\item
{\it matching $f_{\theta_1z}$}
\begin{eqnarray}
\label{f_theta1z}
& & \frac{8 \sqrt{2} \left(9 b^2+1\right)^4 b^{12} \Sigma _1}{243 \pi ^3 \left(1-3 b^2\right)^{10} \left(6 b^2+1\right)^8
   {g_s}^{9/4} \log N ^4 N^{7/6} {N_f}^3 \alpha _{\theta _1}^7 \alpha _{\theta _2}^6 \left(6 a^2
   {r_0}^3-27 a^4 {r_0}+{r_0}^5\right)}+{{\cal C}^{\rm th}_{\theta_1 z}} \nonumber \\
   & &
   =\frac{539 \pi ^3 N^{2/5} \alpha _{\theta _2}^2
   {f_{x^{10}x^{10}}}({r_0})}{1728 {g_s}^3 M^2 {N_f}^2 \log ^2(r)}.
\end{eqnarray}

\item
{\it matching $f_{zz}$}
\begin{eqnarray}
\label{f_zz}
& & \hskip -0.7in  {\cal C}_{zz}^{\rm th}-\frac{b^{10} \left(9 b^2+1\right)^4 M {r_0}^2 \Sigma _1 \log ({r_0})}{27 \pi ^{3/2} \left(3
   b^2-1\right)^5 \left(6 b^2+1\right)^4 \sqrt{{g_s}} \log N ^2 N^{23/20} {N_f} \alpha _{\theta
   _2}^5}=\frac{539 \pi ^3 N^{2/5} \alpha _{\theta _2}^2 {f_{x^{10}x^{10}}}({r_0})}{864 {g_s}^3 M^2 {N_f}^2 \log
   ^2({r_0})}.
\end{eqnarray}

\item
One can solve (\ref{f_theta1x}) - (\ref{f_zz}) to obtain:
\begin{eqnarray}
\label{f_zz-f_theta1x-f_theta1z}
& & {\cal C}_{zz}^{\rm th} \sim \frac{ \left(\frac{1}{N}\right)^{3/4} {r_0}^2 \Sigma _1}{\epsilon ^5
   {g_s}^{7/2} \log N ^2 M {N_f}^3 \alpha _{\theta _2}^3 \log ({r_0})}\nonumber\\
& & {\cal C}_{\theta_1z}^{\rm th} \sim \frac{ \left(\frac{1}{N}\right)^{3/4} {r_0}^2 \Sigma _1}{2 \epsilon ^5
   {g_s}^{7/2} \log N ^2 M {N_f}^3 \alpha _{\theta _2}^3 \log ({r_0})}\nonumber\\
& &{\cal C}_{\theta_1x}^{\rm th} \sim \frac{\left(\frac{1}{N}\right)^{7/6} \Sigma _1}{ \epsilon ^{11} {g_s}^{9/4}
   \log N ^4 {N_f}^3 {r_0}^5 \alpha _{\theta _1}^7 \alpha _{\theta _2}^6}. 
\end{eqnarray}
This thus confirms that ${\cal C}_{zz}^{\rm th} - 2 {\cal C}_{\theta_1z}^{\rm th} + 2{\cal C}_{\theta_1x}^{\rm th}  = 2{\cal C}_{\theta_1x}^{\rm th}< 0$ (as $\Sigma_1<0$), which in \cite{MChPT} was argued by requiring compatibility with phenomenological value of the 1-loop renormalized LEC appearing in the
${\cal O}(p^4)$ $SU(3)\ \chi$PT $\left(\nabla_\mu U^\dagger\nabla_\mu U\right)^2$. 
\end{itemize}

\end{document}